\documentclass[11pt]{article}
\usepackage{amsmath, amsfonts, stmaryrd,setspace, url, hyperref, verbatim, xcolor, cite, graphicx, todonotes, cancel, slashed, mathtools,breqn}
\usepackage[T1]{fontenc}
\usepackage[utf8]{inputenc}
\usepackage{fullpage}
\usepackage{tikz}
\usetikzlibrary{arrows,positioning,fit,shapes,snakes,backgrounds,tikzmark}
\usepackage{pgfplots}
\usetikzlibrary{shapes.geometric,calc}

\newcommand{\dd}{\mathrm{d}}
\newcommand{\D}{\mathcal{D}}
\newcommand{\be}{\begin{equation}}
\newcommand{\ee}{\end{equation}}
\numberwithin{equation}{section}

\newcommand{\hmu}{\hat{\mu}}
\newcommand{\hnu}{\hat{\nu}} 
\newcommand{\hrho}{\hat{\rho}}
\newcommand{\hlambda}{\hat{\lambda}}
\newcommand{\hsigma}{\hat{\sigma}}

\newcommand{\gM}{\mathcal{M}}
\newcommand{\cH}{\mathcal{H}}

\newcommand{\Cdef}{\boldsymbol{C}}
\newcommand{\Fdef}{\boldsymbol{F}} 
\newcommand{\Celeven}{\hat{C}}
\newcommand{\Feleven}{\hat{F}}
\newcommand{\geleven}{\hat{g}}

\newcommand{\Aa}{\mathcal{A}}
\newcommand{\Ab}{\mathcal{B}}
\newcommand{\Ac}{\mathcal{C}}
\newcommand{\Ad}{\mathcal{D}}

\newcommand{\Fa}{\mathcal{F}}
\newcommand{\Fb}{\mathcal{H}}
\newcommand{\Fc}{\mathcal{J}}
\newcommand{\Fd}{\mathcal{K}}

\newcommand{\Gthree}{\mathrm{SL}(3) \times \mathrm{SL}(2)}
\newcommand{\Gfour}{\mathrm{SL}(5)}

\newcommand{\Gsix}{E_{6(6)}}

\newcommand{\Geight}{E_{8(8)}}
\newcommand{\Edd}{E_{d(d)}}

\newcommand{\fM}{\mathcal{M}}
\newcommand{\fN}{\mathcal{N}}
\newcommand{\fP}{\mathcal{P}}
\newcommand{\fQ}{\mathcal{Q}}
\newcommand{\fK}{\mathcal{K}}
\newcommand{\fL}{\mathcal{L}}

\definecolor{vub}{RGB}{0,52,154}
\definecolor{vubo}{RGB}{255,102,0}
\definecolor{redd}{RGB}{255,40,40}
\definecolor{r}{RGB}{228,32,20}
\definecolor{o}{RGB}{238,69,4}
\definecolor{y}{RGB}{253,228,1}
\definecolor{g}{RGB}{108,160,0}
\definecolor{b}{RGB}{0,162,203}
\definecolor{i}{RGB}{120,42,117}

\usepackage[utf8]{inputenc}
\usepackage[framemethod=tikz]{mdframed}
\usetikzlibrary{calc}
\usepackage{chngcntr}

\counterwithin{ex}{section}

\makeatletter
\def\mdf@@mynote{}
\define@key{mdf}{mynote}{\def\mdf@@mynote{#1}}

\mdfdefinestyle{mystate}{
skipabove={1\baselineskip},
linewidth=0.5pt,
innertopmargin=1\baselineskip,
roundcorner=10pt,
backgroundcolor=black!0,
linecolor=black,
singleextra={
  \node[xshift=10pt,draw=black,fill=white,rounded corners,anchor=west] at (P-|O) %
  {\strut{\itshape  }\ifdefempty{\mdf@@mynote}{}{\itshape\bfseries \mdf@@mynote}};
},
firstextra={
  \node[xshift=10pt,draw=black,fill=white,rounded corners,anchor=west] at (P-|O) %
  {\strut{\itshape  }\ifdefempty{\mdf@@mynote}{}{\itshape\bfseries \mdf@@mynote}};
}
}

\mdfdefinestyle{mystateb}{
skipabove={1\baselineskip},
linewidth=1pt,
innertopmargin=1\baselineskip,
roundcorner=10pt,
backgroundcolor=white!5,
linecolor=vub!50,
singleextra={
  \node[xshift=10pt,thick,draw=vub!50,fill=b!0,rounded corners,anchor=west] at (P-|O) %
  {\strut{\itshape  }\ifdefempty{\mdf@@mynote}{}{\bf\mdf@@mynote}};
},
firstextra={
  \node[xshift=10pt,thick,draw=vub!50,fill=b!0,rounded corners,anchor=west] at (P-|O) %
  {\strut{\itshape  }\ifdefempty{\mdf@@mynote}{}{\bf\mdf@@mynote}};
}
}

\mdfdefinestyle{mystater}{
skipabove={1\baselineskip},
linewidth=1pt,
innertopmargin=1\baselineskip,
roundcorner=10pt,
backgroundcolor=white!5,
linecolor=r!50,
singleextra={
  \node[xshift=10pt,thick,draw=r!50,fill=b!0,rounded corners,anchor=west] at (P-|O) %
  {\strut{\itshape  }\ifdefempty{\mdf@@mynote}{}{\bf\mdf@@mynote}};
},
firstextra={
  \node[xshift=10pt,thick,draw=r!50,fill=b!0,rounded corners,anchor=west] at (P-|O) %
  {\strut{\itshape  }\ifdefempty{\mdf@@mynote}{}{\bf\mdf@@mynote}};
}
}

\mdfdefinestyle{mystateg}{
skipabove={1\baselineskip},
linewidth=1pt,
innertopmargin=1\baselineskip,
roundcorner=10pt,
backgroundcolor=white!5,
linecolor=g!50,
singleextra={
  \node[xshift=10pt,thick,draw=g!50,fill=g!0,rounded corners,anchor=west] at (P-|O) %
  {\strut{\itshape  }\ifdefempty{\mdf@@mynote}{}{\bf\mdf@@mynote}};
},
firstextra={
  \node[xshift=10pt,thick,draw=g!50,fill=g!0,rounded corners,anchor=west] at (P-|O) %
  {\strut{\itshape  }\ifdefempty{\mdf@@mynote}{}{\bf\mdf@@mynote}};
}
}

\newmdenv[style=mystate,nobreak=true]{state}
\newmdenv[style=mystater,nobreak=true]{stater}
\newmdenv[style=mystateg,nobreak=true]{stateg}
\newmdenv[style=mystateb,nobreak=true]{stateb}

\makeatother

\makeatother

\newcommand{\uA}{\mathsf{A}}
\newcommand{\uB}{\mathsf{B}}
\newcommand{\uC}{\mathsf{C}}

\def\bri{\bar{\imath}}
\def\brj{\bar{\jmath}}

\begin{document}

\title{
\bf A non-relativistic limit of M-theory and 11-dimensional membrane Newton-Cartan geometry\\
}

\author{ Chris\, D.  A.\,  Blair${}^{1}$\,, Domingo Gallegos${}^{2}$\,, Natale Zinnato${}^{2}$}
\date{ }
\maketitle
\begin{center}
{\textit{
${}^{1}$Theoretische Natuurkunde, Vrije Universiteit Brussel, and the International Solvay
Institutes, \\Pleinlaan 2, B-1050 Brussels, Belgium\\
\vspace{3pt}
${}^{2}$Institute for Theoretical Physics, Utrecht University,
Leuvenlaan 4, 3584 CE Utrecht, The Netherlands
\vspace{3pt}
~\\}}
\texttt{christopher.blair@vub.be\,,\, a.d.gallegospazos@uu.nl\,,\, n.zinnato@uu.nl }\\
\end{center}
\begin{abstract}
\noindent
We consider a non-relativistic limit of the bosonic sector of eleven-dimensional supergravity, leading to a theory based on a covariant `membrane Newton-Cartan' (MNC) geometry.
The local tangent space is split into three `longitudinal' and eight `transverse' directions, related only by Galilean rather than Lorentzian symmetries.
This generalises the ten-dimensional stringy Newton-Cartan (SNC) theory.
In order to obtain a finite limit, the field strength of the eleven-dimensional four-form is required to obey a transverse self-duality constraint, ultimately due to the presence of the Chern-Simons term in eleven dimensions.
The finite action then gives a set of equations that is invariant under longitudinal and transverse rotations, Galilean boosts and local dilatations. We supplement these equations with an extra Poisson equation, coming from the subleading action. 
Reduction along a longitudinal direction gives the known SNC theory with the addition of RR gauge fields, while reducing along a transverse direction yields a new non-relativistic theory associated to D2 branes. 
We further show that the MNC theory can be embedded in the U-duality symmetric formulation of exceptional field theory, demonstrating that it shares the same exceptional Lie algebraic symmetries as the relativistic supergravity, and providing an alternative derivation of the extra Poisson equation.
\end{abstract}

\maketitle

\tableofcontents

\section{Introduction and summary}

A great deal has been learned about string theory from the exploration of special limits of the theory.
There are many examples. 
In the $\alpha^\prime \rightarrow 0$ limit, string theory predicts Einstein gravity, extended to supergravity in ten dimensions, via the 1-loop beta functionals of the worldsheet \cite{Callan:1985ia}.
When compactified on a circle of radius $R$, T-duality relates the $R \rightarrow 0$ limit of one string theory to the $R \rightarrow \infty$ limit of another.
The strong coupling limit of the type IIA theory leads to the eleven-dimensional description in terms of M-theory, from the perspective of which we can view all the different dual versions of 10-dimensional string theories again as different limits \cite{Hull:1994ys, Witten:1995ex}. Another limit of M-theory is its low energy effective theory, the eleven-dimensional supergravity \cite{Cremmer:1978km}.

Another interesting class of limits are those which decouple degrees of freedom, and which may again lead to new geometric perspectives or to different dual descriptions (the most famous example being the original derivation of the AdS/CFT correspondence \cite{Maldacena:1997re}).
In this paper, we will focus on `critical' limits of string theory and M-theory, in which the 10- or 11-dimensional geometry becomes \emph{non-relativistic} \cite{Gomis:2000bd, Danielsson:2000gi,Gomis:2004pw,Gomis:2005pg}. 
Our goal is to study the low energy effective description of M-theory in such a limit.

This builds on a recent revival of interest in non-relativistic versions of string theory, see e.g. \cite{Andringa:2012uz,Christensen:2013lma,Christensen:2013rfa,Hartong:2015zia,Batlle:2016iel,Harmark:2017rpg,Bergshoeff:2018yvt,Harmark:2018cdl,Gomis:2019zyu,Gallegos:2019icg,Harmark:2019upf,Bergshoeff:2019pij,Blair:2019qwi,Yan:2019xsf,Roychowdhury:2019qmp,Gomis:2020fui,Gallegos:2020egk,Bergshoeff:2021bmc}. This has been inspired in part by holographic set-ups in which non-relativistic geometries make an appearance, but also intrinsically motivated by the desire to explore new, and potentially simpler, corners of the space of possible quantum gravitational theories.
The target space geometry that appears extends the generally covariant but non-relativistic `Newton-Cartan' (NC) geometry \cite{Cartan1,Cartan2} to what can be called a `stringy Newton-Cartan' (SNC) geometry.
The full ten-dimensional Lorentz symmetry is absent, and there is instead a split into two `longitudinal' directions (including time) and eight purely spatial `transverse' directions which transform into each other only under Galilean boosts. 
Correspondingly one can describe the target space geometry in terms of a pair of mutually orthogonal vielbeins, $\tau_\mu{}^{\uA}$,  $h^\mu{}_a$, such that $\tau_\mu{}^{\uA} h^\mu{}_a = 0$, where $\uA=0,1$ indexes the longitudinal tangent space directions and $a=2,\dots,9$ indexes the transverse tangent space directions.
Introducing longitudinal and transverse flat metrics, one can instead work with degenerate metric-like objects.
In addition, the usual gauge fields such as the string two-form will propagate.
The beta functionals, (bosonic) background field equations and target space action of these geometries have been studied in \cite{Gomis:2019zyu,Yan:2019xsf,Bergshoeff:2019pij,Gallegos:2019icg,Gallegos:2020egk,Bergshoeff:2021bmc}. 

Non-relativistic stringy geometries may be related via T-duality on a longitudinal direction to relativistic string theory with a null isometry \cite{Bergshoeff:2018yvt,Harmark:2018cdl}.
This fact underlies the existence of a straightforward embedding of these theories in the formally T-duality symmetric formulation of supergravity known as double field theory (DFT) \cite{Siegel:1993xq,Siegel:1993th, Hull:2009mi}.
Here the metric and $B$-field are treated as components of a \emph{generalised metric}. It was previously realised in \cite{Lee:2013hma,Ko:2015rha,Morand:2017fnv} that this generalised metric admits `non-Riemannian parametrisations' in which, instead of an invertible metric, a degenerate metric structure (for instance of Newton-Cartan type) appears along with the $B$-field.
This was exploited in \cite{Cho:2019ofr,Gallegos:2020egk} to study the equations of motion and action of non-relativistic strings and other theories of non-Riemannian nature.

Now, the initial investigations of stringy non-relativistic limits \cite{Gomis:2000bd, Danielsson:2000gi,Gomis:2004pw,Kamimura:2005rz} were firmly embedded within the M-theoretic duality web, and provided a variety of non-relativistic limits involving different branes as well as the duality relationships between them. 
We would expect to be able to understand this more completely by constructing the full covariant extended Newton-Cartan geometries and low energy effective dynamics i.e. the non-relativistic limits of 11- and 10-dimensional supergravity.
Note that lower dimensional examples of non-relativistic supergravities have been studied in \cite{Andringa:2013mma,Bergshoeff:2015uaa}.

From the perspective of duality symmetric formulations, the route to eleven-dimensional non-relativistic supergravity was partially provided in \cite{Berman:2019izh} using exceptional field theory (ExFT) \cite{Berman:2010is,Hohm:2013vpa, Hohm:2013uia,Hohm:2014fxa} (see \cite{Berman:2020tqn} for a recent review of both ExFT and DFT). ExFT provides a formally U-duality symmetric formulation of 11- and 10-dimensional SUGRA, and generalises much of the structure of DFT.
In \cite{Berman:2019izh} examples were provided of non-Riemannian parametrisations of generalised metrics that would describe non-relativistic 11-dimensional geometries, however with only a partial analysis of the resulting dynamics.
Separately, more general $p$-brane Newton-Cartan geometries have been studied from the worldvolume perspective in \cite{Kamimura:2005rz,Kluson:2017djw,Kluson:2017abm,Kluson:2019uza}, with in particular \cite{Kamimura:2005rz,Kluson:2019uza} focusing on the M2 case that reappears in this paper (also considered in passing in \cite{Blair:2020ops} using insights from ExFT). A recent study of spacetime aspects is \cite{Pereniguez:2019eoq}.

In this paper we will restrict to the bosonic sector of 11-dimensional SUGRA, and exhibit a non-relativistic limit giving a `membrane Newton-Cartan' (MNC) geometry in eleven dimensions (this is the covariant version of the flat space `Galilean membrane' (GM) limit of \cite{Gomis:2000bd}).
Although we were initially inspired by the ExFT perspective of \cite{Berman:2019izh}, deriving this geometry and the (bosonic) dynamics of the theory turns out to be remarkably straightforward if one starts with an appropriate ansatz for the metric and three-form, inspired by the form a $1/c$ expansion, and generalising both the flat non-relativistic membrane limit of \cite{Gomis:2000bd} and the very recent construction of the NSNS SNC spacetime action in \cite{Bergshoeff:2021bmc}.

The eleven-dimensional theory we construct has a number of interesting features:
\begin{itemize}
\item \emph{Membrane Newton-Cartan geometry (see section \ref{expandSetup}).} The geometry has three `longitudinal' and eight `transverse' directions, which we can describe in terms of an eleven-dimensional Newton-Cartan metric structure. 
This appears by taking the eleven-dimensional metric and its inverse to have the form
\be
\begin{split}
\label{MNCexp_split}
\geleven_{\mu \nu}& = c^2 \eta_{AB}  \tau_{\mu}{}^A \tau_{\nu}{}^B +  c^{-1}  H^{}_{\mu \nu} 
\,,\\
\geleven^{\mu\nu}  & = c H^{\mu\nu} + c^{-2} \eta^{AB} \tau^\mu{}_A \tau^\nu{}_B 
\,,\\
\end{split}
\ee
where $A=0,1,2$ labels the \emph{longitudinal} Newton-Cartan vielbeins, or clock forms, $\tau_\mu{}^A$, and $H^{\mu\nu}$ and $\tau^\mu{}_A$ are projective inverses obeying the Newton-Cartan completeness relations
\be
H^{\mu\rho} H_{\rho \nu} + \tau^\mu{}_A \tau_\nu{}^A = \delta^\mu_\nu
\,\quad
H^{\mu\nu} \tau_\nu{}^A = 0 = H_{\mu\nu} \tau^\nu{}_A 
\,,\quad
\tau^\mu{}_A \tau_\mu{}^B = \delta_A^B \,.
\label{intro_complete}
\ee
We also expand the three-form as
\be
\Celeven_{\mu \nu \rho}  = -c^3 \epsilon_{A B C}  \tau_{\mu}{}^A  \tau_{\nu}{}^B  \tau_{\rho}{}^C  +   C_{\mu \nu \rho} 
+ c^{-3} \widetilde C_{\mu\nu\rho} \,.
\label{Cexp}
\ee
Here $c$ is a dimensionless parameter whose $c\rightarrow \infty$ limit can be interpreted as a non-relativistic limit. It is the geometry that results from this limit that we refer to as membrane Newton-Cartan.
The powers of $c$ in \eqref{MNCexp_split}, along with the leading order power in \eqref{Cexp}, follow the pattern of the powers of the harmonic function in the M2 brane supergravity solution, which is a generic feature of $p$-brane non-relativistic limits \cite{Gomis:2000bd,Gomis:2004pw}.
(The minus sign in the $c^3$ term in \eqref{Cexp} is a choice of convention, and matches with e.g. expressions in the SNC literature on dimensional reduction \cite{Bergshoeff:2021bmc}.)

\item \emph{Transverse self-duality (see section \ref{expandAction}).} Requiring singular terms to cancel in the $c\rightarrow \infty$ limit requires that the finite part $F_{\mu\nu\rho\sigma}=4\partial_{[\mu}C_{\nu\rho\sigma]}$ of the four-form field strength obey a \emph{self-duality constraint} in the eight-dimensional transverse space. This is a consequence of the presence of the Chern-Simons term in the eleven-dimensional action.

\item \emph{Dual degrees of freedom (see section \ref{dualF7}).} The subleading part $\widetilde C_{\mu\nu\rho}$ of the three-form in the expansion appears in the dynamics with its equation of motion imposing the self-duality constraint. The anti-self-dual transverse projection of the field strength $\widetilde F_{\mu\nu\rho\sigma}=4\partial_{[\mu}\widetilde C_{\nu\rho\sigma]}$ of this subleading part can be identified with the totally longitudinal part of the seven-form field strength dual to $F_{\mu\nu\rho\sigma}$. 
Hence the non-relativistic limit involves what would normally be physical and dual degrees of freedom, however, rather than being related to each other as would usually be the case, these degrees of freedom get reorganised into separately self- and anti-self-dual parts. 

\item \emph{Dilatation invariance and a `missing' equation of motion (see section \ref{DilMissing}).} The eleven-dimensional theory is invariant under a `dilatation' symmetry which scales each field with a weight inherited from the power of $c$ that accompanies them in the initial expansion. This is an `emergent' local symmetry \cite{Bergshoeff:2021bmc} and it has the effect of removing a variational degree of freedom when we vary the finite part of the action. 
Hence, at this order, there is a `missing' equation of motion.
This is a familiar feature of non-relativistic theories, with the naively missing equation corresponding to the Poisson equation for the Newtonian gravitational potential. However, we can identify this missing equation by looking at the next order in the $1/c$ expansion \cite{VandenBleeken:2017rij,Hansen:2018ofj,Hansen:2020pqs,Bergshoeff:2021bmc}.
Indeed, here we identify this missing equation by extracting it from the dilatation variation of the action at the next subleading order. In parallel with the situation in the DFT description of the NSNS sector \cite{Gallegos:2020egk}, we also find it directly from the equations of motion of the exceptional field theory description. 

\item \emph{Boost invariance  (see section \ref{11DBoost}).} The eleven-dimensional theory is also invariant under Galilean boost transformations of the form
\be 
\delta H_{\mu\nu}   = 2 \Lambda_{(\mu}{}^A \tau_{\nu) A}\,,  \quad \delta \tau^\mu{}_A   = - H^{\mu\nu} \Lambda_{\nu A}\,,  \quad\delta C_{\mu\nu\rho} = - 3 \epsilon_{ABC}\Lambda_{[\mu}{}^A \tau_\nu{}^B \tau_{\rho]}{}^C\,, 
\label{boosts}
\ee
where the (infinitesimal) boost parameter $\Lambda_{\mu}{}^A$ satisfies $\tau^\mu{}_A\Lambda_{\mu}{}^B = 0$.
The slightly unusual feature here is the transformation of the three-form itself.
This transformation \eqref{boosts} is to be expected based on similar observations in the case of stringy Newton-Cartan. There one can either introduce additional one-form gauge fields transforming under boosts, and treat the two-form gauge field as invariant, or else absorb the former into the latter via a sort of Stueckelberg gauge fixing \cite{Harmark:2019upf, Bergshoeff:2021bmc}.
We do not introduce additional one-forms and so generalise this second picture. 

\item \emph{Reduction to type IIA SNC  (see section \ref{IIASNC}).} Reduction of the theory on a longitudinal isometry direction produces the full type IIA SNC theory, coupling the known NSNS sector to RR fields.
\item \emph{Reduction to type IIA D2NC (see section \ref{D2NC})} Reduction of the theory on a transverse isometry directions produces a novel type IIA non-relativistic theory, that can be associated to D2 branes rather than strings.

\item \emph{Exceptional field theory embedding and U-duality (see section \ref{AndExFT}).} 
Finally, the 11-dimensional MNC theory can be very naturally embedded within exceptional field theory (which also manifestly breaks Lorentz invariance and treats original and dual degrees of freedom together), demonstrating that the same exceptional Lie algebraic structures that appear in the relativistic theory are preserved by the limit.
Furthermore, we can easily use ExFT to study transformations between relativistic and non-relativistic geometries, and to obtain equations of motion which are otherwise missing from the action of the non-relativistic theory.
The achievement of ExFT is to present a unified treatment of both 11- and 10-dimensional supergravities in which $\Edd$ symmetry is manifest. 
The metric and gauge field degrees of freedom are reorganised into $\Edd$ multiplets.
For instance, the wholly $d$-dimensional components of the metric and three-form (and possibly also of the dual six-form) appear in a \emph{generalised metric}.
For the cases $d=3,4$, this has an expression 
\be
\gM_{MN} = |\hat g|^{1/(9-d)}
\begin{pmatrix} 
\hat g_{ij} + \frac{1}{2} \hat C_i{}^{pq} \hat C_{jpq} &\hat C_i{}^{kl} \\
 \hat C_k{}^{ij} &  2 \hat g^{i[k} \hat g^{l]j} 
\end{pmatrix}\,.
\ee
If we adopt the same expansion as in equation \eqref{MNCexp_split}, then in the limit $c \rightarrow \infty$, we obtain an alternative \emph{non-relativistic} or \emph{non-Riemannian} parametrisation\footnote{The flat space limit of \eqref{MNCexp_split} was already studied in exceptional field theory in \cite{Berman:2019izh}, and the general non-Riemannian parametrisation of the $\Gfour$ generalised metric worked out - this can be shown to be equivalent to \eqref{genmetnonrie_firstlook}. However a full analysis of the Newton-Cartan interpretation and dynamics was not carried out.}
\be
\small
\gM_{MN} = \Omega^{\tfrac{2}{9-d}}
\begin{pmatrix} 
H_{ij} - \epsilon_{ABC} \tau_{(i|}{}^A C_{j)kl} \tau^{k B} \tau^{l C} +  C_{ikl} C_{jmn} H^{ km} \tau^{ln}  & - \epsilon_{ABC} \tau_i{}^A \tau^{k B} \tau^{l C} + 2 C_{ipq} H^{p[k} \tau^{l]q}\\
 - \epsilon_{ABC} \tau_k{}^A \tau^{i B} \tau^{j C} + 2 C_{kpq} H^{p[i} \tau^{l]j} & 2H^{ i[k} \tau^{l]j} + 2\tau^{i[k} H^{ l]j}
\end{pmatrix} 
\label{genmetnonrie_firstlook} 
\ee
where $\Omega$ is a measure factor, and $\tau^{ij} \equiv \tau^i{}_A \tau^j{}_B \eta^{AB}$.
This alternative parametrisation then changes the nature of the duality relationships encoded by the dynamics of the generalised gauge fields of exceptional field theory. 
This allows the exceptional field theory formulation to automatically capture the interesting reorganisation of degrees of freedom implied by the non-relativistic limit. 
In addition, the missing equation of motion is associated to variations which do not preserve the non-relativistic nature of the parametrisation \eqref{genmetnonrie_firstlook} of the generalised metric.

\end{itemize}

The outline of this paper is very simple.
In section \ref{expansion} we carry out the expansion at the level of the bosonic action.
In section \ref{11DEom} we discuss the equations of motion and symmetries.
In section \ref{dimRed}, we carry out dimensional reductions to type IIA. 
In section \ref{AndExFT}, we discuss the embedding in ExFT. 
We discuss our conclusions and conclude our discussions in section \ref{Conc}.
The appendix contains additional calculational details.

\section{Membrane Newton-Cartan limit and 11-dimensional SUGRA}
\label{expansion}

\subsection{Setting up the expansion} 
\label{expandSetup}

\paragraph{Metric}

We start by writing the 11-dimensional metric and its inverse as
\be
\geleven_{\mu\nu}  = c^2 \tau_{\mu\nu} + c^{-1} H_{\mu\nu}\,,\quad
\geleven^{\mu\nu}  = c H^{\mu\nu} + c^{-2} \tau^{\mu\nu}  
\,.
\label{gExp}
\ee
We can view this simply as a field redefinition which introduces the 11-dimensional Newton-Cartan variables alongside the (dimensionless) parameter $c$. We will seek to send $c$ to infinity and interpret the result as a non-relativistic limit. In principle, we can also think of this ansatz as containing the first terms in an infinite expansion in $c^{-3}$, and we will occasionally allow such a perspective to influence our presentation. However, we leave the development of the full non-relativistic expansion to future work.
To see that the field redefinition \eqref{gExp} makes sense in Newton-Cartan terms we look at the condition $\delta_\mu^\nu = \geleven_{\mu\rho} \geleven^{\rho\nu}$, which gives at order $c^3$, $c^0$ and $c^{-3}$ respectively the following three conditions:  
\be
\tau_{\mu \rho} H^{\rho \nu} = 0 \,,\quad
\tau_{\mu\rho} \tau^{\rho \nu} + H_{\mu\rho} H^{\rho \nu} = \delta^\nu_\mu \,,\quad H_{\mu\rho} \tau^{\rho \nu} = 0\,.
\label{complete}
\ee
We view these as the defining conditions for $\tau_{\mu\nu}$, viewed as a longitudinal Newton-Cartan metric (of Lorentzian signature), and $H^{\mu\nu}$, viewed as the corresponding orthogonal transverse Newton-Cartan metric (of Euclidean signature).\footnote{As in the stringy Newton-Cartan case, we could choose to include additional one-forms in the expansion \eqref{gExp}, however these can be eliminated by a Stueckelberg gauge fixing \cite{Harmark:2019upf, Bergshoeff:2021bmc}.}
Letting $A=0,1,2$ and $a=1,\dots 8$ denote longitudinal and transverse flat indices, respectively, we can introduce projective vielbeins such that
\be
\tau_{\mu\nu} \equiv \tau_\mu{}^A \tau_\nu{}^B \eta_{AB} \,,\quad
\tau^{\mu\nu} \equiv \tau^\mu{}_A \tau^\nu{}_B \eta^{AB} \,,\quad \tau^\mu{}_A \tau_\mu{}^B = \delta_A^B \,,
\ee
\be
H^{\mu\nu} \equiv h^\mu{}_a h^\nu{}_b \delta^{ab}\,,\quad
H_{\mu\nu} \equiv h^a{}_\mu h^b{}_\nu \delta_{ab} \,, \quad h^\mu{}_a h^b{}_\mu = \delta_a^b\,,
\ee
and hence obeying the Newton-Cartan completeness relations following from \eqref{complete}.
Here $\eta_{AB}$ is the flat three-dimensional Minkowski metric and $\delta_{ab}$ is the flat Euclidean 8-dimensional metric.
We can then compute the determinant of the 11-dimensional metric:
\be
\det \geleven_{\mu\nu}
 =- c^{-2} \Omega^2\,, \quad
\Omega^2  \equiv - \tfrac{1}{3! 8!}  \epsilon^{\mu_1 \dots \mu_{11}} \epsilon^{\nu_1 \dots \nu_{11}} \tau_{\mu_1 \nu_1 } \tau_{\mu_2\nu_2} \tau_{\mu_3 \nu_3} H_{\mu_4\nu_4} \dots H_{\mu_{11} \nu_{11}} \,,
\label{defOmega}
\ee
where $\epsilon^{\mu_1 \dots \mu_{11}}$ denotes the 11-dimensional Levi-Civita symbol.
Hence $\sqrt{-\geleven} = c^{-1}\Omega$ and it is $\Omega$ which will be used as the measure factor in the non-relativistic action.
In terms of the vielbeins, we can write
\be
\Omega = \Big|\tfrac{1}{3!8!} \epsilon^{\mu \nu \rho \sigma_1 \dots \sigma_{8}} \epsilon_{ABC} \epsilon_{a_1 \dots a_8} \tau_{\mu}{}^A \tau_\nu{}^B \tau_\rho{}^C h^{a_1}{}_{\sigma_1} \dots h^{a_8}{}_{\sigma_8}\Big|
\label{defOmegaViel}
\ee
and note that
\be
\partial_\mu \ln\Omega = \tau^\nu{}_A \partial_\mu \tau_\nu{}^A + h^\nu{}_a \partial_\mu h^a{}_\nu\,.
\label{derivOmega}
\ee
We can obtain further useful identities by substituting the expressions \eqref{gExp} into contractions of the Levi-Civita symbol and the metric.
One that we will use later is 
\be
n! H^{[\mu_1| \nu_1} \dots H^{|\mu_n] \nu_n} 
= - \tfrac{\epsilon^{\mu_1 \dots \mu_n \lambda_1 \dots \lambda_{11-n}} \epsilon^{\nu_1 \dots \nu_n \sigma_1 \dots \sigma_{11-n}}}{3! (8-n)! \Omega^2}  \tau_{\lambda_1 \sigma_1} \dots \tau_{\lambda_3\sigma_3} H_{\lambda_4 \sigma_4} \dots H_{\lambda_{11-n} \sigma_{11-n}} \,.
\label{generalHn}
\ee

\paragraph{Three-form} For the three-form, let
\be
\Celeven_3 =  C_3 - \tfrac{1}{6} c^3 \epsilon_{ABC} \tau^A \wedge \tau^B \wedge \tau^C + c^{-3}  \widetilde C_3 \,,
\ee
so that
\be \label{FExp}
\Feleven_4 =  F_4 -  \tfrac{1}{2} c^3 \epsilon_{ABC} d \tau^A \wedge \tau^B \wedge \tau^C + c^{-3} \widetilde F_4 \,,
\ee
where 
\be
 F_4 \equiv d C_3\,,\quad 
\widetilde F_4 \equiv d \widetilde C_3 \,.
\ee
Although $\widetilde C_3$ is subleading, it will explicitly appear in the action and dynamics of the non-relativistic limit.
Its equation of motion will impose a self-duality constraint on $F_4$, and we will be able to identify a certain projection of its field strength with the totally longitudinal components of the dual seven-form field strength.
We can therefore interpret the subleading part of $\hat C_3$ as being `dual' to the finite part. 
This is clearly a general fact: the Hodge star itself has an expansion in powers of $c$ and so inevitably mixes up the terms at different powers of $c$ in any $p$-form it acts on.
What is non-trivial is that the Chern-Simons term of the 11-dimensional theory will lead to both $C_3$ and $\widetilde C_3$ playing a role in the non-relativistic limit.

\subsection{Expanding the action}
\label{expandAction}

The action for the eleven-dimensional metric and three-form is
\be
\begin{split}
S = \int \dd^{11} x  &\left( \sqrt{|\geleven|}\left[ \hat R( \geleven ) - \tfrac{1}{48} \Feleven^{\mu\nu\rho\sigma} \Feleven_{\mu\nu\rho\sigma} \right] + \tfrac{1}{144^2} \epsilon^{\mu_1\dots\mu_{11}} \Feleven_{\mu_1\dots\mu_4} \Feleven_{\mu_5\dots\mu_8} \Celeven_{\mu_9\mu_{10} \mu_{11} }\right)\,.
\end{split} 
\label{Seleven}
\ee
Here $\Feleven_4 = d \Celeven_3$.
In form notation the Chern-Simons term is $\tfrac{1}{6}\Feleven_4 \wedge \Feleven_4 \wedge \Celeven_3$, the equation of motion of the three-form is $d \hat\star \Feleven_4 = \tfrac{1}{2} \Feleven_4 \wedge \Feleven_4$ and its Bianchi identity is $d \Feleven_4 = 0$. 
The Hodge dual field strength is $\Feleven_7 = \hat\star \Feleven_4$, which obeys the Bianchi identity $d \Feleven_7 = \tfrac{1}{2} \Feleven_4 \wedge \Feleven_4$ and the equation of motion $d \hat\star \Feleven_7 = 0$.

\paragraph{Chern-Simons term}

We start with the expansion of the Chern-Simons term.
Leaving wedge products implicit, we can simply compute
\be
\begin{split}
\tfrac{1}{6} \Feleven_4  \Feleven_4  \Celeven_3 & = 
\tfrac{1}{6}  F_4  F_4  C_3 -\tfrac{1}{6}(3 c^3  F_4  F_4  + 6  F_4 \widetilde F_4 ) \tfrac{1}{6} \epsilon_{ABC} \tau^A  \tau^B  \tau^C 
\\ & \qquad - \tfrac{1}{3} d \left( c^3  F_4  C_3  \tfrac{1}{6} \epsilon_{ABC} \tau^A  \tau^B  \tau^C + \tfrac{1}{6} \epsilon_{ABC} \tau^A  \tau^B  \tau^C(   F_4 \widetilde C_3 +  C_3 \widetilde F_4 ) \right) 
+ \mathcal{O}(c^{-3}) \,.
\end{split}
\label{expandCS}
\ee
We drop the total derivative.

\paragraph{Kinetic term for three-form}

First, let's write the component expression
\be
\hat F_{\mu_1\mu_2\mu_3\mu_4} = - 6 c^3 T_{[\mu_1 \mu_2}{}^A \tau_{\mu_3}{}^B \tau_{\mu_4]}{}^C\epsilon_{ABC} + F_{\mu_1\mu_2\mu_3\mu_4} +c^{-3} \widetilde F_{\mu_1\mu_2\mu_3\mu_4} 
\label{Findex}
\ee  
where we introduce the Newton-Cartan torsion
\be
T_{\mu\nu}{}^A \equiv 2 \partial_{[\mu} \tau_{\nu]}{}^A \,.
\label{defTorsion}
\ee
Any term involving three $H^{\mu\nu}$ contracting the first term in \eqref{Findex} vanishes as one $H^{\mu\nu}$ must necessarily contract a $\tau_\mu{}^A$.
As a result,
\be
\begin{split}
& \sqrt{|\geleven|} \geleven^{\mu_1 \mu_4} \dots \geleven^{\mu_4 \nu_4} \Feleven_{\mu_1 \dots \mu_4} \Feleven_{\nu_1 \dots \nu_4} 
 \\ & 
 = 
  \Omega 
 \Big(
c^{3} \left( H^{\mu_1 \nu_1} H^{\mu_2 \nu_2} H^{	\mu_3 \nu_3} H^{\mu_4 \nu_4}  F_{\mu_1 \mu_2 \mu_3 \mu_4}  F_{\nu_1 \nu_2 \nu_3 \nu_4}
- 12 H^{\mu_1\nu_1} H^{\mu_2\nu_2} T_{\mu_1\mu_2}{}^A T_{\nu_1\nu_2}{}^B \eta_{AB}\right)
\\ &\qquad\qquad\quad
- 24 H^{\mu\nu} T_{\mu\rho}{}^A T_{\nu\sigma}{}^B \tau^\rho{}_A \tau^\sigma{}_B 
-12H^{\mu_1 \nu_1} H^{\mu_2 \nu_2}  F_{\mu_1 \mu_2 \mu_3 \mu_4} T_{\nu_1 \nu_2}{}^A \tau^{\mu_3 B} \tau^{\mu_4 C} \epsilon_{ABC}
\\ &\qquad\qquad\quad
+4 H^{\mu_1 \nu_1} H^{\mu_2 \nu_2} H^{\mu_3 \nu_3} \tau^{\mu_4 \nu_4}  F_{\mu_1 \mu_2 \mu_3 \mu_4}  F_{\nu_1 \nu_2 \nu_3 \nu_4}
\\ & \qquad\qquad\qquad +2 H^{\mu_1 \nu_1} H^{\mu_2 \nu_2} H^{\mu_3 \nu_3} H^{\mu_4 \nu_4}  F_{\mu_1 \mu_2 \mu_3 \mu_4} \widetilde F_{\nu_1 \nu_2 \nu_3 \nu_4}
\Big)+ \mathcal{O}(c^{-3})\,.
\end{split}
\label{expandF2}
\ee

\paragraph{Kinetic term/Chern-Simons cancellations and self-duality} 

We now examine the $\mathcal{O}(c^3)$ terms in \eqref{expandCS} and \eqref{expandF2} which involve a field strength $F_4$, as well as the $\mathcal{O}(c^0)$ terms involving the subleading $\widetilde F_4$.
These cannot possibly be cancelled by a contribution from the expansion of the Ricci scalar.
The relevant terms are:
\be
\begin{split} 
-\tfrac{1}{2 \cdot 4!}&\Omega H^{\mu_1 \nu_1} H^{\mu_2 \nu_2} H^{	\mu_3 \nu_3} H^{\mu_4 \nu_4}  F_{\mu_1 \mu_2 \mu_3 \mu_4}( c^{3}   F_{\nu_1 \nu_2 \nu_3 \nu_4} + 2\widetilde F_{\nu_1 \nu_2\nu_3\nu_4} )
\\ & 
- \tfrac{1}{2\cdot 4!4! 3!} \epsilon^{\mu_1 \dots \dots \mu_{11}}   F_{\mu_1 \mu_2 \mu_3 \mu_4} ( c^3   F_{\mu_5 \mu_6 \mu_7 \mu_8} + 2 \widetilde F_{\mu_5 \mu_6 \mu_7 \mu_8} ) \epsilon_{ABC} \tau_{\mu_9}{}^A \tau_{\mu_{10}}{}^B \tau_{\mu_{11}}{}^C 
\end{split}
\label{singularF}
\ee
To cancel the terms at order $c^3$, we are led to require the following constraint:
\be
\Omega H^{\mu_1 \nu_1} H^{\mu_2 \nu_2} H^{	\mu_3 \nu_3} H^{\mu_4 \nu_4}  F_{\nu_1 \nu_2 \nu_3 \nu_4}
= 
-  \tfrac{1}{4! 3!}\epsilon^{\mu_1 \dots \dots \mu_{11}}  F_{\mu_5 \mu_6 \mu_7 \mu_8}  \epsilon_{ABC} \tau_{\mu_9}{}^A \tau_{\mu_{10}}{}^B \tau_{\mu_{11}}{}^C 
\,.
\label{constraint_appear}
\ee
This says that the totally transverse part of $ F_{\mu\nu\rho\sigma}$ is self-dual (or anti-self-dual).
This is self-consistent thanks to \eqref{generalHn}.
We will refer to this as the self-duality constraint.


\paragraph{Three-form equation of motion}

As a sanity check that requiring the constraint \eqref{constraint_appear} is sensible and necessary, let us at this point also take the limit at the level of the equation of motion of the three-form gauge field. 
We will revisit the equations of motion, including that of the metric, in more detail in section \ref{11DEom}. 
For the three-form, we have originally:
\be
\partial_\sigma ( \sqrt{|\geleven|}\geleven^{\mu \lambda_1 } \geleven^{\nu \lambda_2} \geleven^{\rho \lambda_3} \geleven^{\sigma \lambda_4} \Feleven_{\lambda_1 \dots \lambda_4} )
 = \tfrac{1}{2 \cdot 4! 4!} \epsilon^{\mu\nu\rho \sigma_1 \dots \sigma_8} \Feleven_{\sigma_1 \dots \sigma_4}\Feleven_{\sigma_5 \dots \sigma_8} \,.
\ee
Inserting the expansion, one has firstly at $\mathcal{O}(c^{3})$ that
\be
\partial_\sigma \left(
\Omega
 H^{\mu \lambda_1} H^{\nu \lambda_2} H^{\rho \lambda_3} H^{\sigma \lambda_4}   F_{\lambda_1 \dots \lambda_4}  
\right)
 = - \tfrac{1}{3! 4!} \epsilon^{\mu\nu\rho\sigma\sigma_1\dots \sigma_7}\partial_\sigma ( F_{\sigma_1 \dots \sigma_4} \epsilon_{ABC} \tau_{\sigma_5}{}^A \tau_{\sigma_6}{}^B \tau_{\sigma_7}{}^C )\,,
\label{derivConstraint}
\ee
which is the duality relation \eqref{constraint_appear} under a derivative.

At $\mathcal{O}(c^0)$ we have the finite equation of motion 
\be
\begin{split} 
\partial_\sigma & \Big(
\Omega \big( 
4 H^{[\mu| \lambda_1} H^{|\nu| \lambda_2} H^{|\rho|\lambda_3} \tau^{|\sigma] \lambda_4}  F_{\lambda_1 \dots \lambda_4} 
 - 6 H^{[\mu|\lambda_1} H^{|\nu|\lambda_2}  \tau^{|\rho |B} \tau^{|\sigma] C} T_{\lambda_1 \lambda_2}{}^A \epsilon_{ABC}  
\\ & \qquad\qquad+ H^{\mu \lambda_1} H^{\nu \lambda_2} H^{\rho \lambda_3} H^{\sigma \lambda_4} \widetilde F_{\lambda_1 \dots \lambda_4} 
\big) 
\Big) 
\\ & 
=  \tfrac{1}{2 \cdot 4! 4!} \epsilon^{\mu\nu\rho \sigma_1 \dots \sigma_8} (  F_{\sigma_1 \dots \sigma_4}  F_{\sigma_5 \dots \sigma_8} 
- 12 \epsilon_{ABC} T_{\sigma_1 \sigma_2}{}^A \tau_{\sigma_3}{}^B \tau_{\sigma_4}{}^C \widetilde F_{\sigma_5 \dots \sigma_8} 
) 
\,.
\end{split}
\label{3eomFinite}
\ee
This will be reproduced from the action that we find below.

\paragraph{Ricci scalar} 

Now we come to the Ricci scalar.
A very quick way to take the limit is to start with the explicit expression for the Ricci scalar in terms of the metric and its derivatives:
\be
\begin{split}
\hat R & = 
\frac{1}{4} \geleven^{\mu \nu} \partial_\mu \geleven_{\rho \sigma} \partial_\nu \geleven^{\rho \sigma} 
- \frac{1}{2} \geleven^{\mu \nu} \partial_\nu \geleven^{\rho \sigma} \partial_{\rho} \geleven_{\mu \sigma}\\
& \qquad- \frac{1}{4} \geleven^{\mu \nu} \partial_\mu \ln \geleven \, \partial_\nu \ln \geleven
- \geleven^{\mu \nu} \partial_\mu \partial_\nu \ln \geleven - \partial_\mu \ln \geleven \, \partial_\nu \geleven^{\mu \nu}
- \partial_\mu \partial_\nu \geleven^{\mu \nu} \,.
\end{split}
\ee
Calculating the expansion is trivial. One has $\hat R  = c^4 R^{(4)} + c R^{(0)} + \mathcal{O}(c^{-2})$ with
\be
\begin{split}
R^{(4)} & = 
\frac{1}{4} H^{\mu \nu}  \partial_\nu H^{\rho \sigma} \partial_\mu \tau_{\rho \sigma}
- \frac{1}{2} H^{\mu \nu} \partial_\nu H^{\rho \sigma} \partial_{\rho} \tau_{\mu \sigma}
\,,\\
R^{(0)} & = 
\frac{1}{4} H^{\mu \nu} ( \partial_\mu \tau_{\rho \sigma} \partial_\nu \tau^{\rho \sigma} + \partial_\mu H_{\rho \sigma} \partial_\nu H^{\rho \sigma}) 
+ \frac{1}{4} \tau^{\mu\nu} \partial_\mu \tau_{\rho \sigma} \partial_\nu H^{\rho \sigma} 
\\ & \qquad\quad
- \frac{1}{2} H^{\mu \nu} \partial_\nu \tau^{\rho \sigma} \partial_{\rho} \tau_{\mu \sigma}
- \frac{1}{2} H^{\mu \nu} \partial_\nu H^{\rho \sigma} \partial_{\rho}H_{\mu \sigma}
- \frac{1}{2} \tau^{\mu \nu} \partial_\nu H^{\rho \sigma} \partial_{\rho} \tau_{\mu \sigma}
\\ & \qquad\quad-  H^{\mu \nu} \partial_\mu \ln \Omega \, \partial_\nu \ln \Omega
-  2 H^{\mu \nu} \partial_\mu \partial_\nu \ln \Omega - 2 \partial_\mu \ln \Omega \, \partial_\nu H^{\mu \nu}
- \partial_\mu \partial_\nu H^{\mu \nu} \,.
\label{expandRicci}
\end{split}
\ee
Recall that the measure $\sqrt{-\geleven}$ introduces a further power of $c^{-1}$.
The singular piece can be easily rewritten as
\be
\begin{split} 
R^{(4)} &= 
   - \tfrac{1}{2} H^{\mu\nu} H^{\rho\sigma} (  \partial_\mu \tau_\rho{}^A \partial_\nu \tau_\sigma{}^B - \partial_\rho \tau_\mu{}^A \partial_\nu \tau_\sigma{}^B )\eta_{AB} 
= -  \tfrac{1}{4} H^{\mu\nu} H^{\rho \sigma} T_{\mu\rho}{}^A T_{\nu\sigma}{}^B \eta_{AB}\,.
\end{split} 
\label{RicciSing}
\ee
This cancels exactly the remaining singular term appearing in the expansion \eqref{expandF2} of the kinetic term for the three-form.
An entirely similar cancellation appeared in the NSNS sector expansion of \cite{Bergshoeff:2021bmc}, and as noted there is reminiscent of what happens when taking the Gomis-Ooguri limit on the string worldsheet.
In the conclusions in section \ref{Conc} we discuss the comparison with this limit in more detail.

\subsection{Result of expansion and covariant formulation} 

\paragraph{Action and constraint} 

Combining \eqref{expandCS}, \eqref{expandF2} and \eqref{expandRicci} we obtain the expansion of the 11-dimensional SUGRA action in the form $S = c^3 S^{(3)} + c^0 S^{(0)} + \dots$.
The singular part is:
\be
S^{(3)} =  -\int \dd^{11} x \, \tfrac{1}{2 \cdot 4!} F_{\mu_1 \dots \mu_4} \left( \Omega H^{\mu_1 \nu_1} \dots H^{\mu_4 \nu_4} + \tfrac{1}{4!3!} \epsilon^{\mu_1 \dots \mu_4 \nu_1 \dots \nu_7} \epsilon_{ABC} \tau_{\nu_5}{}^A \tau_{\nu_6}{}^B \tau_{\nu_7}{}^C \right) F_{\nu_1 \dots \nu_4} \,,
\label{S3}
\ee
and in order to have a good $c \rightarrow \infty$ limit, we impose the constraint
\be
\Omega H^{\mu_1 \nu_1} H^{\mu_2 \nu_2} H^{	\mu_3 \nu_3} H^{\mu_4 \nu_4}  F_{\nu_1 \nu_2 \nu_3 \nu_4}
= 
-  \tfrac{1}{4! 3!}\epsilon^{\mu_1 \dots \dots \mu_{11}}  F_{\mu_5 \mu_6 \mu_7 \mu_8}  \epsilon_{ABC} \tau_{\mu_9}{}^A \tau_{\mu_{10}}{}^B \tau_{\mu_{11}}{}^C \,,
\label{magic}
\ee
to ensure that $S^{(3)}$ vanishes.\footnote{Strictly speaking this is a sufficient condition for the vanishing of $S^{(3)}$, as we could alternatively integrate by parts and use \eqref{derivConstraint}. However the full constraint \eqref{magic} will follow from the expansion of the metric equations of motion that we discuss in section \ref{ExpandEom}, as well as in the expansion of the dual field strength discussed in section \ref{dualF7}, and also follows directly from the exceptional field theory formulation of section \ref{AndExFT}.} 
The finite part of the action is: 
\be
\begin{split}
S^{(0)} = \int \dd^{11} x \,\Omega &
\Big( R^{(0)}  
+ \tfrac{1}{2} H^{\mu\nu} T_{\mu\rho}{}^A T_{\nu\sigma}{}^B \tau^\rho{}_A \tau^\sigma{}_B 
\\ &\quad
- \tfrac{1}{12} H^{\mu_1 \nu_1} H^{\mu_2 \nu_2} H^{\mu_3 \nu_3} \tau^{\mu_4 \nu_4}  F_{\mu_1 \mu_2 \mu_3 \mu_4}  F_{\nu_1 \nu_2 \nu_3 \nu_4}
\\ &\quad
 + \tfrac{1}{4}H^{\mu_1 \nu_1} H^{\mu_2 \nu_2}  F_{\mu_1 \dots \mu_4} \epsilon_{ABC} T_{\nu_1\nu_2}{}^A \tau^{\mu_3 B} \tau^{\mu_4 C} 
\\ & \quad 
-\tfrac{1}{\cdot 4!} \widetilde F_{\nu_1 \nu_2\nu_3\nu_4}
\big(  H^{\mu_1 \nu_1} H^{\mu_2 \nu_2} H^{\mu_3 \nu_3} H^{\mu_4 \nu_4}  F_{\mu_1 \mu_2 \mu_3 \mu_4}
\\ & \qquad\qquad\qquad\qquad + \tfrac{1}{4! 3!\Omega} \epsilon^{\nu_1\nu_2\nu_3\nu_4 \mu_1 \dots  \mu_{7}}   F_{\mu_1 \mu_2 \mu_3 \mu_4}  \epsilon_{ABC} \tau_{\mu_5}{}^A \tau_{\mu_{6}}{}^B \tau_{\mu_{7}}{}^C 
\big)
\Big) 
\\ & + \tfrac{1}{6}  F_4 \wedge  F_4 \wedge  C_3 \,,
\end{split}
\label{resultaction}
\ee
where $R^{(0)}$ is as defined in \eqref{expandRicci}.
The equation of motion of $C_{\mu\nu\rho}$ gives exactly \eqref{3eomFinite}, and we will discuss the equations of motion of the Newton-Cartan fields in detail in section \ref{11DEom}.
The equation of motion of $\widetilde C_{\mu\nu\rho}$ is \eqref{derivConstraint}, giving the constraint under a derivative. 
Alternatively if we were just to take the action \eqref{resultaction} at face value, forgetting about its origin via an expansion of the three-form, we could make the choice to view $\widetilde F_{\mu\nu\rho\sigma}$ as an independent field, serving as a Lagrange multiplier imposing the constraint in its form \eqref{magic}.

\paragraph{Symmetries} 

The action is diffeomorphism invariant (as follows from the covariant rewriting we carry out below), as well as gauge invariant under $\delta C_3 = d \lambda_2$, $\delta \widetilde C_3 = d \widetilde \lambda_2$.
The vielbeins $h^a{}_\mu$ and $\tau^A{}_\mu$ transform under $\mathrm{SO}(8)$ and $\mathrm{SO}(1,2)$ rotational symmetries respectively, which are also symmetries of the action. 
The non-relativistic theory is also invariant under Galilean boosts and a dilatation symmetry. 

The Galilean boosts mix the longitudinal and transverse degrees of freedom.
The parameter for such a boost is denoted $\Lambda_a{}^A$. Letting $\Lambda_\mu{}^A \equiv h^a{}_\mu \Lambda_a{}^A$ such that $\tau^{\mu}{}_A \Lambda_\mu{}^B = 0$, we can give the (infinitesimal) action of these symmetries as
\be
\delta_\Lambda H_{\mu\nu}  = 2 \Lambda_{(\mu}{}^A \tau_{\nu) A}\,,\quad 
\delta_\Lambda \tau^\mu{}_A  = - H^{\mu\nu} \Lambda_{\nu A} \,,\quad
\delta_\Lambda C_{\mu\nu\rho} =  - 3 \epsilon_{ABC}\Lambda_{[\mu}{}^A \tau_\nu{}^B \tau_{\rho]}{}^C \,.
\label{action_boost}
\ee
The action $S^{(0)}$ is invariant under these transformations on using the self-duality constraint.
One way for the action to be exactly invariant would be to treat $\widetilde{F}_{\mu\nu\rho\sigma}$ as an independent field transforming as $\delta_\Lambda \widetilde{F}_{\mu\nu\rho\sigma} = - 4 \Lambda_{[\mu}{}^A {F}_{\nu\rho\sigma]\lambda} \tau^\lambda{}_A$, or to have $\widetilde{C}_{\mu\nu\rho}$ transform in a way leading to this transformation.

The dilatations are meanwhile induced by the expansion in powers of $c$, with the dilatation weight of each field equal to the power of $c$ which accompanies it in the expansion.
The (infinitesimal) action of dilatations is hence:
\be
\delta_\lambda H^{  \mu   \nu} =  + \lambda    H^{  \mu   \nu}\,,\quad
\delta_\lambda H_{\mu\nu} = -\lambda H_{\mu\nu} \,,\quad
\delta_\lambda \tau^{  \mu}{}_{  A} =-\lambda    \tau^{  \mu}_{  A}\,,\quad
\delta_\lambda \tau_\mu{}^A = + \lambda \tau_\mu{}^A \,,\quad
\delta_\lambda C_{\mu\nu\rho} =0\,.
\label{dil_all}
\ee 
Note $\delta\Omega = -\lambda   \Omega$.
For $\lambda$ coordinate dependent this is a symmetry of the action $S^{(0)}$ on using the self-duality constraint \eqref{magic}. 
If we treat $\widetilde{F}_{\mu\nu\rho\sigma}$ as an independent field transforming as $\delta_\lambda \widetilde{F}_{\mu\nu\rho\sigma} = - 3 \lambda \widetilde{F}_{\mu\nu\rho\sigma}$, then the action $S^{(0)}$ is exactly invariant.
We will explicitly verify the invariance of the action and study these symmetries in more detail in section \ref{11DEom}.

\paragraph{Newton-Cartan connections and covariant rewriting} 
The way we obtained the action \eqref{resultaction} was by a straightforward computation at the level of the metric and three-form.
To better understand the result, we rewrite the action in a covariant way by introducing the following connection 
\be
\Gamma^\rho_{\mu\nu} = \tau^{\rho}{}_A \partial_\mu \tau_\nu{}^A +\frac{1}{2} H^{\rho\sigma} \left( \partial_{\mu} H_{\sigma\nu}+\partial_{\nu} H_{\mu\sigma}-\partial_{\sigma} H_{\mu \nu}\right)\,,
\label{defConn}
\ee
whose covariant derivative we denote by $\nabla_\mu$.
This satisfies the following metric compatibility conditions:
\be
\nabla_{\rho} H^{\mu\nu} = 0 = \nabla_\rho \tau_{\mu}{}{}^A \,,
\ee
though it is not the unique solution.\footnote{Here $\nabla$ acts only on the curved indices. It would also be possible to define a connection covariant under local $SO(1,2)$ transformations by replacing the partial derivative $\partial_\mu \tau_\nu{}^A$ term with a spin covariant derivative.} 
The antisymmetric component of \eqref{defConn} is the torsion \eqref{defTorsion}:
\be
\Gamma^\rho_{[\mu\nu]} = \frac{1}{2} \tau^\rho{}_A T_{\mu\nu}{}^A\,.
\ee
It is also useful to define the `acceleration' and its trace
\begin{equation} 
 {a}_{  \mu}{}^{  A  B} \equiv   -  {\tau}^{ \rho A}   T_{ \rho \mu}^{  B}\,,\qquad   a_{ \mu}\equiv  {a}_{  \mu}{}^{  A  B}\eta_{  A  B}\,,
\end{equation}
as well as its symmetric traceless component
\begin{equation} 
{a}_{ \mu}{}^{\{A B\}} \equiv  {a}_{\mu}{}^{(A B)} - \frac{1}{d_L}\eta^{ A B}{a}_{ \mu}\,,
\end{equation} 
where $d_L$ is the dimension of the longitudinal space (which is $d_L=3$ here, but we will also use this notation in the reduction to the $d_L=2$ case of SNC in section \ref{IIASNC}). The final tensor that will appear is the extrinsic curvature defined by
\begin{equation}
\mathcal{K}_{\mu\nu A} = \frac{1}{2} \mathcal{L}_{\tau^{\rho A}} H_{\mu\nu} \,, \qquad\qquad \mathcal{K}_A \equiv H^{\mu\nu} \mathcal{K}_{\mu\nu A}\,,
\end{equation} 
and obeying the following useful identities
\begin{equation}
\tau^{\mu (A} \mathcal{K}_{\mu\nu}{}^{B)} =0, \qquad \qquad  \nabla_\mu \tau^{\nu A} = H^{\nu\rho}\mathcal{K}_{\mu\rho}{}^A\,.
\end{equation}
Finally, let's introduce some notation to make the expressions more compact.
Given an arbitrary tensor $M_{\mu\nu}$ carrying lower indices, we will employ for convenience the following short-hand notation:
\be
M^{\mu\nu} \equiv H^{\mu\rho} H^{\nu\sigma} M_{\rho\sigma} \,,\quad M_{AB} \equiv \tau^{\mu}{}_A\tau^{\nu}{}_B M_{\mu\nu} \,,\quad  \nabla_\rho M_{AB} \equiv \nabla_\rho\left( \tau^{\mu}{}_A\tau^{\nu}{}_B M_{\mu\nu} \right)\,,
\ee
and similarly for tensors of arbitrary rank. 
The meaning of raised indices should then hopefully clear from context -- note that e.g. the field strengths, Newton-Cartan torsion and covariant derivative are all naturally defined with lower curved indices so when they appear instead with raised curved or longitudinal flat indices this uses the above notation.

The action can then be written in terms of these manifestly covariant quantities as
\begin{equation}\label{11DAction}
  S = \int \dd^{11} x\,   \Omega \, \left(\mathcal{  L}+\mathcal{  L}_{ \widetilde{F}}+ \Omega^{-1} \mathcal{  L}_{\text{top}}\right),
\end{equation}
with 
\begin{align}
\begin{split} 
\mathcal{  L}&= \mathcal{  R} -   a^{ \mu AB}  a_{  \mu (AB)} +\frac{3}{2}  a^{ \mu}  a_{  \mu} -\frac{1}{12} F ^{ \mu   \nu  \rho   A}F_{ \mu   \nu  \rho  A}  +  \frac{1}{4}\epsilon_{ABC} F^{  A  B  \mu \nu}  T_{ \mu \nu}{}^{C}\,,\\
\mathcal{  L}_{\widetilde{F}}  &=-\frac{1}{4!} \widetilde F_{  \nu_1 \dots \nu_4}\left( F^{  \nu_1\dots \nu_4} +\frac{1}{4!3!  \Omega} \epsilon^{  \nu_1 \dots \nu_4  \mu_1 \dots \mu_7} F_{  \mu_1\dots\mu_4} \epsilon_{  A   B   C}   \tau_{  \mu_5}{}^{  A}  \tau_{  \mu_6}{}^{  B}  \tau_{  \mu_7}{}^{  C}   \right)\,,\\
\mathcal{  L}_{\text{top}} &=\frac{1}{6} F_4 \wedge F_4 \wedge C_3 = \frac{1}{6 } \frac{1}{3! 4!^2} \epsilon^{ \mu_1\dots\mu_{11}} F_{ \mu_1\dots\mu_4} F_{ \mu_5\dots\mu_8} C_{ \mu_9\dots\mu_{11}}\,,
\end{split}
\label{Lbits}
\end{align}
where the Ricci scalar $\mathcal{R}$ is defined in terms of the usual Riemann curvature tensor of the connection \eqref{defConn} via
\be
\mathcal{R}^\rho{}_{\sigma\mu\nu} = \partial_\mu \Gamma^\rho_{\nu\sigma}-\partial_\nu \Gamma^\rho_{\mu\sigma}+ \Gamma^{\rho}_{\mu\lambda}\Gamma^\lambda_{\nu\sigma}-\Gamma^{\rho}_{\nu\lambda}\Gamma^\lambda_{\mu\sigma}
\,,\quad
\mathcal{R}= \mathcal{R}^\rho{}_{\mu\rho\nu} H^{\mu\nu}\,.
\ee

\subsection{Dual field strength}
\label{dualF7}

The appearance of the two field strengths $F_4$ and $\widetilde F_4$ in the finite action \eqref{resultaction} may seem rather exotic.
In fact, we can relate the latter to components of the dual seven-form field strength, revealing that the non-relativistic action involves a partially democratic treatment of what are originally dual degrees of freedom. 
In 11-dimensional SUGRA, we have
\be
\hat F_7 = d \hat C_6 + \tfrac{1}{2} \hat C_3 \wedge \hat F_4 \,, \qquad
\hat F_7 = \hat \star \hat F_4 \,.
\ee
With our expansion, we can compute $\hat \star \hat F_4$ in components:
\be
\begin{split} 
(\hat \star \hat F_4)_{\mu_1 \dots \mu_7} & 
= \Omega \epsilon_{\mu_1 \dots \mu_7 \nu_1 \dots \nu_4} \big(
c^3 H^{\nu_1 \lambda_1}\dots H^{\nu_4 \lambda_4} F_{\rho_1 \dots \rho_4} 
 + H^{\nu_1 \lambda_1} \dots H^{\nu_4 \lambda_4} \widetilde F_{\lambda_1 \dots \lambda_4} 
\\ & \qquad\qquad\qquad\qquad + 4  H^{\nu_1 \lambda_1} \dots H^{\nu_3 \lambda_3} \tau^{\nu_4 \lambda_4}  F_{\lambda_1 \dots \lambda_4}  
\\ & \qquad\qquad\qquad\qquad
 - 6  H^{\nu_1 \lambda_1}  H^{\nu_2 \lambda_2}  T_{\lambda_1\lambda_2}{}^A \tau^{\nu_3 B} \tau^{\nu_4 C} \epsilon_{ABC}
\big)+ \mathcal{O}(c^{-3}) \,.
\end{split}
\label{F7oneway}
\ee
We then search for an expansion of $\hat C_6$ that can reproduce the singular term and lead to a sensible definition of the dual six-form in the non-relativistic theory. 
This is provided by
\be
\hat C_6 = - \tfrac{1}{2} c^3 C_3 \wedge \tfrac{1}{6} \epsilon_{ABC} \tau^A \wedge \tau^B \wedge \tau^C 
+ C_6 - \tfrac{1}{2} \widetilde  C_3 \wedge \tfrac{1}{6} \epsilon_{ABC} \tau^A \wedge \tau^B \wedge \tau^C 
+\mathcal{O}(c^{-3}) \,,
\label{defC6}
\ee
leading to
\be
\hat F_7 =  - \tfrac{1}{6} c^3 \epsilon_{ABC} \tau^A \wedge \tau^B \wedge \tau^C  \wedge F_4 
+ d C_6 + \tfrac{1}{2} C_3 \wedge F_4 
- \tfrac{1}{6}  \epsilon_{ABC} \tau^A \wedge \tau^B \wedge \tau^C  \wedge \widetilde F_4 + \mathcal{O}(c^{-3}) \,.
\label{F7oranother}
\ee
The singular piece in \eqref{F7oranother} agrees with that in \eqref{F7oneway} on using the self-duality constraint \eqref{magic} obeyed by $F_4$.
From the finite terms, we can define in the non-relativistic limit the quantity $F_7 \equiv d C_6 + \tfrac{1}{2} C_3 \wedge F_4$ which obeys again $d F_7 = \frac{1}{2} F_4 \wedge F_4$.
We could also define this quantity directly in the non-relativistic theory after taking the limit by starting with the equation of motion \eqref{3eomFinite} of the gauge field.
In that case, we would define the dual seven-form field strength to be the quantity appearing under the exterior derivative, including all terms on the left-hand side of \eqref{3eomFinite} as well as that involving $d\tau$ on the right-hand side.  
In components, this means
\be
\begin{split} 
 F_{\mu_1 \dots \mu_7} & = \tfrac{1}{4!} \Omega \epsilon_{\mu_1 \dots \mu_7 \nu_1 \dots \nu_4} (
 H^{\nu_1 \lambda_1} \dots H^{\nu_4 \lambda_4} \widetilde F_{\lambda_1 \dots \lambda_4}  + 4  H^{\nu_1 \lambda_1} \dots H^{\nu_3 \lambda_3} \tau^{\nu_4 \lambda_4}  F_{\lambda_1 \dots \lambda_4}  
\\ & \qquad\qquad\qquad\qquad
 - 6  H^{\nu_1 \lambda_1}  H^{\nu_2 \lambda_2}  T_{\lambda_1\lambda_2}{}^A \tau^{\nu_3 B} \tau^{\nu_4 C} \epsilon_{ABC}
\\ & \qquad\qquad\qquad\qquad
+ \tfrac{1}{4!3!} \Omega^{-1} \epsilon^{\nu_1 \dots \nu_4 \lambda_1 \dots \lambda_7} \epsilon_{ABC}\tau_{\lambda_1}{}^A \tau_{\lambda_2}{}^B \tau_{\lambda_3}{}^C  \widetilde F_{\lambda_4 \dots \lambda_7}
)\,.
\end{split}
\ee
Now, we can take the totally longitudinal contraction
\be
\begin{split}
 F_{\mu_1 \dots \mu_4 \sigma_1 \sigma_2 \sigma_3} \tau^{\sigma_1}{}_A \tau^{\sigma_2}{}_B \tau^{\sigma_3}{}_C 
 & = 
\tfrac{1}{4!} \Omega \epsilon_{\mu_1 \dots \mu_4  \nu_1 \dots \nu_4 \sigma_1 \sigma_2 \sigma_3} \tau^{\sigma_1}{}_A \tau^{\sigma_2}{}_B \tau^{\sigma_3}{}_C
 H^{\nu_1 \lambda_1} \dots H^{\nu_4 \lambda_4} \widetilde F_{\lambda_1 \dots \lambda_4}
 \\ & 
 \qquad +  \epsilon_{ABC} \widetilde F_{\mu_1 \dots\mu_4} 
 \,.
 \label{defF7}
\end{split}
\ee
Using \eqref{generalHn}, it can be shown that whereas the transverse part of $F_{\mu\nu\rho\sigma}$ obeys a self-duality constraint, the longitudinal part of $F_{\mu_1\dots \mu_7}$ obeys an anti-self-duality constraint:
\be
\begin{split} 
\Omega H^{\mu_1 \nu_1} \dots H^{\mu_4 \nu_4} &
 F_{\mu_1 \dots \mu_4 \sigma_1 \sigma_2 \sigma_3} \tau^{\sigma_1}{}_A \tau^{\sigma_2}{}_B \tau^{\sigma_3}{}_C
\\ & =
+ \tfrac{1}{4!3!} \epsilon^{\mu_1 \dots \mu_4 \nu_1 \dots \mu_4 \lambda_1 \dots \lambda_3} \epsilon_{DEF}\tau_{\lambda_1}{}^D \tau_{\lambda_2}{}^E \tau_{\lambda_3}{}^F 
 F_{\mu_1 \dots \mu_4 \sigma_1 \sigma_2 \sigma_3} \tau^{\sigma_1}{}_A \tau^{\sigma_2}{}_B \tau^{\sigma_3}{}_C\,.
\end{split}
\label{F7constraint}
\ee
The conclusion is that \eqref{defF7} shows that the totally longitudinal part of $ F_{\mu_1 \dots \mu_7}$ can be identified with the anti-self-dual transverse part of $\widetilde F_{\mu\nu\rho\sigma}$. 
Notice that the longitudinal part of the latter projects trivially out of the action, and in fact it is exactly the projection as on the right-hand side of \eqref{defF7} which appears in \eqref{resultaction}.
Hence we can re-express the terms in the Lagrangian involving $\widetilde F_{\mu\nu\rho\sigma}$ as
\be
\begin{split} 
\mathcal{L}_{\widetilde F} = 
 -\tfrac{1}{2}  \tfrac{1}{4!}  & F_{\mu_1 \dots \mu_4 \lambda_1 \dots \lambda_3}  \tfrac{1}{6}  \epsilon^{ABC}  \tau^{\lambda_1}{}_A \tau^{\lambda_2}{}_B \tau^{\lambda_3}{}_C \times
 \\ & \times \big(
 H^{\mu_1 \nu_1} \dots H^{\mu_4\nu_4} 
  + \tfrac{1}{4! 3!  \Omega} \epsilon^{\mu_1 \dots \mu_4 \nu_1 \dots  \nu_{7}}    \epsilon_{DEF} \tau_{\nu_5}{}^D \tau_{\nu_{6}}{}^E \tau_{\nu_{7}}{}^F
 \big)F_{\nu_1 \nu_2 \nu_3 \nu_4} \,.
\end{split}
\ee
This appearance of (components of) both the four-form and its dual together in the action is again reminiscent of exceptional field theory.

\section{Equations of motion and symmetries} 
\label{11DEom}

We have expanded the action, and now we turn our attention to the equations of motion, and the role played by the non-relativistic dilatation and boost symmetries. 

\subsection{Equations of motion from expansion}
\label{ExpandEom}

To keep track of the equations of motion at each order, we will consider the result of expanding the variation of the action. We will explicitly find that this gives the same results as varying the expansion of the action we considered previously. The reason we take this approach is that it will provide a useful way to keep track of which parts of the expansion of the eleven-dimensional equations of motion appear at which order. 
Recall that we view our non-relativistic limit as arising from a field redefinition, and we do not consider possible subleading terms which would occur in a true non-relativistic expansion. That said, we set up the expansion below in a way that would be reminiscent of such an expansion. 

The relativistic equations of motion are obtained from the variation of the action \eqref{Seleven}:
\be
\delta S = \int \dd^{11}x ( \sqrt{|\hat g|} \delta \hat g^{\mu\nu} \mathcal{G}_{\mu\nu} + \delta \hat C_{\mu\nu\rho} \mathcal{E}^{\mu\nu\rho} ) \,,
\label{deltaSeleven}
\ee
where
\be
\begin{split}
\mathcal{G}_{\mu\nu} & = R_{\mu\nu} - \tfrac{1}{12} \hat F_{\mu \rho_1 \dots \rho_3} \hat F_{\nu}{}^{\rho_1 \dots \rho_3}
- \tfrac{1}{2} \hat g_{\mu\nu} ( R - \tfrac{1}{48} \hat F^{\rho_1 \dots \rho_4} \hat F_{\rho_1 \dots \rho_4})\,,\\
\mathcal{E}^{\mu\nu\rho} & =  -\tfrac{1}{6} \left( \partial_\sigma ( \sqrt{|\hat g|} \hat F^{\mu\nu\rho \sigma} )- \tfrac{1}{2\cdot 4! \cdot 4!} \epsilon^{\mu\nu\rho\sigma_1 \dots \sigma_8} \hat F_{\sigma_1 \dots \sigma_4} \hat F_{\sigma_5 \dots \sigma_8}\right) \,.
\label{GE}
\end{split}
\ee
We consider the non-relativistic expansion of the fields, in the form 
\be
\hat g^{\mu\nu} = c H^{\mu\nu} + c^{-2} \tau^{\mu\nu} \,,\quad
\hat g_{\mu\nu} = c^2 \tau_{\mu\nu} + c^{-1} H_{\mu\nu} \,,\quad
\hat C_{\mu\nu\rho} = c^3 \omega_{\mu\nu\rho} + C_{\mu\nu\rho} + c^{-3} \widetilde C_{\mu\nu\rho} \,,
\label{simplerExpansion}
\ee
where $\omega_{\mu\nu\rho} = - \epsilon_{ABC} \tau_\mu{}^A \tau_\nu{}^B \tau_\rho{}^C$.
Both $\mathcal{G}$ and $\mathcal{E}$ admit an expansion in powers of $c^3$, with
\be
\mathcal{G} = c^6 \mathcal{G}^{(6)} + c^3 \mathcal{G}^{(3)} + c^0 \mathcal{G}^{(0)} + c^{-3} \mathcal{G}^{(-3)} + \dots\,,\quad
\mathcal{E} =  c^3 \mathcal{E}_{(3)} + c^0 \mathcal{E}_{(0)} + c^{-3} \mathcal{E}_{(-3)} + \dots\,.
\label{GEexp}
\ee
We now re-organise the variation of the action that results from \eqref{simplerExpansion}, by inserting the expressions \eqref{simplerExpansion} for the metric and three-form.
We choose to consider the variations of $\tau^\mu{}_A$ and $H^{\mu\nu}$ as independent, in terms of which
\be
\delta \omega_{\mu\nu\rho}=
 - \omega_{\mu\nu\rho}\tau_{\lambda}^D\delta \tau^\lambda{}_D   -3 \omega _{\lambda [\mu\nu} H_{\rho]\kappa}\delta H^{\lambda\kappa} \,.
\ee
The general result at order $c^{3n}$ following from \eqref{deltaSeleven} is that
\be
\begin{split}
\delta S^{(3n)} 
 = \int \dd^{11}x &  \Big[
  \delta H^{\mu\nu} ( \Omega \mathcal{G}_{\mu\nu}^{(3n)} - 3 \omega_{\mu \rho \sigma} H_{\lambda \nu} \mathcal{E}_{(3n-3)}^{\rho\sigma\lambda} )
  \\ & 
  + \delta \tau^\mu{}_A ( 2 \tau^{\nu A} \Omega \mathcal{G}_{\mu\nu}^{(3n+3)} - \tau^A{}_\mu \omega_{\rho\sigma\lambda} \mathcal{E}_{ (3n-3)}^{\rho\sigma\lambda})
  \\ & 
  + \delta C_{\mu\nu\rho} \mathcal{E}_{(3n)}^{\mu\nu\rho} + \delta \widetilde C_{\mu\nu\rho} \mathcal{E}_{(3n+3)}^{\mu\nu\rho} 
 \Big]\,,
\end{split} 
\label{deltaSn}
\ee
using $\sqrt{|\hat g|} = \Omega c^{-1}$. 
Hence, in general, if we expand the theory up to order $3k$, for $k \le n \le 2$, the equations of motion will be 
\begin{align}
\begin{split}\label{EOMexp}
\mathcal{G}^{(3n)}_{\langle\mu\nu\rangle}= 3 H_{\lambda \langle\mu}\omega_{\nu\rangle \rho\sigma }\Omega^{-1} \mathcal{E}^{\lambda \rho \sigma}_{(3n-3)}\,, \quad
2\mathcal{G}^{(3n+3)}_{\mu A}= \tau_{\mu A} \omega_{ \rho\sigma \lambda}\Omega^{-1} \mathcal{E}^{ \rho \sigma \lambda}_{(3n-3)}\,, \quad
\mathcal{E}_{(3n)}^{\mu\nu\rho} = 0\,,
\end{split}
\end{align}
with the understanding that $\mathcal{G}^{(9)} = \mathcal{E}^{(6)} = 0$.
The angle bracket notation takes into account that the variation of $H^{\mu\nu}$ is constrained by $\delta H^{\mu\nu} \tau_\mu{}^A \tau_\nu{}^B = 0$. 
We can solve this constraint by letting $\delta H^{\mu\nu} = H^{\rho(\mu} H_{\rho\sigma} M^{\nu)\sigma}$ such that the naive variation $\delta H^{\mu\nu} T_{\mu\nu}$ implies instead the equation of motion
\be
T_{\langle \mu \nu \rangle}= \tfrac{1}{2} (  H_{\mu\rho} H^{\rho\sigma} T_{(\sigma \nu)} + H_{\nu\rho} H^{\rho\sigma} T_{(\mu \sigma)} ) 
\label{langlebrackets}
\ee
which is symmetric and obeys $\tau^\mu{}_A \tau^\nu{}_B T_{\langle\mu\nu\rangle} = 0$.
Note that the equation of motion for $\widetilde C$ at each order is exactly that of $C$ at the previous order.

We should contrast the equations of motion \eqref{EOMexp} with the result of independently expanding $\mathcal{G}$ and $\mathcal{E}$ .
If we naively set each other of the expansion of the latter to zero, we would find the equations $\mathcal{G}^{(3n)} = 0 =\mathcal{E}^{(3n)}$ at any given order. 
However, in the non-relativistic expansion, treating $\tau^\mu{}_A$ and $H^{\mu\nu}$ as independent fields, then equation \eqref{EOMexp} says that we cannot simply expand the relativistic equations and set each order independently to zero unless we consider the full expansion (potentially infinite if treating subleading terms). A similar subtlety is the question of which equations of motion we are meant to expand. For instance, in the relativistic theory both $\mathcal{E}^{\mu\nu\rho} = 0$ and $g_{\mu\sigma} g_{\rho\kappa}g_{\sigma \lambda} \mathcal{E}^{\sigma\kappa\lambda} = 0$ are equivalent, but lead to different truncations to finite order in the $1/c$ expansion. Here we have made the choice to expand the equations of motion that appear conjugate to the variations $\delta g^{\mu\nu}$ and $\delta C_{\mu\nu\rho}$.

Let us look for example at the first two orders, $c^6$ and $c^3$. If we simply wanted to expand the theory up to order $c^6$ we would find the equation $ \left(\mathcal{G}^{(6)} -3 \omega  \mathcal{E}^{(3)}H\right)_{\langle \mu \nu\rangle} =0$, however if we proceed with expanding up to order $c^3$ we find that the equation for the 3-form tells us that $\mathcal{E}^{(3)}=0$, so that we can safely impose the two equations $\mathcal{G}^{(6)}_{\langle \mu \nu\rangle}= \mathcal{E}^{(3)}=0$ \textit{independently}.

Matters are further complicated by a number of `off-shell' identities obeyed by the terms appearing in the expansion of $\mathcal{G}$ and $\mathcal{E}$.
These identities will feature heavily below, and in fact are crucial for the consistency and symmetries of the non-relativistic truncation.

To put all these ideas together, we now look in detail at the first orders of the expansion of \eqref{deltaSeleven}.

\paragraph{Terms at $\mathcal{O}(c^6)$}
Here we encounter the leading terms in the expansions of $\mathcal{G}$ and $\mathcal{E}$.
First of all, we have
\be
\begin{split}
\mathcal{G}_{\mu\nu}^{(6)} & = 
\tfrac{1}{2} \tau_{\mu\nu} \left(
\tfrac{1}{2} T_{\rho_1 \sigma_1}^A T_{\rho_2 \sigma_2}^B \eta_{AB} H^{\rho_1 \sigma_1} H^{\rho_2 \sigma_2} 
+ \tfrac{1}{48} H^{\rho_1 \sigma_1} \dots H^{\rho_4 \sigma_4} F_{\rho_1 \dots \rho_4} F_{\sigma_1 \dots \sigma_4} 
\right) 
\end{split} 
\label{G6}
\ee
which obeys $\mathcal{G}_{\langle \mu \nu \rangle} = 0$ automatically. 
Hence the $\delta H^{\mu\nu}$ variation at order $c^6$ does not imply an actual equation of motion. 
One also has
\be
\mathcal{E}_{(3)}^{\mu\nu\rho}  =  - \tfrac{1}{6} \partial_\sigma \left(
\Omega
 H^{\mu \lambda_1} H^{\nu \lambda_2} H^{\rho \lambda_3} H^{\sigma \lambda_4}   F_{\lambda_1 \dots \lambda_4}  
+ \tfrac{1}{3! 4!} \epsilon^{\mu\nu\rho\sigma\sigma_1\dots \sigma_7}  F_{\sigma_1 \dots \sigma_4} \epsilon_{ABC} \tau_{\sigma_5}{}^A \tau_{\sigma_6}{}^B \tau_{\sigma_7}{}^C \right) \,.
\ee
This is the self-duality constraint under a derivative. 
It obeys $\tau_\mu{}^A \tau_\nu{}^B \mathcal{E}_{(3)}^{\mu\nu\rho} = 0$, and so also the $\delta \tau$ variation at order $c^6$ vanishes identically.
This is however necessary for consistency: the expansion of the action itself started only at order $c^3$, i.e. $S^{(6)} \equiv 0$.
Hence at this order we do not obtain any equations of motion. 

\paragraph{Terms at $\mathcal{O}(c^3)$}
At this order, there was a non-zero $S^{(3)}$ given by \eqref{S3}, for which we required the self-duality constraint \eqref{magic} to set to zero. 
Let us see how this information is reproduced. 
First of all, the variation of $C_3$ coming from \eqref{deltaSn} at this order implies $\mathcal{E}_{(3)}=0$.
The variation of $\tau^\mu{}_A$ involves a contribution from $\mathcal{E}_{(0)}$, which can be read off from the finite part of the expansion of the three-form equation of motion, which was \eqref{3eomFinite}.
For convenience, we repeat this here:
\be
\begin{split} 
\mathcal{E}_{(0)}^{\mu\nu\rho} = - \tfrac{1}{6} \partial_\sigma & \Big(
\Omega \big( 
4 H^{[\mu| \lambda_1} H^{|\nu| \lambda_2} H^{|\rho|\lambda_3} \tau^{|\sigma] \lambda_4}  F_{\lambda_1 \dots \lambda_4} 
 - 6 H^{[\mu|\lambda_1} H^{|\nu|\lambda_2}  \tau^{|\rho |B} \tau^{|\sigma] C} T_{\lambda_1 \lambda_2}{}^A \epsilon_{ABC}  
\\ & \qquad\qquad+ H^{\mu \lambda_1} H^{\nu \lambda_2} H^{\rho \lambda_3} H^{\sigma \lambda_4} \widetilde F_{\lambda_1 \dots \lambda_4} 
\big) 
\Big) 
\\ & 
+ \tfrac{1}{2 \cdot 3! 4! 4!} \epsilon^{\mu\nu\rho \sigma_1 \dots \sigma_8} (  F_{\sigma_1 \dots \sigma_4}  F_{\sigma_5 \dots \sigma_8} 
- 12 \epsilon_{ABC} T_{\sigma_1 \sigma_2}{}^A \tau_{\sigma_3}{}^B \tau_{\sigma_4}{}^C \widetilde F_{\sigma_5 \dots \sigma_8} 
) 
\,.
\end{split}
\label{E0}
\ee
What one finds then is that  
\be
\begin{split} 
2 &\tau^{\nu A} \Omega \mathcal{G}_{\mu\nu}^{(6)}- \tau_\mu{}^A \omega_{\rho\sigma\lambda} \mathcal{E}^{(0) \rho \sigma \lambda} 
\\ & =
\tfrac{1}{2 \cdot 4!} \tau_\mu{}^A \Omega F_{\nu_1 \dots \nu_4} \left( H^{\nu_1\rho_1} \dots H^{\nu_4 \rho_4} F_{\rho_1 \dots \rho_4} 
+ \tfrac{1}{\Omega 3! 4!} \epsilon^{\nu_1 \dots \nu_4 \rho_1 \dots \rho_7} F_{\rho_1 \dots \rho_4} \epsilon_{ABC} \tau_{\rho_5}{}^A \tau_{\rho_6}{}^B \tau_{\rho_7}{}^C 
\right)\,,
\end{split}
\label{G6E0}
\ee
which is proportional to the self-duality constraint. 
For the terms accompanying the $\delta H^{\mu\nu}$ variation one finds
\be
\begin{split} 
\delta H^{\mu\nu}& (\Omega \mathcal{G}_{\mu\nu}^{(3)} -3 \omega_{(\mu|\rho\sigma} H_{\lambda| \nu)} \mathcal{E}_{(0)}^{\rho\sigma\lambda} )
\\
 & = \delta H^{\mu\nu} \big( \tfrac{1}{4 \cdot 4!^2}  \epsilon_{ABC} H_{\lambda_1 (\mu}\tau_{\nu)}{}^A\tau_{\lambda_2}{}^B\tau_{\lambda_3}{}^C F_{\sigma_1 \dots\sigma_4} F_{\sigma_5\dots\sigma_8}\epsilon^{\lambda_1 \dots\lambda_3 \sigma_1\dots\sigma_8}
\\ & \qquad\qquad
- \tfrac{\Omega}{12} F_{\mu \rho_1 \dots \rho_3} F_\nu{}^{\rho_1 \dots \rho_3} 
 + \tfrac{\Omega}{96} H_{\mu\nu} F^2 \big) 
\end{split}
\label{G3E0}
\ee
such that after projecting using \eqref{langlebrackets}
\be
\begin{split} 
\Omega  \mathcal{G}_{\langle\mu\nu\rangle}^{(3)} -3 \omega_{\langle\mu|\rho\sigma} H_{\lambda| \nu\rangle} \mathcal{E}_{(0)}^{\rho\sigma\lambda} 
&=
\tfrac{ 1}{8 \cdot 4!^2} \epsilon_{ABC} H_{\lambda_1 (\mu}\tau_{\nu)}{}^A\tau_{\lambda_2}{}^B\tau_{\lambda_3}{}^C F_{\sigma_1 \dots\sigma_4} F_{\sigma_5\dots\sigma_8}\epsilon^{\lambda_1 \dots\lambda_3 \sigma_1\dots\sigma_8}
\\ & \qquad 
+\tfrac{\Omega}{96} H_{\mu\nu}F^2 -\tfrac{\Omega}{12}H_{\kappa(\mu}F_{\nu) \rho\sigma \lambda}F^{\kappa \rho\sigma \lambda} \,,
\end{split}
\label{G3E0_proj}
\ee
using the obvious shorthand for raised indices and $F^2$ instead of writing $H^{\mu\nu}$ multiple times.
This exactly reproduces the variation $\delta S^{(3)}$ of the leading part of the expansion of the action \eqref{S3}. 
Then, after projecting and using the Schouten identity \eqref{G3E0} or \eqref{G3E0_proj} can be shown to again be proportional to the self-duality constraint (specifically: the time-space projection of the first term combines with the time-space projection of the third term, and the space-space projection of the second term combines with the space-space projection of the third term).

Hence the sole equation of motion we obtain at this order is the self-duality constraint. This is consistent with what we required from the expansion of the action.

\paragraph{Terms at $\mathcal{O}(c^0)$}

We next consider \eqref{deltaSn} with $n=0$.
First of all, the equation of motion of $C$ indeed gives $\mathcal{E}_{(0)}$, as in \eqref{E0}, while that of $\widetilde{C}$ gives the constraint in the form $\mathcal{E}_{(3)}$. 
This is exactly what we obtain from varying the finite action $S^{(0)}$ directly. 
Note that the longitudinal projection of $\mathcal{E}_{(0)}$ in conjunction with the self-duality constraint implies the equation
\be
\tfrac{1}{2} \eta_{AB} H^{\mu\rho} H^{\nu \sigma} T_{\mu\nu}{}^A T_{\rho \sigma}{}^B = - \tfrac{1}{48} 
 H^{\mu_1 \nu_1} \dots H^{\mu_4 \nu_4} F_{\mu_1 \dots \mu_4} F_{\nu_1 \dots \nu_4} \,,
\label{folcon}
\ee
thereby reproducing the equation we would get by setting $\mathcal{G}^{(6)}=0$ (compare \eqref{G6}). Hence although we could not set $\mathcal{G}^{(6)}=0$ previously, the non-relativistic theory is not missing this equation.
Note that for generic non-vanishing $F_4$, equation \eqref{folcon} is incompatible with imposing foliation-type constraints on the MNC torsion such that the left-hand side vanishes, however if $F_4$ is also restricted to vanish (for example) one could require such constraints (as is always possible in the NSNS sector case \cite{Bergshoeff:2021bmc}).

Now we turn to the equations of motion following from the variations of $\tau$ and $H$.
For simplicity, we present here the independent equations of motion after projecting onto longitudinal (time) and transverse (space) components.
The temporal and spatial projectors are defined as
\be
\left(\Delta_T\right)^\mu{}_\nu = \tau^\mu{}_A \tau_\nu{}^A, \qquad \left(\Delta_S\right)^\mu{}_\nu = H^{\mu\rho}H_{\rho\nu},\qquad \left(\Delta_T\right)^\mu{}_\nu+\left(\Delta_S\right)^\mu{}_\nu = \delta^\mu_\nu .
\ee
We start with the equations of motion of $\tau$.
The trace of the time projection gives an equation involving the Ricci scalar:
\begin{equation} 
\begin{split} \label{REQ}
\mathcal{R} &= \tfrac{7}{3} \nabla^{  \mu }   a_{  \mu} + a^{\mu \{AB\}}a_{\mu AB}+\tfrac{7}{6}a^2 + \tfrac{1}{36} F_{  A   \nu   \rho   \sigma} F^{  A  \nu   \rho   \sigma} -\tfrac{1}{6}\epsilon_{  A   B   C} F^{  AB \rho\sigma} T_{  \rho   \sigma }  {}^{C} 
\\ & \qquad
+ \tfrac{1}{4!} \widetilde{F}_{\mu\nu\rho\sigma}
\left(
F^{\mu\nu\rho\sigma} + \tfrac{1}{ \Omega 4! 3!} \epsilon_{ABC} \epsilon^{\mu\nu\rho\sigma \lambda_1 \dots \lambda_7} F_{\lambda_1 \dots \lambda_4} \tau_{\lambda_5}{}^A \tau_{\lambda_6}{}^B \tau_{\lambda_7}{}^C 
\right)\,.
\end{split}
\end{equation}
The traceless part of the time-time projection is:
\begin{align}
\begin{split}
  \nabla^{\mu }&a_{\mu \{AB\}}+a^{\mu }a_{\mu\{ AB\}}  + a_{\mu [C (A]} a^{\mu}{}_ {\{B) D\}}\eta^{CD} 
  \\& =   -  \tfrac{1}{12} F_{A}{}^{ \mu\nu\rho} F_{B \mu\nu\rho} +  \epsilon_{(A|CD} F_{|B) }{}^{C \mu\nu } T_{\mu\nu }{}^{D}
  -\tfrac{\eta_{AB}}{3}\left(-\tfrac{1}{12} F^{C \mu\nu\rho} F_{C\mu\nu\rho}+\epsilon_{CDE} F^{ \mu\nu CD} T_{\mu\nu}{}^{E}  \right) \,.
\end{split}
\end{align}
The space projection is
\begin{align}
\begin{split} \label{dTs}
\nabla_{\rho }T^{\mu \rho}{}_{  A}+a_{\rho A C} T^{\mu \rho C} &=\tfrac{1}{6}F^{\mu \nu \rho \sigma} F_{A \nu\rho\sigma}-\tfrac{1}{2} \epsilon_{ABC}F^{\mu \rho \sigma B}T_{\rho\sigma}{}^C\\ 
\end{split}
\end{align}
Finally, consider the equations of motion of $H$.
The space-space projection is:
\begin{align}
\begin{split}
\mathcal{R}^{(\mu\nu)}
-&a^{\mu AB} a^{\nu}{}_{\{AB\}}
+\tfrac{1}{6}\left( a^\mu a^\nu 
- a^2 H^{\mu\nu}\right) 
\\ &  =\tau^{\rho A}\nabla^{(\mu}T^{\nu)}{}_{\rho A}
+  \tfrac{1}{6} H^{\mu\nu} \nabla^\rho a_\rho
+  \tfrac{1}{4}  F^{\mu  \rho\sigma A} F^{\nu}{}_{\rho \sigma A}
-\tfrac{1}{36}H^{\mu\nu} F^{A \rho \sigma\lambda}F_{A \rho \sigma\lambda}
\\&\quad - \tfrac{1}{2} \epsilon_{ABC}  F^{(\mu| \rho AB } T^{|\nu)}{}_{\rho}{}^{C}
+\tfrac{1}{24}H^{\mu\nu}\epsilon_{ABC}  F^{\rho\sigma AB } T_{\rho\sigma}{}^{C} 
\\&\quad
+  \tfrac{1}{6}  F^{(\mu|\rho\sigma \lambda} \widetilde{F}^{|\nu)}{}_{\rho\sigma \lambda}
-\tfrac{1}{48}H^{\mu\nu}F^{\rho\sigma\lambda \kappa}\widetilde{F}_{\rho\sigma\lambda \kappa}
\\ &
\quad + \tfrac{1}{2} H^{\mu\nu} \big(
- \mathcal{R} + \tfrac{7}{3} \nabla^{  \mu }   a_{  \mu} + a^{\mu \{AB\}}a_{\mu AB}+\tfrac{7}{6}a^2 + \tfrac{1}{36} F_{  A   \nu   \rho   \sigma} F^{  A  \nu   \rho   \sigma} -\tfrac{1}{6}\epsilon_{  A   B   C} F^{  AB \rho\sigma} T_{  \rho   \sigma }  {}^{C}
\big)\,.
\end{split}
\label{Hspacespace}
\end{align}
Combining the trace of \eqref{Hspacespace} with \eqref{REQ} we find that the self-duality constraint \eqref{magic} appears (contracted with $\widetilde{F}_{\mu\nu\rho\sigma}$).

The time-space projection is (with $\epsilon^{A}{}_{BC} \equiv \eta^{AD} \epsilon_{DBC}$)
\begin{align}
\begin{split}
\mathcal{R}^{(\mu A)} 
-& a^\mu{}_{BC} a^{A (BC)}
+\tfrac{1}{2} a_B a^{\mu BA} 
\\ & = 
\tfrac{1}{4} \epsilon^{A}{}_{BC}\nabla^{\rho} F^\mu{}_{\rho}{}^{BC} 
+  \tfrac{1}{4}\epsilon^{A}{}_{BC}  a_\rho F^{\mu \rho}{}^{BC}
+  \tfrac{1}{4}  \epsilon_{BCD} a_{\rho}{}^{AB}F^{\mu \rho CD} \\
&\qquad
+\tfrac{1}{4}F^{AB \rho\sigma}F^{\mu}{}_{B \rho\sigma}
+\tfrac{1}{4}\epsilon_{BCD}F^{ABC \rho}T_{\rho}{}^{\mu D}
\\ &\qquad
+  \tfrac{1}{2} a^{\rho BA}  \nabla_{\rho }\tau^{\mu}{}_{B} 
-\tfrac{1}{2}\nabla^2 \tau^{\mu  A}
- a^\rho \nabla_\rho \tau^{\mu A}
- \tfrac{1}{2} a^{\mu BA} \mathcal{K}_{ B} 
+\tfrac{1}{2} a^\mu \mathcal{K}^A 
\\ & \qquad
-\tfrac{1}{2}  \nabla_{B}a^{\mu B A} 
+  \nabla^A a^\mu  
+\tfrac{1}{2} T^{\mu}{}_{\sigma B}  \nabla^B \tau^{\sigma A} 
+  \tfrac{1}{2}\nabla_{\rho }\nabla^\mu \tau^{\rho A} 
-\tfrac{1}{2}\tau^\rho{}_B \nabla^{\mu} a_\rho{}^{AB} \\ 
& \qquad
+ \tfrac{1}{6}F^{(\mu}{}_{\nu\rho\sigma} \widetilde{F}^{A)\nu\rho\sigma} 
- \tfrac{1}{4\cdot 4!^2 \Omega} \epsilon^{A}{}_{BC} \tau_{\lambda_2}{}^B\tau_{\lambda_3}{}^C H^{\mu\kappa}H_{\kappa \lambda_1}  F_{\sigma_1\dots\sigma_4}\widetilde{F}_{\sigma_5 \dots\sigma_8} \epsilon^{\lambda_1 \dots\lambda_3\sigma_1\dots\sigma_8}\, .
\end{split}
\label{HTS}
\end{align} 
We have verified that these are indeed exactly the equations of motions that one gets by varying the finite part of the action, $S^{(0)}$, given in \eqref{resultaction}.

\subsection{Dilatations and a `missing' equation of motion}
\label{DilMissing}

We already mentioned the existence of a dilatation transformation given by \eqref{dil_all}, whose origin lay in the expansion in powers of $c$.
There is evidently a freedom to rescale $c$ by some constant while simultaneously rescaling the component fields such that the eleven-dimensional fields are unchanged.
This \emph{rigid dilatation} leaves the full action invariant.
Hence for an infinitesimal dilatation, with $\delta_\lambda c = - \lambda c$, we have the transformations \eqref{dil_all}, and clearly order-by-order for the action we should have
\begin{align}
\label{dil1}
\delta_\lambda S^{(6)} = 6\lambda S^{(6)}\,,\quad
\delta_\lambda S^{(3)} = 3\lambda S^{(3)}\,,\quad
\delta_\lambda S^{(0)} = 0\cdot\lambda S^{(0)}\,,\quad
\delta_\lambda S^{(-3)} = -3\lambda S^{(-3)}\,,\dots
\end{align}
Recall that $S^{(6)}$ and $\delta S^{(6)}$ vanish identically, so the first of these is just $0=0$.

A powerful consequence of the rigid dilatations is that if we know the equations of motion for the action $S^{(3k)}$ at a given order $k \neq 0$ we can immediately write down an action that produces them (which will agree up to total derivatives with that arising from the expansion).
This works by applying the formula \eqref{deltaSn} for the variation and specialising to the dilatation variation.
This is guaranteed to produce $3k S^{(3k)}$. 
This singles out the finite order action as being special, as here knowing the equations of motion and dilatation symmetry is not enough to determine its form.
Furthermore, for this case we can promote the dilatation parameter to be coordinate dependent, and obtain a \emph{local dilatation} symmetry.

Let's verify these statements.
Under a rigid dilatation with parameter $\lambda $, the variation of the $c^3$ part of the action is
\begin{align}
\delta_\lambda{S}^{(3)} = \int \dd^{11}x \,\Omega \left( \lambda \mathcal{G}^{(3)}_{\mu\nu}H^{\mu\nu} - \lambda \left(2(\mathcal{G}^{(6)})^A{}_A + 3 \epsilon_{ABC}\Omega^{-1}\mathcal{E}_{(0)}^{ABC}\right) \right) \,, 
\end{align}
where $\mathcal{E}^{ABC} \equiv \tau_\mu{}^A \tau_\nu{}^B \tau_\rho{}^C \mathcal{E}^{\mu\nu\rho}$. 
It can be checked that $\mathcal{G}^{(3)}_{\mu\nu}H^{\mu\nu} =0$. Then, if we denote the self-duality constraint by
\be
\Theta^{\mu_1 \dots \mu_4} \equiv  H^{\mu_1\rho_1} \dots H^{\mu_4 \rho_4} F_{\rho_1 \dots \rho_4} 
+ \tfrac{1}{\Omega 3! 4!} \epsilon^{\mu_1 \dots \mu_4 \rho_1 \dots \rho_7} F_{\rho_1 \dots \rho_4} \epsilon_{ABC} \tau_{\rho_5}{}^A \tau_{\rho_6}{}^B \tau_{\rho_7}{}^C 
\ee
we have 
\be
2(\mathcal{G}^{(6)})^A{}_A + 3 \epsilon_{ABC}\Omega^{-1}\mathcal{E}_{(0)}^{ABC}   = 
3\tfrac{1}{2 \cdot 4!}   F_{\mu_1 \dots \mu_4} \Theta^{\mu_1 \dots \mu_4}
\,,
\label{trace3}
\ee
hence indeed referring to \eqref{S3} for $S^{(3)}$ we indeed have
\be
\delta_\lambda S^{(3)} = 3 \lambda S^{(3)} \,.
\ee
Next consider the finite part of the action, with:
\be
\delta_\lambda{S}^{(0)} =  \int \dd^{11}x \,\Omega \left( \lambda \mathcal{G}^{(0)}_{\mu\nu}H^{\mu\nu} - \lambda \left(2(\mathcal{G}^{(3)})^A{}_A + 3 \epsilon_{ABC}\Omega^{-1}\mathcal{E}_{(-3)}^{ABC}\right) +\Omega^{-1}  \mathcal{E}^{(3) \mu\nu\rho} \delta_\lambda \widetilde{C}_{\mu\nu\rho}\right) \,.
\ee
Now we can show that
\be
\mathcal{G}^{(0)}_{\mu\nu}H^{\mu\nu} -  \left(2(\mathcal{G}^{(3)})^A{}_A + 3 \epsilon_{ABC}\Omega^{-1}\mathcal{E}_{(-3)}^{ABC}\right)
=  - \tfrac{1}{8} \widetilde F_{\mu_1 \dots \mu_4} \Theta^{\mu_1 \dots \mu_4} \,,
\label{trace0}
\ee
such that using $\mathcal{E}^{\mu\nu\rho}_{(3)} = - \tfrac{1}{6} \partial_\sigma \Theta^{\mu\nu\rho\sigma}$ we have 
\be
\begin{split} 
\delta_\lambda{S}^{(0)} & = \int \dd^{11}x \, ( - \tfrac{1}{8} \lambda \widetilde F_{\mu\nu\rho \sigma} \Theta^{\mu\nu\rho\sigma} - \tfrac{1}{6} \partial_\sigma \Theta^{\mu\nu\rho\sigma} \delta_\lambda \widetilde{C}_{\mu\nu\rho})  \,,\\
& = \int \dd^{11}x \, ( - \tfrac{1}{8} \lambda \widetilde F_{\mu\nu\rho \sigma} \Theta^{\mu\nu\rho\sigma} - \tfrac{1}{24} \Theta^{\mu\nu\rho\sigma} \delta_\lambda \widetilde{F}_{\mu\nu\rho\sigma} )\,,
\end{split}
\label{deltaS0dil}
\ee
after integrating by parts. 
For arbitrary local $\lambda$, we therefore have $\delta_\lambda S^{(0)} = 0$ \emph{on imposing the self-duality constraint}, irrespective of the transformation of $\widetilde C_{\mu\nu\rho}$.
Alternatively, if we require that
\be
\delta_\lambda \widetilde{F}_{\mu\nu\rho\sigma} = - 3 \lambda \widetilde{F}_{\mu\nu\rho\sigma} \,,
\label{FtildeDil}
\ee
then \eqref{deltaS0dil} vanishes identically without use of the constraint.
This would mean accepting a non-local transformation for $\widetilde{C}_{\mu\nu\rho}$ itself, which is not completely outlandish given the discussion in section \ref{dualF7} suggests we may think of it as being a dual degree of freedom to $C_3$.

What this means in practice is that the action $S^{(0)}$ is invariant under variations of $H^{\mu\nu}$ and $\tau^\mu{}_A$ of the form \eqref{dil_all}.
This implies that there is a `direction' in the space of variations which leaves the action $S^{(0)}$ unchanged (or at best produces the self-duality constraint, which is not an independent equation of motion).
Hence if we vary $S^{(0)}$ to obtain the equations of motion of $H^{\mu\nu}$ and $\tau^\mu{}_A$, we will find that we are `missing' an equation of motion.
This is exactly as in the NSNS sector case \cite{Gallegos:2020egk, Bergshoeff:2021bmc} and reflects a known difficulty, even in the purely gravitational context, of obtaining the Poisson equation from an action principle for non-relativistic theories \cite{Hansen:2018ofj,Hansen:2020pqs}, at least at first order.

Thus, in order to obtain an equation of motion for this missing variation, we go one step further in the expansion.
The variation of $S^{(-3)}$, from \eqref{deltaSn}, is:
\be
\begin{split}
\delta S^{(-3)} 
 = \int \dd^{11}x &  \Big[
  \delta H^{\mu\nu} ( \Omega \mathcal{G}_{\mu\nu}^{(-3)} - 3 \omega_{\mu \rho \sigma} H_{\lambda \nu} \mathcal{E}_{(-6)}^{\rho\sigma\lambda} )
  + \delta \tau^\mu{}_A ( 2 \tau^{\nu A} \Omega \mathcal{G}_{\mu\nu}^{(0)} - \tau^A{}_\mu \omega_{\rho\sigma\lambda} \mathcal{E}_{ (-6)}^{\rho\sigma\lambda})
  \\ & 
  + \delta C_{\mu\nu\rho} \mathcal{E}_{(-3)}^{\mu\nu\rho} + \delta \widetilde C_{\mu\nu\rho} \mathcal{E}_{(0)}^{\mu\nu\rho} 
 \Big]\,,
\end{split} 
\label{deltaSminus3}
\ee
For dilatations we have
\be
\begin{split}
\delta_\lambda S^{(-3)} 
 = \int \dd^{11}x &  \Big[
    \lambda\left( H^{\mu\nu}  \Omega \mathcal{G}_{\mu\nu}^{(-3)}
  - 2 \Omega (\mathcal{G}^{(0)})_A{}^A - 3 \epsilon_{ABC} \mathcal{E}_{(-6)}^{ABC}
  \right)
  +  \delta_\lambda \widetilde C_{\mu\nu\rho} \mathcal{E}_{(0)}^{\mu\nu\rho} 
 \Big]\,.
\end{split} 
\label{deltaSminus3_dil}
\ee
With constant $\lambda$, equation \eqref{dil1} implies that 
\be
S^{(-3)} = \int \dd^{11} x \,( \Omega \,\mathcal{N} + \widetilde C_{\mu\nu\rho} \mathcal{E}^{\mu\nu\rho}_{(0)})\,,
\label{Sminus3}
\ee
where we defined the combination 
\be
\mathcal{N} \equiv 
\tfrac{1}{3} ( -H^{\mu\nu} \mathcal{G}_{\mu\nu}^{(-3)}  + 2  (\mathcal{G}^{(0)})_A{}^A) 
+  \epsilon_{ABC}\Omega^{-1} \mathcal{E}_{(-6)}^{ABC}
\label{missingFromDil} \,.
\ee
Crucially, \eqref{missingFromDil} does not vanish on applying the self-duality constraint, unlike the combination of terms \eqref{trace3} and \eqref{trace0} which appeared at the previous orders, and nor is it a combination of any other equations of motion resulting from the finite action. 
It can therefore be used as the equation of motion of the `dilatation mode'.
(We are not really interested in the $\widetilde C$ variation appearing in \eqref{deltaSminus3_dil}, which multiplies something we have already taken into account as an equation of motion.)
It involves the fully longitudinal part of $\mathcal{G}^{(0)}$, which has not yet appeared in the equations of motion.
Hence, we identify it with the `Poisson equation', in which the longitudinal part of $C_{\mu\nu\rho}$ plays the role of the Newton potential (as did the longitudinal part of the $B$-field in the Stueckelberg gauge-fixed NSNS sector).
This is because $\mathcal{E}_{(-6)}^{}$ is the first equation of motion which contains two derivatives acting on the former. Explicitly,
\be
\begin{split}
\mathcal{E}_{(-6)}^{\mu\nu\rho} & = 
- \tfrac{1}{6} \partial_\sigma \Big(
\Omega \big(
4 H^{[\mu| \lambda_1} \tau^{|\nu| \lambda_2} \tau^{|\rho|\lambda_3} \tau^{|\sigma] \lambda_4}   F_{\lambda_1 \dots \lambda_4} 
 + 6 H^{[\mu|\lambda_1} H^{|\nu|\lambda_2}  \tau^{|\rho |\lambda_3} \tau^{|\sigma] \lambda_4} \widetilde F_{\lambda_1 \dots \lambda_4}\Big)
\\ & \quad + \tfrac{1}{2\cdot 4! 4! 3!} \epsilon^{\mu\nu\rho \sigma_1 \dots \sigma_8}   \widetilde F_{\sigma_1 \dots \sigma_4}\widetilde F_{\sigma_5 \dots \sigma_8}  \,.
\end{split}
\ee
Intriguingly, the combination of $\mathcal{G}^{(-3)}$ and $\mathcal{G}^{(0)}$ appearing in \eqref{missingFromDil} has a somewhat murky relationship to the `trace-reversed' version of the metric equation of motion.
The equation $\mathcal{G}_{\mu\nu}=0$ in the original 11-dimensional theory can be simplified somewhat by taking its trace and solving that for the Ricci scalar.
This trace is \be
\hat g^{\mu\nu} \mathcal{G}_{\mu\nu} = - \tfrac{9}{2} R + \tfrac{1}{32}\hat  F^2 
\ee
and the equation of motion without the Ricci scalar is 
\be
\begin{split} 
\bar{\mathcal{G}}_{\mu\nu} & \equiv \mathcal{G}_{\mu\nu} - \tfrac{1}{9} \hat g_{\mu\nu} \hat g^{\rho \sigma} \mathcal{G}_{\rho \sigma}
 = R_{\mu\nu} - \tfrac{1}{12} \hat F_{\mu}{}^{\rho\sigma\lambda} \hat F_{\nu \rho\sigma \lambda} + \tfrac{1}{144} \hat g_{\mu\nu} \hat F^2 \,,
\end{split} 
\ee
for which
\be
\tau^{\mu\nu} \bar{\mathcal{G}}_{\mu\nu}^{(0)}
= \tfrac{1}{3} ( 2 \tau^{\mu\nu} \mathcal{G}_{\mu\nu}^{(0)} - H^{\mu\nu} \mathcal{G}_{\mu\nu}^{(-3)} )\,,
\ee
which is exactly the combination appearing in \eqref{missingFromDil}. Note the relative numerical factors here are the same as the relative numerical factors in the powers of $c$ in the expansion.

Now, what exactly is the equation \eqref{missingFromDil}?
Expanding the metric equation contributions and covariantising everything, one arrives at
\be
\begin{split}
\tau^{\mu\nu} \bar{\mathcal{G}}_{\mu\nu}^{(0)}
 & =  2 \tau^{\mu A}  \nabla^{\rho }\mathcal{K}_{\mu \rho A} -     \nabla^A\mathcal{K}_{ A} 
 -\tfrac{1}{4} a^{ABC}a_{ABC}
  -  \tfrac{1}{2} a^{ABC} a_{ACB} -  a^A a_A\\
&  \quad - \epsilon_{ABC} F^{D AB \rho}a_{\rho D}{}^C
 - \tfrac{1}{8}F^{A B \mu\nu} F_{AB \mu\nu}+\tfrac{1}{48} \widetilde{F}^{\mu\nu\rho\sigma}\widetilde{F}_{\mu\nu\rho\sigma}+\tfrac{1}{4} \epsilon_{ABC}\widetilde{F}^{\mu\nu AB}T_{\mu\nu}{}^C \\
&\quad -  a^A \mathcal{K}_{A} +\mathcal{K}^{\mu \nu A} \mathcal{K}_{\mu\nu A} -2\tau^{\mu A} \tau^{\nu B}\nabla_\nu a_{\mu [AB]} -\tau^{\mu\nu} \nabla_\mu a_\nu
\,,\end{split}
\label{metriclong}
\ee
\be
\begin{split}
\epsilon_{ABC} \tau_\mu{}^A \tau_\nu{}^B \tau_\rho{}^C \Omega^{-1} \mathcal{E}^{\mu\nu\rho}_{(-6)} 
& = 
- \tfrac{1}{6}  \epsilon_{ABC} \nabla^\mu F^{ABC}{}_{\mu} 
 - \tfrac{1}{4} \epsilon_{ABC} \widetilde{F}^{AB \mu\nu}T_{\mu\nu}{}^C 
 \\ & \quad + \tfrac{1}{2\cdot 4!^2\Omega}\tfrac{1}{6}\epsilon^{\lambda_1 \dots\sigma_1\dots\sigma_8}\widetilde{F}_{\sigma_1\dots\sigma_4}\widetilde{F}_{\sigma_5\dots\sigma_8} \epsilon_{ABC}\tau_{\lambda_1}{}^A\tau_{\lambda_2}{}^B\tau_{\lambda_3}{}^C\,,
\end{split}
\label{Clong}
\ee
hence the covariant Poisson equation is 
\be
\begin{split}
\mathcal{N} & =
 - \tfrac{1}{6}  \epsilon_{ABC} ( \nabla^\mu F^{ABC}{}_{\mu}  +   a_\mu F^{ABC \mu} + 3 a_\mu{}_D{}^{A} F^{BCD \mu} )
- \tfrac{1}{8}F^{A B \mu\nu} F_{AB \mu\nu}
\\ &\quad +\tfrac{1}{48} \widetilde{F}^{\mu\nu\rho\sigma}\widetilde{F}_{\mu\nu\rho\sigma}
 + \tfrac{\Omega^{-1}}{2\cdot 4!^2 3!}\epsilon^{\lambda_1 \dots\lambda_3\sigma_1\dots\sigma_8}\widetilde{F}_{\sigma_1\dots\sigma_4}\widetilde{F}_{\sigma_5\dots\sigma_8} \epsilon_{ABC}\tau_{\lambda_1}{}^A\tau_{\lambda_2}{}^B\tau_{\lambda_3}{}^C
\\ & \quad
  -     \nabla^A\mathcal{K}_{ A} -  a^A \mathcal{K}_{A} -\mathcal{K}^{\mu \nu A} \mathcal{K}_{\mu\nu A} -2a^{\mu[AB]} \mathcal{K}_{\mu AB} -2\tau^{\mu\nu} \nabla_\mu a_\nu
\\ & \quad
- a^{ABC} ( \tfrac{1}{4} a_{ABC} + \tfrac{1}{2} a_{ACB}  + \eta_{BC} a_A )
\\ & = 0 \,.
\end{split} 
\label{CovariantPoisson}
\ee
Note that this expression could equivalently be rewritten in terms of the Ricci tensor, using the following identity:
\be
\mathcal{R}^A{}_A =\tau^{\mu\nu} \mathcal{R}_{\mu\nu}= -\nabla^A \mathcal{K}_A -\mathcal{K}^{\mu\nu A}\mathcal{K}_{\mu\nu A} -a^{\mu AB} \mathcal{K}_{\mu AB}\,.
\label{RtoK11}
\ee
Remarkably, equation \eqref{CovariantPoisson} transforms covariantly under \emph{local} dilatations.
Exactly this equation will also be selected by the exceptional field theory description as an `extra' equation of motion that one can not find from the variation of the finite part of the action.
Furthermore, under Galilean boosts (discussed in next subsection), it transforms into the other equations of motions.
All this is in keeping with the properties of the missing Poisson equation in the NSNS sector \cite{Gallegos:2020egk, Bergshoeff:2021bmc} and supports including equation \eqref{CovariantPoisson} as an equation of motion of the non-relativistic theory.

If we think in terms of the expansion it might seem strange to find the rest of the equations of motion from the expansion at order $c^0$ and this extra equation from order $c^{-3}$.
Clearly, if we would vary the action $S^{(-3)}$ we would find additional $\mathcal{O}(c^{-3})$ contributions to the finite equations of motion, and if we would vary the action $S^{(-6)}$ we would find additional $\mathcal{O}(c^{-3})$ contributions to the equation of motion \eqref{CovariantPoisson}, i.e. it would become $\mathcal{N} = \mathcal{O}(c^{-3})$. 
The guiding philosophy is to find the lowest order non-zero equation of motion resulting from the variations of the action. For the Poisson equation associated to the degree of freedom that disappears into dilatations at the level of $S^{(0)}$, this happens to arise at lower order than the other equations of motion.

As a final remark, just as in the NSNS sector case \cite{Bergshoeff:2021bmc}, it is also possible to define a covariant derivative that is covariant with respect to dilatations.
Letting $b_\mu$ denote this dilatation connection, and simultaneously introducing $\omega_{\mu}{}^{AB}$ as the longitudinal spin connection, we this new affine connection is defined by the following metric compatibility conditions
\begin{align}
\widetilde\nabla_\mu \tau_\nu{}^A &= \partial_\mu \tau_\nu{}^A -\omega_\mu{}^{AB}\tau_{\nu B} -b_{\mu}\tau_\nu{}^A -\widetilde\Gamma^\rho_{\mu\nu}\tau_\rho{}^A = 0\,,\\
\widetilde\nabla_\mu H^{\rho\sigma} &= \partial_\mu H^{\rho\sigma} -b_{\mu}H^{\rho\sigma} +\widetilde\Gamma^\rho_{\mu\lambda}H^{\lambda\sigma} +\widetilde\Gamma^\sigma_{\mu\lambda}H^{\rho\lambda} = 0\,.
\end{align}
The solution to these equations is 
\begin{align}
\widetilde\Gamma^\rho_{\mu\nu} = \Gamma^\rho_{\mu\nu}-\tau^\rho{}_A \left(b_\mu \tau_\nu{}^A +\omega_\mu{}^{AB}\tau_\nu{}_B\right)-\frac{1}{2} H^{\rho\sigma}\left(b_\mu H_{\nu\rho}+b_\nu H_{\mu\rho}-b_\rho H_{\mu\nu}\right)\, 
\end{align}
where the dilatation and spin connections are explicitly given by
\be
b_{\mu} = \frac{1}{3} a_\mu + \frac{1}{6} \tau_\mu{}^A a_A\, , \quad
\omega_\mu{}^{AB} = -a_\mu{}^{[AB]} + \frac{1}{2} \tau_\mu{}^C a^{AB}{}_C +\tau_\mu{}^{[A}a^{B]} \,.
\ee

\subsection{Boost invariance}
\label{11DBoost}

Now let's consider the boost transformations defined in \eqref{action_boost}. 
The calculations are very similar to those in the previous subsection.
The variation of $S^{(3)}$ under \eqref{action_boost} vanishes identically.
The variation of the finite action gives
\be
\begin{split}
\delta S^{(0)} 
 = \int \dd^{11}x &  \Big[
  -  \Lambda_{\rho}{}^A \left( 2 H^{\mu \rho}   \tau^{\nu}{}_{A} \Omega \mathcal{G}_{\mu\nu}^{(3)}
 +3 \epsilon_{ABC}\tau_\mu{}^B \tau_\nu{}^C \mathcal{E}_{(0)}^{\mu\nu\rho}\right) + \delta_\Lambda \widetilde C_{\mu\nu\rho} \mathcal{E}_{(3)}^{\mu\nu\rho} 
 \Big]\,,
\end{split} 
\label{deltaS0}
\ee
and the combination of $\mathcal{G}$ and $\mathcal{E}$ terms appearing here is
\be
- 2 \Omega \mathcal{G}^{(3)}_{A \mu}\Lambda^{\mu A} -3  \epsilon_{ABC}\mathcal{E}_{(0)}^{\mu AB} \Lambda_{\mu}{}^C  = \tfrac{1}{6} F^{A \mu\nu\rho}\Lambda^\sigma{}_A F_{\sigma \mu\nu\rho} -\tfrac{ \epsilon^{\lambda_1 \dots \lambda_3\sigma_1 \dots\sigma_8}}{4\cdot 4!^2 \Omega} F_{\sigma_1\dots\sigma_4} F_{\sigma_5\dots\sigma_8}\Lambda_{\lambda_1}{}^A\tau_{\lambda_2}{}^B\tau_{\lambda_3}{}^C \epsilon_{ABC}   \,.
\ee
Using $\Lambda_{\mu A} \tau^{\mu}{}_B= 0$ and the Schouten identity this can be shown to be proportional to the self-duality constraint.
Hence the finite action $S^{(0)}$ is invariant under boosts up to a total derivative and the self-duality constraint. 
To make the action boost-invariant off-shell we must improve the transformations \eqref{boosts} by requiring $\widetilde{F}$ to transform as well, similarly to \eqref{FtildeDil}. The improved boost transformations are
\be
\begin{split}
\delta_\Lambda H_{\mu\nu}  & = 2 \Lambda_{(\mu}{}^A \tau_{\nu) A}\,, \quad\quad\quad\quad\quad\,\,\,\,\,\,\,
 \delta_\Lambda \tau^\mu{}_A   = - H^{\mu\nu} \Lambda_{\nu A}\,,  \quad\\
\delta_\Lambda C_{\mu\nu\rho} &= - 3 \epsilon_{ABC}\Lambda_{[\mu}{}^A \tau_\nu{}^B \tau_{\rho]}{}^C\,, \quad
\delta_\Lambda \widetilde{F}_{ \mu\nu\rho\sigma} =-4 \tau^{\lambda}{}_A F_{\lambda [\mu\nu\rho}\Lambda_{\sigma]}{}^A\,.
\end{split}
\label{boostsImp}
\ee
Furthermore, one can then check that the set of equations of motion presented in the previous sections is boost-invariant (i.e. closed under boosts) as expected. 
This includes the extra equation of motion \eqref{CovariantPoisson}, which under boosts transforms into the time-space projection of the equation of motion of $H^{\mu\nu}$, equation \eqref{HTS}, as well as the self-duality constraint. This further implies that it is consistent to include it on the same footing as the remaining equations of motion that can be derived by varying $S^{(0)}$.
Indeed, one can obtain the boost variation directly from that of $S^{(-3)}$, which is:
\be
\begin{split}
\delta S^{(-3)} 
 = \int \dd^{11}x &  \Big[
  -  \Lambda_{\rho}{}^A \left( 2 H^{\mu \rho}   \tau^{\nu}{}_{A} \Omega \mathcal{G}_{\mu\nu}^{(0)}
 +3 \epsilon_{ABC}\tau_\mu{}^B \tau_\nu{}^C \mathcal{E}_{(-3)}^{\mu\nu\rho}\right) + \delta_\Lambda \widetilde C_{\mu\nu\rho} \mathcal{E}_{(0)}^{\mu\nu\rho} 
 \Big]\,.
\end{split} 
\label{deltaSm3boost}
\ee
The quantity in round brackets is exactly the time-space projection of the $H^{\mu\nu}$ equation of motion.
(As a side-remark, note that this means that the boost variation of $S^{(-3)}$ is not identically zero, although it is zero on using the equations of motion following from the finite action.)

\section{Dimensional reductions and type IIA Newton-Cartan} 
\label{dimRed}

In this section we will  propose reductions from the 11-dimensional Newton-Cartan theory to ten-dimensional type IIA Newton-Cartan theories.
We have a choice of whether to reduce on a longitudinal or a transverse direction.
Reducing on a longitudinal direction will lead to type IIA stringy Newton-Cartan with RR fields.
Reducing on a transverse direction will lead to a novel type IIA Newton-Cartan geometry which can be thought of as arising from a non-relativistic limit associated to D2 branes rather than strings. 
Similar reductions have been carried out in \cite{Kamimura:2005rz,Kluson:2019uza} from the M2 worldvolume theory.

For comparison with the reduction ansatzes below, let us record here the usual decomposition of the eleven-dimensional metric and three-form into ten-dimensional fields:
\be
d \hat s_{11}^2  = e^{4\hat \Phi/3} ( dy+ \hat A_1)^2 + e^{ -2\hat \Phi/3} d\hat s_{10}^2 \,,\quad
\hat C_3  = \hat A_3 + \hat B_2 \wedge  dy \,,
\label{usualIIA}
\ee
where $y$ denotes the direction on which we reduce.

\paragraph{Index book-keeping} 

In this section, we denote the 11-dimensional Newton-Cartan fields and curved spacetime indices with hats, thus $\hat h^{a}{}_{\hmu}$, $\hat \tau_{\hmu}{}^{A}$, $\hat \Omega$, and so on such that the 11-dimensional coordinates are $x^{\hmu} = ( x^\mu, y)$, with $\mu=0,\dots,9$. We assume that we have an isometry in the $y$ direction.
The 11-dimensional three-forms are denoted $C_{\hmu\hnu\hrho}$, $\widetilde C_{\hmu\hnu\hrho}$. 

\subsection{Type IIA SNC} 
\label{IIASNC} 

Here we present a reduction ansatz which produces the known Stueckelberg gauge-fixed form of the SNC NSNS sector action, supplemented with RR fields. 

\paragraph{Reduction ansatz} 
We want to reduce on a longitudinal direction.
We therefore split the longitudinal index $A = (\uA,2)$ with $\uA=0,1$.
Then we single out 
\be
\hat \tau^2 \equiv e^{2 \Phi /3} ( dy + A_\mu dx^\mu)\,,
\ee
thereby defining the dilaton $\Phi$ and RR one-form $A_\mu$ that will appear in the reduced theory.
If we take $\hat \tau_2 = e^{-2\Phi/3} \partial_y$ then the remaining pair of Newton-Cartan clock forms and vectors must have the form
\be \label{IIAans1}
\hat \tau^{\uA} = e^{-\Phi/3} \tau_\mu{}^{\uA} dx^\mu \,,\quad
\hat \tau_{\uA} = e^{+\Phi/3} ( \tau^\mu{}_{\uA} \partial_\mu, - \tau^\nu{}_{\uA} A_\nu \partial_y ) \,.
\ee
A compatible ansatz for the transverse vielbein is
\be\label{IIAans2}
\hat h^{ a}{}_{\hmu} = ( e^{- \Phi/3}  h^{ a}{}_{\mu} , 0 ) \,,\quad
\hat h^{\hmu}{}_{ a} = (e^{\Phi/3} h^\mu{}_{ a} , - e^{\Phi/3} h^{\nu}{}_{ a} A_\nu ) \,.
\ee
These are such that $\tau_\mu{}^{\uA}$, $\tau^\mu{}_{\uA}$ and $h^\mu{}_a$, $h^a{}_\mu$ are ten-dimensional fields obeying the usual stringy Newton-Cartan completeness identities.
We can define $\tau_{\mu\nu} \equiv \tau_{\mu}{}^{\uA}\tau_\nu{}^{\uB} \eta_{\uA \uB}$, $H_{\mu\nu} \equiv h^{ a}{}_\mu h^{ b}{}_{\nu} \delta_{ a  b}$, and similarly for the projective inverses.
We also have
\be
\hat \Omega = e^{-8 \Phi/3} \Omega\,,\quad
\Omega \equiv \tfrac{1}{2!8!}  \epsilon^{\mu \nu \sigma_1 \dots \sigma_{8}} \epsilon_{\uA \uB} \epsilon_{ a_1 \dots  a_8} \tau_{\mu}{}^{\uA} \tau_{\nu}{}^{\uB} h^{ a_1}{}_{\sigma_1} \dots h^{ a_8}{}_{\sigma_8}\,.
\ee
Finally, we make the traditional decomposition of the three-form and its field strength:
\be
 C_3 = A_3 + B_2 \wedge dy \,,\quad
F_4 = G_4 + H_3 \wedge (dy+A_1) \,,\quad
G_4 = dA_3 - A_1\wedge \mathcal{H}_3 \,,\quad
\mathcal{H}_3 = d B_2 \,,
\label{FtoIIA}
\ee
where $A_1 \equiv A_\mu dx^\mu$,
along with
\be
\widetilde C_3 = \widetilde A_3 + \widetilde B_2 \wedge dy\,,\quad
\widetilde F_4 = \widetilde G_4 + \widetilde{\mathcal{H}}_3 \wedge (dy+A_1) \,,\quad
\widetilde G_4 = d\widetilde A_3 - A_1 \wedge \widetilde{\mathcal{H}}_3 \,,\quad
\widetilde{\mathcal{H}}_3 = d \widetilde B_2 \,.
\label{tildeFtoIIA}
\ee

\paragraph{Interpretation as an expansion} 

Inserting the above ansatz into the original limit \eqref{MNCexp_split} gives 
\be
\begin{split} 
d\hat s_{11}^2 & = c^2 e^{4\Phi/3} (dy + A_1)^2 + e^{-2\Phi/3} ( c^2 \tau_{\mu\nu} + c^{-1} H_{\mu\nu} )  \,,\\
\hat C_3 & = - c^3 \tfrac{1}{2}  \epsilon_{\uA \uB} \tau^{\uA} \wedge \tau^{\uB} \wedge (dy+A_1) + A_3 + B_2 \wedge dy + c^{-3} ( \widetilde A_3 + \widetilde B_2 \wedge dy ) \,.
\end{split}
\ee
Hence according to \eqref{usualIIA} this translates into the following expansion of the ten-dimensional type IIA string frame metric $\hat g_{\mu\nu}$, NSNS two-form, $\hat B_2$, and dilaton $\hat \Phi$:
\be
\begin{split}
\hat g_{\mu\nu} & = c_{s}^2 \tau_{\mu\nu} + H_{\mu\nu}  \,,\\
\hat B_2 & = - \tfrac12 c_{s}^2  \epsilon_{\uA \uB} \tau^{\uA} \wedge \tau^{\uB}  + B_2 + c_{s}^{-2} \widetilde B_2 \,,\\
e^{\hat \Phi} & = c_{s} e^{\Phi} 
\,,
\end{split} 
\label{SNC1}
\ee
where $c_{s} \equiv c^{3/2}$. This is nothing but the limit leading to stringy Newton-Cartan.
In addition, we have an expansion of the RR fields:
\be
\hat A_3 = A_3 - \tfrac12 c_s^2 \epsilon_{\uA \uB} \tau^{\uA} \wedge \tau^{\uB} \wedge A_1
 + c_{s}^{-2} \widetilde A_3 \,,\quad
\hat A_1 = A_1 \,,
\label{SNC2} 
\ee
It is clear from these expressions that we can equivalently view this reduction as the result of the usual M-theory to type IIA reduction using \eqref{usualIIA} followed by the SNC field redefinitions of \eqref{SNC1} and \eqref{SNC2}.
At first glance, this is not completely general, given that the ansatz for the RR 1-form $A_1$ does not involve a subleading term while the other gauge fields do. 
A justification for the above ansatz is that it correctly produces the NSNS sector dynamics of SNC. Modifications to the ansatz would involve relaxing the implicit Stueckelberg gauge-fixing in 11-dimensions and comparing this to the possible 10-dimensional expansions. We do not consider this in this paper.

\paragraph{Constraint} The constraint \eqref{magic} becomes
\be
\Omega  H^{\mu_1 \nu_1} H^{\mu_2 \nu_2}  H^{	\mu_3 \nu_3}  H^{\mu_4 \nu_4} G_{\nu_1\nu_2\nu_3\nu_4}
= 
-  \tfrac{1}{4! 2!}\epsilon^{\mu_1 \dots \mu_{10}} G_{\mu_5 \mu_6\mu_7\mu_8} \epsilon_{\uA\uB} \tau_{\mu_9}{}^{\uA} \tau_{\mu_{10}}{}^{\uB}
\label{magicIIA}
\ee
and so only involves the RR 4-form field strength. The field strength of the NSNS 2-form is not constrained.
This is to be expected, as the limit of the NSNS sector alone makes sense without any constraint, and in the eleven-dimensional case the constraint arose as a consequence of the Chern-Simons term, which is not present in the truncation to the NSNS sector.

\paragraph{Type IIA SNC with RR fields} 
The action obtained from the reduction ansatz \eqref{IIAans1} and \eqref{IIAans2} is
\begin{equation}
S_{\text{IIA SNC}} = \int \dd^{10}x\, \Omega\, \left(e^{-2\Phi}\mathcal{L} +\mathcal{L}_{\widetilde{G}} + \Omega^{-1} \mathcal{L}_{\text{top}}\right)
\end{equation}
with
\begin{align}
\begin{split}
 \mathcal{L} = & \mathcal{ R} - a^{\mu \uA \uB}a_{ \mu \{\uA \uB\}} + (a^{\mu}-2D^\mu \Phi)( a_{ \mu} - 2D_\mu \Phi ) -\tfrac{1}{12} \mathcal{H}^{\mu\nu\rho}\mathcal{H}_{\mu\nu\rho} -\tfrac{1}{2} \epsilon_{\uA \uB}\tau^{\rho \uA}\mathcal{H}_{\rho\mu\nu} T^{\mu\nu \uB}\\
& -\tfrac{1}{2} e^{2\Phi} {G}^{\mu \uA}{G}_{\mu \uA} -\tfrac{1}{12} e^{2\Phi}{G}^{\mu\nu\rho \uA} {G}_{\mu\nu\rho \uA} +\tfrac{1}{4} e^{2\Phi} \epsilon^{\uA\uB}{G}_{\uA\uB\rho\sigma} {G}^{\rho\sigma},\\
\mathcal{L}_{{\widetilde{G}}} &= -\tfrac{1}{4!}{ \widetilde{G}}_{\nu_1 \dots \nu_4}\left( {G}^{\nu_1 \dots\nu_4}+\tfrac{1}{4!2! \Omega} \epsilon^{\nu_1 \dots\nu_4\mu_1\dots\mu_{6} } {G}_{\mu_1 \dots\mu_4} \epsilon_{\uA \uB} \tau_{\mu_5}{}^{\uA}\tau_{\mu_6}{}^{\uB} \right)\,,\\
\mathcal{L}_{\text{top}}&=  \tfrac{1}{2} dA_3\wedge dA_3 \wedge B_2\,,
\end{split}
\end{align}
using the field strengths defined in \eqref{FtoIIA} and \eqref{tildeFtoIIA} along with $G_{\mu\nu} \equiv 2 \partial_{[\mu} A_{\nu]}$.
As before, we write for convenience $G^{\mu\nu} \equiv H^{\mu \rho} H^{\nu \sigma} G_{\rho \sigma}$.
The Ricci scalar and connection, torsion, acceleration and so on are defined in the same way as before but for the SNC geometry.
If we ignore the RR fields, this is exactly the Stueckelberg gauge fixed action for NSNS SNC (note that the subleading component $\widetilde{B}_2$ only appears in the definition of $\widetilde{G}_4$).
Furthermore, one can check that the reduction of the Poisson equation agrees with the Poisson equation for SNC, with of course additional contributions from the RR sector. The reduced Poisson equation is found to be
\begin{align}
\begin{split} 
- \tfrac{1}{2}&\epsilon_{{\mathsf{A}}{\mathsf{B}}}\nabla_\mu \mathcal{H}^{{\mathsf{A}}{\mathsf{B}} \mu}
 +\nabla^{\mathsf{A}}\mathcal{K}_{\mathsf{A}}
-2 \tau^{\mu\nu} \nabla_\mu \nabla_\nu \Phi
+ 2 \tau^{\mu\nu} \nabla_\mu a_\nu 
+ \epsilon_{{\mathsf{A}}{\mathsf{B}}} \mathcal{H}^{{\mathsf{A}}{\mathsf{B}} \mu} \nabla_\mu \Phi 
- 2  a^{\mathsf{A}} \nabla_{\mathsf{A}} \Phi
\\
& 
+ \mathcal{K}^{\mu\nu \mathsf{A}}\mathcal{K}_{\mu\nu \mathsf{A}}
+a^{\mathsf{A}} \mathcal{K}_{\mathsf{A}} 
+2 a^{\mu [{\mathsf{A}}{\mathsf{B}}]} \mathcal{K}_{\mu\mathsf{A} {\mathsf{B}}}
+a^{{\mathsf{A}}{\mathsf{B}}{\mathsf{C}}}\left(\tfrac{1}{4}a_{\uA \uB \uC}
+\tfrac{1}{2}a_{\uA\uC\uB}
+\eta_{{\mathsf{B}}{\mathsf{C}}}a_{\mathsf{A}}\right)\\
&
+\tfrac{1}{4} \mathcal{H}^{{\mathsf{A}}\mu\nu}\mathcal{H}_{{\mathsf{A}}\mu\nu}- \epsilon_{{\mathsf{A}}{\mathsf{B}}}\mathcal{H}^{{\mathsf{C}}{\mathsf{B}} \mu} \left(a_{\mu {\mathsf{C}}}{}^{\mathsf{A}} 
+\tfrac{1}{2} a_\mu \delta^{\mathsf{A}}_{\mathsf{C}}\right) 
+\tfrac{1}{4} e^{2\Phi}\left( G^{{\mathsf{A}}{\mathsf{B}}}G_{{\mathsf{A}}{\mathsf{B}}} 
+\tfrac{1}{2} G^{{\mathsf{A}}{\mathsf{B}}\mu\nu}G_{{\mathsf{A}}{\mathsf{B}}\mu\nu}
  \right)  \\
&
-  e^{2\Phi}\tfrac{1}{48}\left( \widetilde{G}^{\mu\nu\rho\sigma} \widetilde{G}_{\mu\nu\rho\sigma} 
+\tfrac{1}{48 \Omega}\epsilon^{ \lambda_1\lambda_2  \mu_1 \dots\mu_8} \widetilde{G}_{  \mu_1 \dots\mu_4}\widetilde{G}_{ \mu_5 \dots\mu_8}\epsilon_{{\mathsf{A}}{\mathsf{B}} } \tau_{ \lambda_1}{}^{\mathsf{A}}  \tau_{ \lambda_2}{}^{\mathsf{B}} \right)  
\\ 
& = 0
\,.
\end{split}
\end{align}
In this case \cite{Bergshoeff:2021bmc}, it is the longitudinal components of the NSNS 2-form playing the role of the Newton potential.
It is also interesting to look at the reduction of the equation \eqref{folcon}, which was the equation of motion of the longitudinal components of the three-form.
This reduces to 
\be
\tfrac{1}{2} \eta_{\uA \uB} H^{\mu\rho} H^{\nu\sigma} T_{\mu\nu}{}^{\uA} T_{\rho\sigma}{}^{\uB} 
= - \tfrac{1}{48} e^{2\Phi} H^{\mu_1 \nu_1} \dots H^{\mu_4 \nu_4} G_{\mu_1 \dots \mu_4} G_{\nu_1 \dots \nu_4}  \,,
\ee
and in particular in the truncation to the NSNS sector the right-hand side is zero. This allows imposing foliation constraints on the NSNS sector SNC torsion $T_{\mu\nu}{}^{\uA}$, such as those discussed in \cite{Bergshoeff:2021bmc}.

\subsection{Type IIA D2NC}
\label{D2NC} 

\paragraph{General decompositions breaking local rotational invariance} 

The next reduction we do involves reducing on a transverse reduction.
This breaks part of the local $\mathrm{SO}(8)$ rotational invariance.
Accordingly, write the flat index $a = ( \mathsf{a}, \bri)$, with $\mathsf{a}=1,\dots,8-q$ and $\bri =1\dots q$. 
Simultaneously we can consider a \emph{different} decomposition of the spacetime coordinate index $\hmu=(\mu,i)$ where $\mu$ is $n$-dimensional and $i$ is $(11-n)$-dimensional. 
We then pick a lower triangular form for the vielbein $\hat h^{a}{}_{\hmu}$ such that
\be
\hat h^{a}{}_{\hmu} = \begin{pmatrix}
h^{\mathsf{a}}{}_\mu & 0 \\
A_\mu{}^k h^{\bri}{}_k & h^{\bri}{}_i 
\end{pmatrix} \,.
\label{breakh}
\ee
The condition $\hat h^{a}{}_{\hmu} \hat \tau^{\hmu}{}_{A}=0$ implies
\be
h^{\mathsf{a}}{}_{\mu} \hat \tau^\mu{}_{A} = 0 \,,\quad
h^{\bri}{}_i ( \hat \tau^i{}_{A} + A_\mu{}^i \hat\tau^\mu{}_{A} ) = 0\,.
\label{breakhv}
\ee
The diagonal blocks in \eqref{breakh} will in general not be square.
Two interesting examples however are to take these blocks to be square and invertible.
In this subsection, we will take the lower right block to be a non-zero $1\times 1$ matrix, and perform a reduction to a novel type of type IIA Newton-Cartan geometry associated to D2 branes.
In section \ref{AndExFT}, we will take the upper left block to be an invertible $(11-d)\times (11-d)$ matrix, and offer a description of the M-theory Newton-Cartan theory in terms of exceptional field theory.  

\paragraph{Transverse reduction to type IIA}

The dimensional reduction to type IIA corresponds to taking $n=10$, and $q=1$ above.
We again label the coordinates again as $x^{\hmu} = (x^\mu, y)$.
In this case $h^{\bar y}{}_y$ is a scalar and we can identify it with the dilaton as $h^{\bar y}{}_y \equiv e^{2\Phi/3}$.\footnote{Enthusiasts of non-relativistic geometries could also consider null reductions of the already non-relativistic theory.}
Using the conditions \eqref{breakhv}, the full Kaluza-Klein ansatz is:
\be
\hat h^{a}{}_{\hmu} = \begin{pmatrix}
e^{-\Phi/3} h^{\mathsf{a}}{}_\mu & 0 \\
 e^{2\Phi/3}A_\mu{} & e^{2\Phi/3}
\end{pmatrix} \,,\quad
\hat h^{\hmu}{}_{a}  = \begin{pmatrix}
e^{\Phi/3} h^\mu{}_{\mathsf{a}} & 0 \\
 -e^{\Phi/3} A_\nu h^\nu{}_{\mathsf{a}} & e^{-2\Phi/3}
\end{pmatrix} \,,
\label{breakhDNC}
\ee
\be
\hat \tau_{\hmu}{}^{A} = e^{-\Phi/3} ( \tau_\mu{}^{A},0) \,,\quad
\hat \tau^{\hmu}{}_{A} = e^{+\Phi/3} (  \tau^{\mu}{}_{A}, -A_\nu \tau^{\nu}{}_{A} ) \,,
\ee\label{tauDNC}
plus the same definitions \eqref{FtoIIA} and \eqref{tildeFtoIIA} for the three-forms and field strengths.
We also have
\be
\hat \Omega = e^{-8 \Phi/3} \Omega\,,\quad
\Omega \equiv \tfrac{1}{3!7!}  \epsilon^{\mu \nu \rho \sigma_1 \dots \sigma_{7}} \epsilon_{ABC} \epsilon_{\mathsf{a}_1 \dots \mathsf{a}_7} \tau_{\mu}{}^A \tau_{\nu}{}^B \tau_{\rho}{}^C  h^{\mathsf{a}_1}{}_{\sigma_1} \dots h^{\mathsf{a}_7}{}_{\sigma_1}\,.
\ee

\paragraph{Interpretation as an expansion} 

Inserting the above ansatz into the original limit \eqref{MNCexp_split} gives 
\be
\begin{split} 
d\hat s_{11}^2 & = c^{-1} e^{4\Phi/3} (dy + A_1)^2 + e^{-2\Phi/3} ( c^2 \tau_{\mu\nu} + c^{-1} H_{\mu\nu} )  \,,\\
\hat C_3 & = - c^3 e^{-\Phi} \tfrac{1}{3!}   \epsilon_{A B C} \tau^{A} \wedge \tau^{B} \wedge \tau^{C}+ A_3 + B_2 \wedge dy + c^{-3} ( \widetilde A_3 + \widetilde B_2 \wedge dy ) \,.
\end{split}
\ee
Hence according to \eqref{usualIIA} this translates into the following expansion of the ten-dimensional type IIA string frame metric $\hat g_{\mu\nu}$, RR three-form, $\hat C_2$, and dilaton $\hat \Phi$:
\be
\begin{split}
\hat g_{\mu\nu} & = c_{D}^2 \tau_{\mu\nu} +  c_D^{-2} H_{\mu\nu}  \,,\\
\hat C_3  & = - c_{D}^4  \epsilon_{ABC} e^{-\Phi} \tau^A \wedge \tau^B \wedge \tau^C  + C_3 + c_{D}^{-4} \widetilde C_3 \,,\\
e^{\hat \Phi} & = c_{D}^{-1} e^{\Phi} \,,
\end{split} 
\label{D2exp}
\ee
along with expansions for the NSNS two-form, $\hat B_2$, and RR one-form, $\hat A_1$:
\be
\hat B_2 = B_2 + c_{D}^{-4} \widetilde B_2,,\quad
\hat A_1 = A_1\,,\quad  
\ee
where $c_D \equiv c^{3/4}$.
This is an expansion and non-relativistic limit associated to the D2 brane (the powers of $c_D$ appear in the same way as those of the harmonic function in the D2 brane SUGRA solution). We can refer to it as D2 Newton-Cartan (D2NC).

\paragraph{Constraint} 

The constraint \eqref{magic} becomes
\be
\begin{split}
\Omega  H^{\mu_1 \nu_1} H^{\mu_2 \nu_2}  H^{	\mu_3 \nu_3}  H^{\mu_4 \nu_4} G_{\nu_1\nu_2\nu_3\nu_4}
 & = 
+ \tfrac{1}{3! 3!}e^{-\Phi} \epsilon^{\mu_1 \dots \mu_{10}} \mathcal{}H_{\mu_5 \mu_6\mu_7} \epsilon_{ABC} \tau_{\mu_8}{}^A \tau_{\mu_{9}}{}^B \tau_{\mu_{10}}{}^C
\,,\\
\Omega  e^{-\Phi} H^{\mu_1 \nu_1} H^{\mu_2 \nu_2}  H^{\mu_3 \nu_3} \mathcal{H}_{\nu_1\nu_2\nu_3}
 & = 
+ \tfrac{1}{4! 3!}\epsilon^{\mu_1 \dots \mu_{10}} G_{\mu_4 \mu_5\mu_6 \mu_7} \epsilon_{ABC} \tau_{\mu_8}{}^A \tau_{\mu_{9}}{}^B \tau_{\mu_{10}}{}^C
\,,\\
\end{split} 
\label{magicDIIA}
\ee
which are equivalent. 
So now we have a duality relation between the RR 3-form gauge field and the NSNS 2-form.

\paragraph{Type IIA D2 Newton-Cartan theory}  
The action obtained from the reduction ansatz \eqref{breakhDNC} and \eqref{tauDNC} is
\begin{equation}
S_{\text{D2NC}} = \int \dd^{10}x\, \Omega\, \left(e^{-2\Phi}\mathcal{L} +\mathcal{L}_{\widetilde{G}} + \Omega^{-1} \mathcal{L}_{\text{top}}\right)
\end{equation}
with
\begin{align}
\begin{split}
 \mathcal{L} = &\, \mathcal{ R} - a^{\mu AB}a_{ \mu (AB)} + \tfrac{3}{2}a^{\mu}a_{\mu}- 5a^\mu D_\mu \Phi +\tfrac{9}{2} D^\mu \Phi D_\mu \Phi -\tfrac{1}{4} \mathcal{H}^{\mu\nu A}\mathcal{H}_{\mu\nu A}\\
& -\tfrac{1}{4} e^{2\Phi} {G}^{\mu \nu} {G}_{\mu \nu} -\tfrac{1}{12} e^{2\Phi} {G}^{\mu\nu\rho A} {G} _{\mu\nu\rho A} +\tfrac{1}{4} e^{\Phi} \epsilon^{ABC} {G} _{AB\rho\sigma}T^{\rho\sigma}{}_C,\\
\mathcal{L}_{{\widetilde{G}} }  =&\, -\tfrac{1}{4!}{\widetilde{G}} _{\nu_1\dots\nu_4}\left( {G}^{\nu_1\dots\nu_4}-\tfrac{1}{3!^2 \Omega}e^{-\Phi} \epsilon^{\nu_1\dots\nu_4 \mu_1\dots \mu_6} \mathcal{H}_{\mu_1 \dots\mu_3} \epsilon_{ABC} \tau_{\mu_4}{}^{A}\tau_{\mu_5}{}^{B}\tau_{\mu_6}{}^{C} \right)\,\\
 &-\tfrac{1}{3!}e^{-2\Phi}\mathcal{\widetilde{H}}_{\nu_1\dots\nu_3 }\left( \mathcal{H}^{\nu_1\dots\nu_3}-\tfrac{1}{4!3! \Omega}e^{+\Phi} \epsilon^{\nu_1\dots\nu_3\mu_1\dots\mu_7 } {G} _{\mu_1 \dots\mu_4} \epsilon_{ABC} \tau_{\mu_5}{}^{A}\tau_{\mu_6}{}^{B}\tau_{\mu_7}{}^{C} \right),\\
  = & \, -\tfrac{1}{4!}
 \left( 
  {\widetilde{G}} _{\nu_1\dots\nu_4} - \tfrac{1}{3!} e^{-\Phi} \widetilde{\mathcal{H}}_{\rho_1 \dots \rho_3} \epsilon^{\rho_1 \dots \rho_3 \sigma_1 \dots \sigma_7} \tfrac{1}{3! \Omega} \epsilon_{ABC} H_{\nu_1 \sigma_1} \dots H_{\nu_4 \sigma_4} \tau_{\sigma_5}{}^A \tau_{\sigma_6}{}^B \tau_{\sigma_7}{}^C   \right)\times
 \\ & \qquad \times \left( {G}^{\nu_1\dots\nu_4}-\tfrac{1}{3!^2 \Omega}e^{-\Phi} \epsilon^{\nu_1\dots\nu_4 \mu_1\dots \mu_6} \mathcal{H}_{\mu_1 \dots\mu_3} \epsilon_{ABC} \tau_{\mu_4}{}^{A}\tau_{\mu_5}{}^{B}\tau_{\mu_6}{}^{C} \right)\,,\\
\mathcal{L}_{\text{top}} =&\, \tfrac{1}{2 } dA_3\wedge dA_3 \wedge B_2\,,
\end{split}
\end{align}
where the field strengths are defined as in \eqref{FtoIIA} and \eqref{tildeFtoIIA} with again $G_2 \equiv d A_1$.
Note that we obtain what appears to be an extra contribution to the dilaton kinetic term due to the $e^{-\Phi}$ factor that in the expansion of $\hat C_3$ in \eqref{D2exp}. We could alter this by redefining the RR fields in the reduced theory. 
In addition, the reduction of the Poisson equation \eqref{CovariantPoisson} gives
\begin{align}
\begin{split}
& \tfrac{1}{6}e^{\Phi}\epsilon_{ABC} \left( \nabla_\mu G^{ABC \mu} + a_\mu  G^{ABC \mu}
+3 a_{\mu D}{}^A G^{DBC\mu}\right)
-\tfrac{1}{3}e^{\Phi}\epsilon_{ABC} G^{ABC \mu} \nabla_\mu \Phi 
\\ & 
+\nabla^A \mathcal{K}_A
-3 \tau^{\mu\nu} \nabla_\mu \nabla_\nu \Phi  
- 3  a^A \nabla_A \Phi
+2 \nabla^A\Phi  \nabla_A \Phi
-\mathcal{K}^A \nabla_A \Phi
+2 \tau^{\mu\nu} \nabla_\mu a_\nu 
\\ & 
+\mathcal{K}^{\mu \nu A}\mathcal{K}_{\mu\nu A}
+a^A\mathcal{K}_A 
+ 2a^{\mu [AB]} \mathcal{K}_{\mu AB} 
+a^{{{A}}{{B}}{{C}}}\left(\tfrac{1}{4}a_{A B C}
+\tfrac{1}{2}a_{ACB}
+\eta_{{{B}}{{C}}}a_{{A}}\right)\\
& 
+\tfrac{1}{4} \mathcal{H}^{AB\mu}\mathcal{H}_{AC\mu} 
+\tfrac{1}{8} e^{2\Phi}\left( G^{AB\mu\nu}G_{AB\mu\nu} 
+4 G^{A\mu}G_{A\mu}\right) 
-  e^{2\Phi}\tfrac{1}{48} \widetilde{G}^{\mu\nu\rho\sigma} \widetilde{G}_{\mu\nu\rho\sigma}
- \tfrac{1}{12} \mathcal{\widetilde H}^{\mu\nu\rho}\mathcal{\widetilde H}_{\mu\nu\rho} 
\\ & 
+e^{-\Phi}\tfrac{1}{4! 3! 3! \Omega}\epsilon^{\lambda_1 \lambda_2 \lambda_3 \mu_1\dots \mu_7} \epsilon_{ABC} \tau_{\lambda_1}{}^A  \tau_{\lambda_2}{}^B\tau_{\lambda_3}{}^C \widetilde{G}_{\mu_1\dots\mu_4}\mathcal{\widetilde{H}}_{\mu_5\dots\mu_7}    = 0 \,. 
\end{split}
\end{align}
As in the MNC case, the longitudinal components of the three-form gauge field play the role of the Newton potential.


\section{Dimensional decompositions and exceptional field theory description}
\label{AndExFT}

\subsection{Exceptional field theory} 

We will now discuss the exceptional field theory description of the 11-dimensional MNC theory.
ExFT automatically has a number of features in common with the non-relativistic theory: breaking of 11-dimensional Lorentz symmetry, a geometry arising from mixing metric and form-field components, and the inclusion of dual degrees of freedom.
We will see how it provides a unified framework treating the relativistic and non-relativistic theory on an equal footing, which demonstrates that the same exceptional Lie algebraic structures that underlie the relativistic theory are present in the non-relativistic one.
In addition, the ExFT equations of motion include the additional missing Poisson equation.

We will focus particularly on the relatively unexceptional case of the $\Gthree$ ExFT \cite{Hohm:2015xna}.
This makes use of an $(8+3)$-dimensional split of the 11-dimensional spacetime. 
As such, it is a very natural fit for the $(8+3)$-dimensional split into transverse and longitudinal directions present in the MNC expansion.
The $\Gthree$ ExFT includes a Riemannian metric for the 8-dimensional part of the spacetime, but the 3-dimensional part is described by an `extended geometry' involving an $\Gthree$ symmetric generalised metric. 
By decomposing the 11-dimensional Newton-Cartan theory appropriately, we will replace the transverse Newton-Cartan metric with an invertible 8-dimensional metric, $\hat H^{\hmu\hnu} \rightarrow g^{\mu\nu}$, and the longitudinal metric with an invertible 3-dimensional metric, $\hat \tau_{\hmu\nu} \rightarrow \tau_{ij}$, which will be embedded into the generalised metric description.
This drastic simplification of the geometry is nonetheless sufficient to highlight the key features of the theory.

It would also be interesting to consider for example the opposite $(3+8)$-dimensional split corresponding to the $\Geight$ ExFT, embedding the transverse metric into the $\Geight$ generalised metric.
However as the known formulation of ExFT makes use of a Riemannian metric for the unextended part of the spacetime, this is not immediately available for our purposes. 
Evidently, for any given $\Edd$ ExFT, one can construct or imagine multiple other `hybrid' formulations depending on how one chooses to separate or mix the longitudinal and transverse directions.
More ambitiously, one could choose to work with the recently fully constructed `master' $E_{11}$ ExFT \cite{Bossard:2021ebg}, for which no coordinate decomposition is necessary. Evidently this would eschew the technical difficulties of the latter in favour of the technicalities associated to working with an infinite-dimensional algebra. 
In this paper, although many features that we will see are quite general, we describe the explicit details mainly for the $d \leq 4$ cases.

\paragraph{ExFT ingredients}

The basic idea behind ExFT is to replace $d$-dimensional vectors with \emph{generalised vectors} $V^M$ transforming in a specified representation of $\Edd$.
This representation is such that we can decompose $V^M$ under $\mathrm{GL}(d)$ as $V^M = (V^i, V_{ij}, V_{ijklm}, \dots)$ where $V^i$ is a $d$-dimensional vector, $V_{ij}$ and $V_{ijklm}$ a two- and five-form, and the ellipsis corresponds to higher rank mixed symmetry tensors that appear for $d \geq 7$.\footnote{This decomposition is relevant to the description of 11-dimensional SUGRA. There are also mutually inequivalent $\mathrm{GL}(d-1)$ decompositions relevant to the description of type II SUGRA.}
Generalised vectors are used to provide an $\Edd$-compatible local symmetry of \emph{generalised diffeomorphisms}.
These are defined in terms of a \emph{generalised Lie derivative} which acts on a generalised vector $V^M$ of weight $\lambda_V$ as
\be
\delta_U V^M = 
\mathcal{L}_U V^M \equiv U^N \partial_N V^M - V^N \partial_N U^M + Y^{MN}{}_{PQ} \partial_N U^P V^Q + ( \lambda_V - \tfrac{1}{9-d} ) \partial_N U^N V^M\,.
\label{gld}
\ee
Here $Y^{MN}{}_{PQ}$ is constructed from invariant tensors of $\Edd$. This together with the weight term with coefficient $-1/(9-d)$ appear such that this generalised Lie derivative involves an infinitesimal $\Edd$, rather than $\mathrm{GL}(N)$ transformation.   
The partial derivatives written here formally involve an extended set of coordinates $y^M$.
However, consistency requires the imposition of a constraint $Y^{MN}{}_{PQ} \partial_M \partial_N = 0$ where the derivatives can act on a single field or a product of fields. 
One solution to this constraint is to view the $d$-dimensional partial derivatives as being embedded such that $\partial_M = ( \partial_i , 0 , \dots , 0)$.
We always assume we have made this choice below. (An alternative solution leads to a ten-dimensional type IIB description.)

Given this choice, for the $d\leq 4$ cases we will look at in detail, the action of $U^M = ( u^i, \lambda_{ij})$ on $V^M = ( V^i, V_{ij})$ (both having generalised diffeomorphism weight $1/(9-d)$) is $\mathcal{L}_U V^M = ( L_u V^i , L_u V^{ij} - 3 V^k \partial_{[k} \lambda_{ij]})$, where $L_u$ denotes the usual $d$-dimensional Lie derivative.
Identifying the two-form components $\lambda_{ij}$ with the gauge transformation parameter of a three-form $\hat \Cdef_{ijk}$, this means we can write $V^M = ( V^i, \widetilde V_{ij} - \hat\Cdef_{ijk} V^k)$, with $\widetilde V_{ij}$ gauge invariant. We use this to give explicit parametrisations for the ExFT fields.

The field content of ExFT is as follows. We now let $\mu,\nu,\dots$ be $(11-d)$-dimensional indices.
We then have an $(11-d)$-dimensional metric, $g_{\mu\nu}$, which is a scalar of weight $-2/(9-d)$ under generalised diffeomorphisms.
The $\Edd$ extended geometry is equipped with a generalised metric, $\gM_{MN}$, transforming as a rank two symmetric tensor of weight zero under generalised diffeomorphisms.
In addition, there is a `tensor hierarchy' of gauge fields, starting with an $(11-d)$-dimensional one-form $\Aa_\mu{}^M$, and continuing with $p$-forms $\Ab_{\mu\nu}$, $\Ac_{\mu\nu\rho}$, $\dots$ in particular representations of $\Edd$.
This set of fields mimics and extends what appears in a dimensional decomposition (or reduction) of the bosonic fields of supergravity.

\paragraph{Dimensional decomposition and field redefinitions} 

We describe now the dimensional decomposition used to embed 11-dimensional SUGRA in the ExFT framework.
We split the 11-dimensional coordinates $x^{\hmu} = ( x^\mu, y^i)$, making an $(11-d)+d$ split. 
The supergravity degrees of freedom are then similarly decomposed under this split, classified according to their nature from the point of view of $(11-d)$-dimensional spacetime, and then rearranged into multiplets of the exceptional groups $\Edd$.
We assume no restriction on the coordinate dependence.
This can be viewed as a partial fixing of the local Lorentz symmetry in which we choose the 11-dimensional vielbein $\hat e^{\hat a}{}_{\hmu}$ and hence metric $\hat g_{\hmu \hnu}$ to be
\be
\hat e^{\hat a}{}_{\hmu} = \begin{pmatrix}
|\phi|^{-\tfrac{1}{2(9-d)}} e^{\mathsf{a}}{}_\mu & 0 \\
A_\mu{}^k\phi^{\bri}{}_k & \phi^{\bri}{}_i 
\end{pmatrix} \,,\quad
\hat g_{\hmu\hnu} = 
\begin{pmatrix}
|\phi|^{-\tfrac{1}{9-d}} g_{\mu\nu} + \phi_{kl} A_\mu{}^k A_\nu{}^l & \phi_{ik} A_\nu{}^l \\
\phi_{jk} A_\nu{}^k & \phi_{ij}
\end{pmatrix} \,,\quad
\label{breakeExFT}
\ee
where $e^{\mathsf{a}}{}_\mu$ is a vielbein for an $(11-d)$-dimensional (Einstein frame) metric $g_{\mu\nu}$ and $\phi^{\bri}{}_i$ is a vielbein for a $d$-dimensional metric $\phi_{ij}$, with $|\phi| \equiv |\det (\phi_{ij})|$. 
Normally one takes $g_{\mu\nu}$ to be Lorentzian, such that this corresponds to fixing the Lorentz symmetry as $\mathrm{SO}(1,10) \rightarrow \mathrm{SO}(1,10-d) \times \mathrm{SO}(d)$, however we can also take it to be Euclidean, such that $\mathrm{SO}(1,10) \rightarrow \mathrm{SO}(11-d) \times \mathrm{SO}(1,d-1)$. The latter choice is relevant for the version of ExFT applicable to the non-relativistic theory. 

The `Kaluza-Klein vector' $A_\mu{}^i$ has a field strength defined by
\be
F_{\mu\nu}{}^i = 2 \partial_{[\mu} A_{\nu]}{}^i - 2 A_{[\mu|}{}^j \partial_j A_{|\nu]}{}^i\,.
\label{kkf}
\ee
Letting $L$ denote the $d$-dimensional Lie derivative, the Kaluza-Klein vector also appears as the connection in the derivative $D_\mu = \partial_\mu - L_{A_\mu}$ which is covariant with respect to $d$-dimensional diffeomorphisms, using the transformation $\delta_\Lambda A_\mu{}^i = D_\mu \Lambda^i$ induced by the action of $11$-dimensional diffeomorphisms on \eqref{breakeExFT}.

For the three-form and its field strength, we define a succession of gauge field components (denoted by bold font) via
\be
\hat C_3 = \hat\Cdef_3 + \hat\Cdef_{2 i } Dy^i  + \tfrac{1}{2} \hat\Cdef_{1 i j } Dy^i Dy^j + \tfrac{1}{3!} \hat\Cdef_{ijk} Dy^i Dy^j Dy^k
\label{Cdecomp_compact}
\ee
where $Dy^i \equiv dy^i + A_\mu{}^i dx^\mu$, the subscripts on the right-hand side denote the form degree in $(11-d)$ dimensions, and we omit the implicit wedge products.
Similarly, for $\hat F_4 = d \hat C_3$ we let
\be
 \hat F_4 = \hat\Fdef_4 + \hat\Fdef_{3 i } Dy^i  + \tfrac{1}{2} \hat\Fdef_{2 i j } Dy^i Dy^j + \tfrac{1}{3!} \hat\Fdef_{1 ijk} Dy^i Dy^j Dy^k + \tfrac{1}{4!} \hat\Fdef_{ijkl} Dy^i Dy^j Dy^k Dy^l \,,
\label{Fdecomp_compact}
\ee
The persistence of hats reflects the fact that we still want to take the non-relativistic limit of all these quantities.
Explicit component expressions can be found in appendix \ref{appExDecomp}.
We can make similar redefinitions for the dual six-form and its field strength.

\paragraph{Metric and generalised metrics}

The metric $g_{\mu\nu}$ appearing in \eqref{breakeExFT} is directly used as the $(11-d)$-dimensional ExFT metric (the generalised diffeomorphism weight $-2/(9-d)$ follows from the conformal factor in \eqref{breakeExFT}).

The generalised metric $\gM_{MN}$, or its generalised vielbein, may be defined as an $\Edd$ element valued in a coset $\Edd/H_d$ where $H_d$ is the maximal compact subgroup (in the Euclidean case) or a non-compact version thereof (in the Lorentzian case).
Under generalised diffeomorphisms it transforms as a rank two symmetric tensor of weight zero. 
It is normally parametrised in terms of the wholly $d$-dimensional components of the eleven-dimensional fields, $\phi_{ij}$ and $\hat\Cdef_{ijk}$, in a manner consistent with its transformation under generalised diffeomorphisms.
For $d \geq 6$, this parametrisation also includes internal components of the dual-six form.
For simplicity, we will restrict to $d \leq 4$, in which case the conventional parametrisation of the generalised metric is given by 
\be
\gM_{MN} = |\phi|^{1/(9-d)}
\begin{pmatrix} 
\phi_{ij} + \frac{1}{2} \hat \Cdef_i{}^{pq} \hat \Cdef_{jpq} &\hat \Cdef_i{}^{kl} \\
 \hat \Cdef_k{}^{ij} &  2 \phi^{i[k} \phi^{l]j} 
\end{pmatrix} \,.
\label{genmet_usual}
\ee
The conformal factor here ensures that $|\det \gM| =1$. 

In specific cases, we can find factorisations of the generalised metric leading to simpler expressions.
This includes the $\Gthree$ ExFT.
Here, generalised vectors $V^M = (V^i,  V_{ij})$ transform in the $(\mathbf{3}, \mathbf{2})$ of $\Gthree$, with $i,j,\dots$ three-dimensional.
We can dualise $V_{ij}$ using the three-dimensional epsilon symbol, and define $\widetilde V^{i} \equiv \frac{1}{2} \epsilon^{ijk} \widetilde V_{jk}$. 
Introduce an $\mathrm{SL}(2)$ fundamental index, $\alpha=1,2$, and let $V^M \equiv V^{i\alpha}$ with $V^{i1} \equiv V^i$ and $V^{i2} \equiv \widetilde V^i$.
In terms of this basis we have a factorisation
\be
\gM_{MN} = \gM_{i\alpha,j\beta} = \gM_{ij} \gM_{\alpha \beta} \,,
\label{32genm}
\ee
where $\gM_{ij} = \gM_{ji}$ with $|\det \gM_{ij}|=1$, and $\gM_{\alpha \beta} = \gM_{\beta \alpha}$ with $|\det \gM_{\alpha\beta}|=1$.
When $\phi_{ij}$ has Lorentzian signature, the expressions which reproduce \eqref{genmet_usual} are 
\be
\gM_{ij} = |\phi|^{-1/3} \phi_{ij} \,,\quad
\gM_{\alpha \beta} = \begin{pmatrix}
|\phi|^{1/2} - |\phi|^{-1/2} \hat\Cdef^2 &  - |\phi|^{-1/2}\hat\Cdef \\
- |\phi|^{-1/2} \hat\Cdef & - |\phi|^{-1/2} 
\end{pmatrix} \,,\quad
\hat\Cdef \equiv \tfrac{1}{3!} \epsilon^{ijk} \hat\Cdef_{ijk} \,,
\label{32_factorised}
\ee

\paragraph{Gauge fields and dual degrees of freedom} 
\newcommand{\pmblue}{\color{blue}\pm\color{black}}

Along with the Kaluza-Klein vector, $A_\mu{}^i$, coming from the metric decomposition \eqref{breakeExFT}, the $p$-forms obtained from the decomposition \eqref{Cdecomp_compact} of the three-form fit into $\Edd$-valued multiplets denoted $\Aa_\mu$, $\Ab_{\mu\nu}$, $\Ac_{\mu\nu\rho}$, $\dots$.
Their field strengths are denoted $\Fa_{\mu\nu}$, $\Fb_{\mu\nu\rho}, \Fc_{\mu\nu\rho\sigma}$, $\dots$.
To obtain full $\Edd$ representations, we have to include here the set of $p$-forms obtained by decomposing the dual six-form. 
This is unsurprising from the point of $\Edd$ U-duality transformations, which mix electric and magnetic degrees of freedom (e.g. M2 and M5 branes) coupling respectively to $p$-forms and their duals.

For $d=3$, this works as follows \cite{Hohm:2015xna}.
The ExFT gauge fields $\Aa_\mu{}^{i\alpha}$, $\Ab_{\mu\nu i}$, $\Ac_{\mu\nu\rho}{}^\alpha$, $\Ad_{\mu\nu\rho\sigma}{}^i$ have weights $1/6, 2/6, 3/6, 4/6$ respectively, and their field strengths are denoted $\Fa_{\mu\nu}{}^{i\alpha}$, $\Fb_{\mu\nu\rho i}$, $\Fc_{\mu\nu\rho\sigma}{}^\alpha$ and $\Fd_{\mu\nu\rho\sigma\lambda}{}^i$ (the latter does not appear in the action).
Under generalised diffeomorphisms, $\Fa^{i\alpha}$ transforms as a generalised vector of weight $1/6$, while $\Fb$ and $\Fc$ transform via the generalised Lie derivative acting as
\be
\mathcal{L}_\Lambda \Fb_i = \Lambda^{j\beta} \partial_{j\beta} \Fb_i + \partial_{i \beta} \Lambda^{j\beta} \Fb_j\,,\quad
\mathcal{L}_\Lambda \Fc^\alpha  = \Lambda^{j\beta} \partial_{j \beta} \Fc^\alpha - \partial_{j \beta} \Lambda^{j\alpha} \Fc^\beta + \partial_{j \beta} \Lambda^{j \beta} \Fc^\alpha \,.
\label{32_gldF}
\ee
These field strengths obey Bianchi identities:
\be
3 \D_{[\mu} \Fa_{\nu\rho]}{}^{i\alpha} = \epsilon^{ijk} \epsilon^{\alpha \beta} \partial_{j \beta} \Fb_{\mu\nu\rho k}\,,
\label{32BianchiA}
\ee
\be
4 \D_{[\mu} \Fb_{\nu\rho\sigma] i} + 3 \epsilon_{ijk} \epsilon_{\alpha \beta} \Fa_{[\mu\nu}{}^{j \alpha} \Fa_{\rho\sigma ]}{}^{k\beta} = \partial_{i \alpha} \Fc_{\mu\nu\rho\sigma}{}^{\alpha}\,,
\label{32BianchiB}
\ee
\be
5 \D_{[\mu} \Fc_{\nu\rho\sigma\lambda]}{}^{\alpha} + 10 \Fa_{[\mu\nu}{}^{i\alpha} \cH_{\rho\sigma \lambda ]i}= \epsilon^{\alpha \beta} \partial_{i \beta} \Fd_{\mu\nu\rho\sigma\lambda}{}^i\,,
\label{32BianchiC}
\ee
where $\D_\mu \equiv \partial_\mu -\mathcal{L}_{\Aa_\mu}$.
The ExFT one-form can be simply identified as $\Aa_\mu{}^M = ( A_\mu{}^i, \tfrac{1}{2} \epsilon^{ijk} \Cdef_{\mu jk} )$.
The two-form $\Ab_{\mu\nu i}$ transforms in the $(\mathbf{\bar{3}}, \mathbf{1})$ of $\Gthree$ and is identified (up to a further field redefinition) with $\hat\Cdef_{\mu\nu i}$. 
However, rather than give the precise field redefinitions for the potentials, it is simpler to work at the level of the field strengths. These are all tensors under generalised diffeomorphisms, meaining in particular that they transform in a particular way under $d$-dimensional three-form gauge transformations.
This allows us to decompose in terms of manifestly gauge invariant combinations
\be
\mathcal{F}_{\mu\nu}{}^{i1} \equiv F_{\mu\nu}{}^i\,,\quad
\mathcal{F}_{\mu\nu}{}^{i2} \equiv \frac{1}{2}\epsilon^{ijk} ( \hat\Fdef_{\mu\nu jk} - \hat \Cdef_{jkl} \hat\Fdef_{\mu\nu}{}^l ) 
\,,\quad
\Fb_{\mu\nu\rho i} \equiv - \hat\Fdef_{\mu\nu\rho i} \,,
\label{defFH32}
\ee
where $F_{\mu\nu}{}^i$,  $\hat\Fdef_{\mu\nu\rho i}$ and $\hat\Fdef_{\mu\nu jk}$ are gauge invariant and can be exactly identified with the quantities defined in \eqref{Fdecomp_compact} with $F_{\mu\nu}{}^i$ as in \eqref{kkf}.\footnote{The minus sign in $\Fb_{\mu\nu\rho i}$ ensures that the ExFT Bianchi identities \eqref{32BianchiB} and \eqref{32BianchiC} reproduce those coming from SUGRA in \eqref{FBI_mod} and is otherwise simply a matter of convention in terms of what we call $\Ab_{\mu\nu i}$.}

The three-form situation is then where it gets interesting.
There is a single $8$-dimensional three-form $\hat \Cdef_{\mu\nu\rho}$ obtained from the 11-dimensional one.
There is also a single three-form $\hat \Cdef_{\mu\nu\rho ijk}$ coming from the 11-dimensional six-form. 
Together these form an $\mathrm{SL}(3)$ singlet and $\mathrm{SL}(2)$ doublet, for which the field strength obeys a self-duality constraint reproducing (in the relativistic case!) the correct duality relationship between the field strengths $\hat \Fdef_{\mu\nu\rho\sigma}$ and $\hat\Fdef_{\mu\nu\rho\sigma ijk}$. 
This duality constraint, which has to be imposed by hand, involves the eight-dimensional Hodge star acting on the $8$-dimensional indices and the $\mathrm{SL}(2)$ generalised metric acting on the $\mathrm{SL}(2)$ indices:
\be
\sqrt{|g|} 
\gM_{\alpha \beta} \Fc^{\mu\nu\rho\sigma\beta} = 
- 48 \kappa \epsilon_{\alpha \beta} \epsilon^{\mu\nu\rho\sigma\lambda_1\dots\lambda_4} \Fc_{\lambda_1\dots \lambda_4}{}^\beta \,.
\label{constraintJ}
\ee
The coefficient $\kappa$  is fixed via the self-consistency of \eqref{constraintJ} (in both the cases where $g_{\mu\nu}$ has Lorentzian or Euclidean signature, with $\gM_{\alpha\beta}$ having the opposite) to be
$\kappa = \pm \tfrac{1}{2 \cdot (24)^2}$, with the choice of sign being a matter of convention (equivalent to changing the sign of the three-form in eleven dimensions).
This is consistent with decomposing the $\mathrm{SL}(2)$ doublet of four-form field strengths as
\be
\mathcal{J}_{\mu\nu\rho\sigma}{}^1 \equiv\hat \Fdef_{\mu\nu\rho\sigma}
\,,\quad
\mathcal{J}_{\mu\nu\rho\sigma}{}^2 \equiv \tfrac{1}{6} \epsilon^{ijk} ( \hat \Fdef_{\mu\nu\rho\sigma ijk} - \hat\Cdef_{ijk} \hat\Fdef_{\mu\nu\rho\sigma} )\,.
\label{defJ32}
\ee
Thus in general, ExFT treats simultaneously degrees of freedom coming from the three-form with dual degrees of freedom coming from the six-form, encoding the duality relations between them in its dynamics.

\paragraph{Dynamics: $\Gthree$ ExFT pseudo-action}

The ExFT Lagrangian can be uniquely fixed by the requirement of invariance under the local symmetries (generalised diffeomorphisms, gauge transformations of the tensor hierarchy, and finally $(11-d)$-dimensional diffeomorphisms). When $11-d$ is even, this gives a pseudo-action which must be accompanied by a self-duality constraint such as \eqref{constraintJ}.
This includes the case $d=3$. The pseudo-action in this case can be written as $S= \int \dd^8x\, \dd^{6} y \sqrt{|g|} \mathcal{L}_{\text{ExFT}}$ where the Lagrangian has the (quite general) expression
\be
\begin{split}
\mathcal{L}_{\text{ExFT}} =  R_{\text{ext}}(g) + \mathcal{L}_{\text{kin}} + \mathcal{L}_{\text{int}} + \sqrt{|g|}^{-1} \mathcal{L}_{\text{top}}
\,,
\label{Lexft}
\end{split} 
\ee
Here, with $\D_\mu = \partial_\mu - \mathcal{L}_{\Aa_\mu}$, we have
\begin{align}
R_{\text{ext}}(g)  & = 
\frac{1}{4} g^{\mu \nu} \D_\mu g_{\rho \sigma} \D_\nu g^{\rho \sigma} 
- \frac{1}{2} g^{\mu \nu} \D_\mu g^{\rho \sigma} \D_\rho g_{\nu \sigma}
+ \frac{1}{4} g^{\mu \nu} \D_\mu \ln g \D_\nu \ln g 
+ \frac{1}{2} \D_\mu \ln g \D_\nu g^{\mu \nu} \,,
\label{Rext32}
\\
\mathcal{L}_{\text{kin}} & = \frac{1}{4} \D_\mu \gM^{ij} \D^\mu \gM_{ij} + \frac{1}{4} \D_\mu \gM_{\alpha \beta} \D^\mu \gM^{\alpha \beta} \nonumber
\\ & \qquad
- \frac{1}{4} \gM_{ij} \gM_{\alpha \beta} \Fa_{\mu\nu}{}^{i\alpha} \Fa^{\mu\nu j \beta} 
- \frac{1}{12} \gM^{ij} \Fb_{\mu\nu\rho i} \Fb^{\mu\nu\rho}{}_j 
- \frac{1}{96} \gM_{\alpha \beta} \Fc_{\mu\nu\rho\sigma}{}^\alpha \Fc^{\mu\nu\rho\sigma \beta} \,,
\label{Lkin32}
\\
\mathcal{L}_{\text{int}} &=
\frac{1}{4} \gM^{MN} \partial_M \gM^{kl} \partial_N \gM_{kl}
+\frac{1}{4} \gM^{MN} \partial_M \gM^{\alpha\beta} \partial_N \gM_{\alpha \beta}
 - \frac{1}{2} \gM^{MN} \partial_M \gM^{KL} \partial_K \gM_{LN} \nonumber
 \\ & \qquad
+ \frac{1}{2} \partial_M \gM^{MN} \partial_N \ln |g|
+ \frac{1}{4} \gM^{MN} \left( \partial_M g_{\mu\nu} \partial_N g^{\mu\nu} + \partial_M \ln |g| \partial_N \ln |g| \right)  \,.
\label{Lint32}
\end{align}
The topological (Chern-Simons) term can be defined 
via its variation:
\be
\begin{split} 
\delta \mathcal{L}_{\text{top}} = 
\kappa
\epsilon^{\mu_1 \dots \mu_8} \Big(&
-
 \delta \Aa_{\mu_1}{}^{i\alpha}\epsilon_{\alpha \beta}
\Fc_{\mu_2 \dots \mu_5}{}^\beta  \Fb_{\mu_6\mu_7\mu_8 i} 
\\ & + 
6 \Delta \Ab_{\mu_1 \mu_2 i} \left(
\epsilon_{\alpha \beta} \Fa_{\mu_3\mu_4}{}^{i\alpha} \Fc_{\mu_5\dots \mu_8}{}^\beta
- \tfrac{4}{9} \epsilon^{ijk} \Fb_{\mu_3 \mu_4\mu_5 j} \Fb_{\mu_6\mu_7\mu_8 k} 
\right)
\\ & 
+ 4 \Delta \Ac_{\mu_1\mu_2\mu_3}{}^\alpha \epsilon_{\alpha \beta}
\left(
\D_{\mu_4} \Fc_{\mu_5 \dots \mu_8}{}^\beta + 4 \Fa_{\mu_4\mu_5}{}^{i\beta} \Fb_{\mu_6 \dots \mu_8 i}
\right)
\\ & 
- 
 \partial_{i\alpha} \Delta \Ad_{\mu_1 \dots \mu_4}{}^i \Fc_{\mu_5 \dots \mu_8}{}^\alpha
\Big) \,,
\end{split}
\ee
where the `improved' $\Delta$ variation includes by definition contributions of variations of lower rank gauge fields, for explicit expressions (which we do not require) see \cite{Hohm:2015xna}.
Finally, we must impose the constraint \eqref{constraintJ} after varying the above pseudo-action.

\subsection{Obtaining the 11-dimensional Newton-Cartan theory via ExFT} 
\label{ExFTLimit}

In this subsection, we perform a dimensional decomposition of the 11-dimensional MNC variables, and use this to explain how exceptional field theory describes this theory.

\paragraph{Dimensional decomposition of 11-dimensional Newton-Cartan theory}

We start with the 11-dimensional coordinates labelled as $x^{\hmu} = (x^\mu, y^i)$ with $\mu=1,\dots,11-d$ and $i=1\dots,d$.
We keep all coordinate dependence on $y^i$ throughout.
Thus this is a decomposition rather than a reduction.
In terms of the vielbein decomposition \eqref{breakh}, we take $q=d-3$ and $n=11-d$.
The flat indices are $\mathsf{a}=1,\dots,11-d$ and $\bri = 1,\dots,d-3$.
Explicitly, we take the $\mathrm{SO}(8)$ vielbein to have the form
\be
\hat h^{a}{}_{\hmu} = \begin{pmatrix}
\Omega^{-\tfrac{1}{9-d}} e^{\mathsf{a}}{}_\mu & 0 \\
A_\mu{}^k h^{\bri}{}_k & h^{\bri}{}_i 
\end{pmatrix} \,,\quad
\hat h^{\hmu}{}_{a} = \begin{pmatrix}
\Omega^{\tfrac{1}{9-d}} e^\mu{}_{\mathsf{a}} & 0 \\
- \Omega^{\tfrac{1}{9-d}} e^{\rho}{}_{\mathsf{a}} A_\rho{}^k & h^i{}_{\bri} 
\end{pmatrix} \,,
\label{breakhExFT}
\ee
with $e^{\mathsf{a}}{}_\mu$ an invertible vielbein for an $(11-d)$-dimensional metric, $g_{\mu\nu} = e^{\mathsf{a}}{}_\mu e^{\mathsf{b}}{}_\nu \delta_{\mathsf{a}\mathsf{b}}$.
We also have to take 
\be
\hat \tau_{\hmu}{}^A = ( A_\mu{}^i \tau_i{}^A , \tau_i{}^{A} ) \,,\quad
\hat \tau^{\hmu}{}_A = ( 0 , \tau^i{}_A )\,.
\label{MNC_splitTau}
\ee
where $\tau_{ij} =\tau_i{}^A \tau_j{}^B \eta_{AB}$, with $A=0,1,2$ as before.
The conformal factor $\Omega$ appearing in \eqref{breakhExFT} is defined by
\be
\Omega^2 = - \tfrac{1}{3! (d-3)!} \epsilon^{i_1 \dots i_d} \epsilon^{j_1 \dots j_d} \tau_{i_1j_1} \tau_{i_2 j_2} \tau_{i_3 j_3} H_{i_4j_4} \dots H_{i_d j_d}\,,
\label{MNC_Omega_ExFT}
\ee
and related to that of the 11-dimensional theory by $\hat \Omega = (\det e) \Omega^{-\tfrac{2}{9-d}}$.
It is useful to write down the full transverse and longitudinal metrics:
\be
\begin{split}
\hat H_{\hmu \hnu} &= \begin{pmatrix}
\Omega^{-\tfrac{2}{9-d} }  g_{\mu\nu} + H_{kl} A_\mu{}^k A_\nu{}^l & H_{j k} A_\mu{}^k \\
H_{ik} A_\nu{}^k  & H_{ij} 
\end{pmatrix} \,,\quad \hat\tau_{\hmu\hnu} = \begin{pmatrix}
A_\mu{}^k A_\nu{}^l \tau_{kl} & A_\mu{}^k \tau_{kj} \\
A_\nu{}^k \tau_{ki} & \tau_{ij} 
\end{pmatrix}\,,\\
\hat H^{\hmu \hnu} & = \begin{pmatrix}
\Omega^{\tfrac{2}{9-d} } g^{\mu\nu} & - \Omega^{\tfrac{2}{9-d} }  g^{\mu \rho} A_\rho{}^j \\
- \Omega^{\tfrac{2}{9-d} }  g^{\nu \sigma} A_\sigma{}^i & H^{ij} + \Omega^{\tfrac{2}{9-d} }  g^{\rho\sigma} A_\rho{}^i A_\sigma{}^j 
\end{pmatrix} \,,\quad
 \hat\tau^{\hmu\hnu} 
=\begin{pmatrix}
0 & 0 \\
0 & \tau^{ij} 
\end{pmatrix} \,.
\end{split}
\label{MNC_splitH_ExFT}
\ee
In this way all the degenerate structure is encoded in the $d$-dimensional part of the spacetime, with a degenerate $d$-dimensional metric $H_{ij}\equiv h^{\bri}{}_i h^{\brj}{}_j \delta_{\bri \brj}$. 
This ensures that the metric $g_{\mu\nu}$ can be identified with the metric appearing in exceptional field theory, while the degenerate Newton-Cartan metric structure will appear in the generalised metric.
In addition, we redefine the three-form and its field strength according to \eqref{Cdecomp_compact} and \eqref{Fdecomp_compact}, now without hats:
\be
 C_3 = \Cdef_3 + \Cdef_{2 i } Dy^i  + \tfrac{1}{2} \Cdef_{1 i j } Dy^i Dy^j + \tfrac{1}{3!} \Cdef_{ijk} Dy^i Dy^j Dy^k\,,
\label{Cdecomp_compactNR}
\ee
\be
  F_4 = \Fdef_4 + \Fdef_{3 i } Dy^i  + \tfrac{1}{2} \Fdef_{2 i j } Dy^i Dy^j + \tfrac{1}{3!} \Fdef_{1 ijk} Dy^i Dy^j Dy^k + \tfrac{1}{4!} \Fdef_{ijkl} Dy^i Dy^j Dy^k Dy^l \,,
\label{Fdecomp_compactNR}
\ee
where again $Dy^i \equiv dy^i + A_\mu{}^i dx^\mu$.
We carry out an analogous decomposition for $\widetilde C_3$ and $\widetilde F_4$, and for $C_6$ and $F_7$.
Finally, we can consider the Newton-Cartan torsion: with $\hat T_{\hmu\hnu}{}^A \equiv 2 \partial_{[\hmu} \hat\tau_{\hnu]}{}^A$ we have
\be
\begin{split}
T_{ij}{}^A & \equiv \hat T_{ij}{}^A = 2 \partial_{[i} \tau_{j]}{}^A \,, \quad T_{\mu i}{}^A  \equiv \hat T_{\mu i}{}^A - A_\mu{}^j \hat T_{ji} = D_\mu \tau_i{}^A  \,, \\ 
T_{\mu\nu}{}^A & \equiv \hat T_{\mu \nu}{}^A - 2 \hat T_{[\mu | i|}{}^A A_{\nu]}{}^i + A_\mu{}^i A_\nu{}^j \hat T_{ij}  =F_{\mu\nu}{}^j \tau_j{}^A \,.
\end{split}
\label{Tdecomp}
\ee

\paragraph{Embedding the limit in ExFT}

Let's start by considering the expansions \eqref{MNCexp_split} and \eqref{Cexp} of the original 11-dimensional metric and three-form.
We make use of the decompositions \eqref{MNC_splitH_ExFT} and \eqref{Cdecomp_compactNR} for the Newton-Cartan variables and three-form appearing in the decomposition, and then use these to work out the decomposition \eqref{breakeExFT} of the 11-dimensional metric and that \eqref{Cdecomp_compact} of the three-form. 
The potentially singular terms as $c\rightarrow \infty$ then appear in the $d$-dimensional components of the metric and of the three-form, with
\be
\phi_{ij} = c^2 \tau_{ij} + c^{-1} H_{ij} 
\,,\quad
\hat \Cdef_{ijk} = - c^3 \epsilon_{ABC} \tau_i{}^A \tau_j{}^B \tau_k{}^C + \Cdef_{ijk} + c^{-3} \widetilde \Cdef_{ijk}\,.
\label{simplerlimit} 
\ee
The metric $g_{\mu\nu}$ and Kaluza-Klein vector $A_\mu{}^i$ appearing in \eqref{breakeExFT} are then exactly those appearing in $\hat H_{\mu\nu}$ in \eqref{MNC_splitH_ExFT}.
The redefined form components carrying an $(11-d)$-dimensional index are all non-singular, so $\hat\Cdef_{\mu ij} = \Cdef_{\mu ij} + \mathcal{O}(c^{-3})$, and so on.
One point of danger is that $\hat \Cdef_{ijk}$ still appears in the field strengths \eqref{Fdecomp_compact} of these fields.
However, consulting the more explicit expressions \eqref{F=dA}, one sees that the field strength $\Fa_{\mu\nu}{}^M$ appearing in ExFT in fact involves the combination $\Fa_{\mu\nu ij} = \hat \Fdef_{\mu \nu ij} - \hat\Cdef_{ijk} F_{\mu\nu}{}^k$, which is in fact independent of $\hat\Cdef_{ijk}$, such that $ \hat \Fdef_{\mu \nu ij} - \hat\Cdef_{ijk} F_{\mu\nu}{}^k=  \Fdef_{\mu \nu ij} - \Cdef_{ijk} F_{\mu\nu}{}^k$.

For the generalised metric \eqref{genmet_usual}, inserting the expressions \eqref{simplerlimit} one finds that all terms at leading order in $c$ cancel, and sending $c \rightarrow \infty$ one has a manifestly finite and boost invariant expression:\footnote{Proving this requires the fact that $H^{ i[k} H^{ l ] j} =0$ when $H^{ ij}$ has rank $1$. For $d>4$ this would suggest we would have problems, however starting at $d=5$ the representation on which $\gM_{MN}$ acts enlarges and the structure of the generalised metric therefore changes. Note for $d>5$ it will also explicitly contain components of the dual six-form.}
\be\small
\gM_{MN} = \Omega^{\tfrac{2}{9-d}}
\begin{pmatrix} 
H_{ij} - \epsilon_{ABC} \tau_{(i|}{}^A \Cdef_{j)kl} \tau^{k B} \tau^{l C} +  \Cdef_{ikl} \Cdef_{jmn} H^{ km} \tau^{ln}  & - \epsilon_{ABC} \tau_i{}^A \tau^{k B} \tau^{l C} + 2 \Cdef_{ipq} H^{p[k} \tau^{l]q}\\
 - \epsilon_{ABC} \tau_k{}^A \tau^{i B} \tau^{j C} + 2 \Cdef_{kpq} H^{p[i} \tau^{l]j} & 2H^{ i[k} \tau^{l]j} + 2\tau^{i[k} H^{ l]j}
\end{pmatrix} \,.
\label{genmetnonrie} 
\ee
The parametrisation \eqref{genmetnonrie} can be viewed as a \emph{non-Riemannian parametrisation} of the generalised metric, and viewed simply as an alternative possibility to taking the usual form \eqref{genmet_usual}. 
The reason why this is a \emph{non-Riemannian parametrisation} is most clearly seen by looking at the inverse generalised metric $\gM^{MN}$.
In the Riemannian case, the parametrisation \eqref{genmet_usual} implies that the $d \times d$ block $\gM^{ij}$ is given by $\gM^{ij} = |\hat \phi|^{-1/(9-d)} \hat \phi^{ij}$ and therefore corresponds to the inverse spacetime metric. Assuming this block is invertible then uniquely fixes (given the definition of the generalised metric as a particular coset element obeying certain properties) the rest of the parametrisation.
In the non-Riemannian case, we instead have $\gM^{ij} = \Omega^{-\tfrac{2}{9-d}} H^{ij}$, which is non-invertible.
This leads instead to an alternative parametrisation. 
This is exactly as in the DFT case \cite{Morand:2017fnv}, which was generalised to ExFT in \cite{Berman:2019izh}. The expression \eqref{genmetnonrie} can be checked to be equivalent to the non-Riemannian $\Gfour$ generalised metric worked out from first principles in \cite{Berman:2019izh}.
In fact, from this point of view, one need not even go through the complications of taking the limit, but simply write down \eqref{genmetnonrie}, insert it into the ExFT and study the resulting dynamics. 

Returning to the embedding of the expansion in ExFT, we also need to worry about the singular pieces in the expansion of the dual gauge field $\hat C_6$.
This inevitably appears in the tensor hierarchy for all exceptional field theories.
From \eqref{defC6}, we have $\hat C_6 \sim c^3  C_3 \wedge \tau \wedge \tau \wedge \tau+ \dots$, and so given the decomposition according to \eqref{MNC_splitTau} and \eqref{Cdecomp_compactNR}, any component of $\hat C_6$ carrying three $d$-dimensional indices will be singular, i.e. $\hat\Cdef_{\mu\nu\rho ijk}$, $\hat\Cdef_{\mu\nu ij kl}$, $\hat\Cdef_{\mu ijklm}$, $\hat\Cdef_{ijklmn}$.
The claim is that, remarkably, all such singularities cancel automatically thanks to the precise combinations of $\hat C_6$ and $\hat C_3$ that appear in the ExFT fields. 
For $d=3,4$, this is most straightforwardly checked at the level of the ExFT field strengths.\footnote{Only the field strengths appear in the equations of motion, and the action can also be defined solely in terms of the field strengths by rewriting the Chern-Simons term in a standard way as an integral over a higher-dimensional spacetime.}.
One sees from \eqref{defJ32} for $\Gthree$ (and from \eqref{deftildeF} for $\Gfour$) that the components of $\hat F_7$ always appear in the combinations $\hat\Fdef_{\mu\nu\rho\sigma ijk} - \hat \Cdef_{ijk} \hat\Fdef_{\mu\nu\rho\sigma}$ and $\hat\Fdef_{\mu\nu\rho ijkl} + 4 \hat\Cdef_{[ijk} \hat \Fdef_{|\mu\nu\rho\sigma|l]}$ exactly such that the singularity coming from $\hat\Cdef_{ijk}$ cancels that coming from $\hat F_7$, which was written down in \eqref{F7oranother}.
That the ExFT gauge potentials themselves are non-singular can further be verified by hunting down the correct field redefinitions that relate the ExFT gauge fields to the 11-dimensional ones. 
Note that for $d\geq 6$ the components $\hat\Cdef_{ijklmn}$ are present and appear in the generalised metric itself: we have not verified explicitly but the expectation would be that it does so in a way that ensures the generalised metric remains finite.

\paragraph{Summary}
From the above we can conclude that the fields used in ExFT are manifestly non-singular in the non-relativistic limit (equivalently this shows that the fields which are U-duality covariant in a genuine dimensional reduction are non-singular).
We can also view the distinction between the relativistic and the non-relativistic 11-dimensional theory as being solely governed by the choice of parametrisation of the generalised metric. Having picked a generalised metric parametrisation, it is then consistent to directly identify the ExFT gauge fields and metric $g_{\mu\nu}$ with the gauge field components and metric of the decomposed relativistic \emph{or} non-relativistic theory.

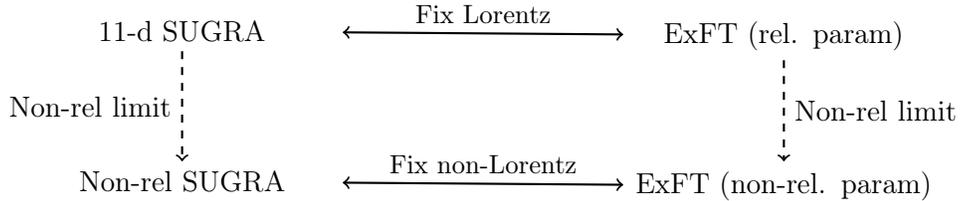
\begin{figure}[h]
\centering
\begin{tikzpicture}
\draw (-4,1) node (sugra) [text width=4cm,align=center,anchor=north] {11-d SUGRA};
\draw (4,1) node (dft) [text width=4cm,align=center,anchor=north] {ExFT (rel. param)};
\draw (4,-1) node (split) [text width=4cm,align=center,anchor=north] {ExFT (non-rel. param)};
\draw (-4,-1) node (nonrel) [text width=4cm,align=center,anchor=north] {Non-rel SUGRA};
\draw [thick,<->] (sugra) to node [midway,above] {\small Fix Lorentz} (dft) ;
\draw [thick,<->] (nonrel) to  node [midway,above] {\small Fix non-Lorentz} (split);
\draw [thick,dashed,->] (dft) to  node [midway,right] {Non-rel limit} (split);
\draw [thick,dashed,->] (sugra) to node [midway,left] {Non-rel limit} (nonrel) ;
\end{tikzpicture}
\caption{Relationship between non-relativistic limit and non-relativistic parametrisation of ExFT}.
\label{diagram}
\end{figure}

This is summarised in figure \ref{diagram}.
The upper triangular half of this diagram corresponds to first embedding the relativistic fields in ExFT in the usual manner, with a Riemannian parametrisation of the generalised metric, and then taking the non-relativistic limit giving a non-Riemannian parametrisation.
The lower triangular half corresponds to first taking the non-relativistic limit for the original 11-dimensional fields, and then embedding these into ExFT, giving the same non-Riemannian parametrisation.
In both cases, one needs to make the appropriate dimensional decomposition of the fields of the Newton-Cartan theory, corresponding to fixing the local tangent space (non-Lorentzian) symmetry.

Inserting the non-Riemannian parametrisation into the ExFT action or equations of motion will then reproduce the finite action and equations of motion results from taking the limit, after decomposing. 
For the action, we calculate this decomposition in appendix \ref{appExDecomp}.
What we will show next is that, remarkably, the ExFT equations of motion also automatically reproduce the Poisson equation \eqref{CovariantPoisson}.

\subsection{Generalised metric and equations of motion}
\label{genmeteom}

We now take a closer look at the consequences of using the non-relativistic parametrisation of the generalised metric.
We focus on the $d=3$ $\Gthree$ ExFT. 
For the $d=3$ Newton-Cartan geometry, $H^{ij}$ and $H_{ij}$ have rank zero and so are identically zero.
The longitudinal metric $\tau_{ij}$ is a three-by-three matrix and in fact invertible, with $\Omega^2 =- \det \tau$.
The resulting non-Riemannian parametrisation of the generalised metric \eqref{32genm} is
\be
\gM_{ij} = \Omega^{-2/3} \tau_{ij} \,,\quad
\gM_{\alpha \beta} = \begin{pmatrix}
2\varphi  &  1 \\
1& 0  
\end{pmatrix} \,,\quad
\varphi \equiv \tfrac{1}{3!} \epsilon^{ijk} \Cdef_{ijk} \,,
\label{32_nonrelM}
\ee
Comparing \eqref{32_nonrelM} and \eqref{32_factorised}, we can note that \eqref{32_nonrelM} is the most general possible $\mathrm{SL}(2)$ non-Riemannian parametrisation (up to the sign of the off-diagonal components), as this is completely fixed by requiring $\gM_{22} = 0$ which prevents us from interpreting that component as the determinant of a standard three-dimensional spacetime metric.

Normally, the generalised metric $\gM_{\alpha\beta}$ encodes two degrees of freedom.
It is clear that the non-Riemannian parametrisation given by \eqref{32_nonrelM} is restricted and is missing one degree of freedom. 
We may identify this missing degree of freedom with the overall scale of the longitudinal metric, as the latter only appears in the combination $|\det \tau|^{-1/3} \tau_{ij}$, which is conformally invariant.
This makes the dilatation invariance trivial in this formulation.

If we insert this parametrisation into the $\Gthree$ pseudo-action, with Lagrangian \eqref{Lexft}, we find that $\mathcal{L}_{\text{int}}$ as defined in \eqref{Lint32} vanishes, while
\be
\frac{1}{4} \D_\mu \gM^{ij} \D^\mu \gM_{ij} + \frac{1}{4} \D_\mu \gM_{\alpha \beta} \D^\mu \gM^{\alpha \beta} 
= \frac{1}{4} D_\mu (\Omega^{2/3} \tau^{ij} ) D^\mu (\Omega^{-2/3} \tau_{ij} )\,.  
\ee
This reproduces exactly the expected terms in the $d=3$ case of \eqref{11kin} and \eqref{11int}.

Notice that the kinetic terms for $\gM_{\alpha \beta}$ completely drop out.
So if we insert the non-relativistic parametrisation into the action, and then vary with respect to $\varphi$, we will never find an equation involving $\D^\mu \D_\mu \varphi$, i.e the Poisson equation.
However, instead we can consider the equations of motion of the generalised metric, which can be evaluated independently of its choice of parametrisation. 
These will provide the missing Poisson equation.
This is exactly analogous to the situation in DFT, see the discussions in \cite{Cho:2019ofr,Gallegos:2020egk}. 
One has to make a choice about whether you allow the equations of motion that follow from variations of the generalised metric that do not preserve the non-Riemannian parametrisation.
In both the DFT SNC case, and the present case, there is exactly one such independent variation, which provides an additional equation of motion beyond what is obtained by varying the fields of the parametrisation themselves.

Let's see how this works. Naively, the result of varying the generalised metric $\gM_{\alpha \beta}$ in the action is 
\be
\delta S = \int \dd^8x \,\dd^6Y\, \sqrt{g} \delta \gM^{\alpha \beta} \mathcal{K}_{\alpha \beta} \,,
\ee
with
\be
\begin{split}
\mathcal{K}_{\alpha\beta} & = 
- \tfrac{1}{4} \tfrac{1}{\sqrt{g}} \left( \D_\mu ( \sqrt{g} \D^\mu \gM_{\alpha\beta} ) 
- \gM_{\alpha\gamma} \gM_{\beta\delta} \D_\mu ( \sqrt{g} \D^\mu \gM^{\gamma\delta} ) \right)
\\ & \qquad
+ \frac{1}{4} \gM_{\alpha \gamma}\gM_{\beta\delta} \gM_{ij} \Fa_{\mu\nu}{}^{i\gamma} \Fa^{\mu\nu j \delta} 
+ \frac{1}{96} \gM_{\alpha \gamma}\gM_{\beta \delta} \Fc_{\mu\nu\rho\sigma}{}^\gamma \Fc^{\mu\nu\rho\sigma \delta} 
\\ & \qquad
+ \tfrac{1}{4} \gM^{ij} \left( \partial_{i(\alpha|} \gM^{kl} \partial_{j|\beta)} \gM_{kl} + \partial_{i(\alpha|} \gM^{\gamma\delta} \partial_{j|\beta)} \gM_{\gamma\delta}
+\partial_{i(\alpha|} g_{\mu\nu} \partial_{j|\beta)} g^{\mu\nu} \right)
\\ & \qquad
- \tfrac{1}{2} \gM^{ij} \partial_{i\alpha} \partial_{j\beta} \ln g + \tfrac{1}{\sqrt{g}} \partial_{i(\alpha|} ( \sqrt{g} \partial_{j |\beta)} \gM^{ij} )
\\ & \qquad 
- \tfrac{1}{2} \gM^{ij}\left( \partial_{i(\alpha|} \gM^{kl} \partial_{k|\beta)} \gM_{lj} 
+ \partial_{i(\alpha|} \gM^{\gamma\delta}\partial_{j\gamma} \gM_{|\beta) \delta}\right)
\\ & \qquad 
+ \tfrac{1}{2\sqrt{g}} ( \partial_{i\gamma} ( \sqrt{g} \gM^{ij} \gM^{\gamma \delta} \partial_{j(\alpha} \gM_{\beta)\delta} )
- \gM_{\gamma(\alpha} \gM_{\beta)\delta} \partial_{j \kappa} ( \sqrt{g} \gM^{ij} \gM^{\epsilon \gamma} \partial_{i \epsilon} \gM^{\kappa \delta} )
\\ & \qquad
- \tfrac{1}{4 \sqrt{g}} ( \partial_{i \gamma} ( \sqrt{g} \gM^{ij} \gM^{\gamma \delta} \partial_{j\delta} \gM_{\alpha \beta} )
- \gM_{\alpha \gamma} \gM_{\beta \delta} \partial_{i\epsilon} ( \sqrt{g} \gM^{ij} \gM^{\epsilon \kappa} \partial_{j \kappa} \gM^{\gamma \delta} 
)\,.
\end{split}
\label{K32_2} 
\ee
Now, the variation $\delta \gM^{\alpha \beta}$ cannot be arbitrary but must preserve that $|\det \gM|=1$. This ensures that one gets two rather than three independent equations, corresponding to the usual two degrees of freedom encoded in $\gM_{\alpha \beta}$.
The true equation of motion taking this into account is:
\be
\label{R32}
\mathcal{R}_{\alpha\beta}\equiv \mathcal{K}_{\alpha\beta} - \tfrac{1}{2} \gM_{\alpha\beta} \gM^{\gamma\delta} \mathcal{K}_{\gamma\delta} = 0 \,.
\ee
This can be thought of as the vanishing of a generalised Ricci tensor, $\mathcal{R}_{\alpha \beta}$.
For the non-Riemannian parametrisation \eqref{32_nonrelM}, the two independent equations are
\be
\mathcal{R}_{22}  = \mathcal{K}_{22}  = 0 \,,\quad
\mathcal{R}_{11}  - 2 \varphi \mathcal{R}_{22} = \mathcal{K}_{11} - 2 \varphi K_{12} =0 \,.
\ee
Setting $\partial_{i1} \equiv \partial_i$, $\partial_{i2} = 0$, we have explicitly that
\be
\begin{split}
\mathcal{K}_{22} & = 
 + \tfrac{1}{4}\gM_{ij} F_{\mu\nu}{}^{i} F^{\mu\nu j} 
+ \tfrac{1}{96} \Fdef_{\mu\nu\rho\sigma} \Fdef^{\mu\nu\rho\sigma }
=0\,.
\end{split}
\label{K22_nonrie} 
\ee
This is the equation of motion \eqref{folcon} arising as the totally longitudinal part of the equation of motion of the three-form.
This is consistent with its appearance here as the equation of motion of $\varphi$, which is indeed the totally longitudinal part of the three-form.

The other equation of motion is (after using \eqref{K22_nonrie}) 
\be
\begin{split}
0 & = \mathcal{K}_{11} - 2 \varphi K_{12}  
\\  &= 
-  \tfrac{1}{\sqrt{g}} \tfrac{1}{6} \epsilon^{ijk}  D_\mu ( \sqrt{g} g^{\mu\nu} \Fdef_{\nu ijk} ) 
\\ & \qquad - \tfrac{1}{8}  \gM^{km} \gM^{ln} \Fdef_{\mu \nu kl} \Fdef^{\mu\nu}{}_{mn} 
+ \tfrac{1}{96}  \Fdef_{\mu\nu\rho\sigma ijk} \Fdef^{\mu\nu\rho \sigma}{}_{lmn} \tfrac{1}{3! 3!} \epsilon^{ijk} \epsilon^{lmn}
\\ & \qquad
+ \tfrac{1}{4} \gM^{ij} \left( \partial_{i} \gM^{kl} \partial_{j} \gM_{kl} 
+\partial_{i} g_{\mu\nu} \partial_{j} g^{\mu\nu} \right)
-\tfrac{1}{2}\gM^{ij} \partial_{i}  \gM^{kl}   \partial_{k } \gM_{jl} 
\\ & \qquad
- \tfrac{1}{2} \gM^{ij} \partial_i \partial_j \ln g
- \tfrac{1}{\sqrt{g}} \partial_i ( \sqrt{g} \partial_j \gM^{ij} )
\,.\end{split}
\label{K11} 
\ee
Here we have $\Fdef_{\mu ijk} = D_\mu \Cdef_{ijk} - 3 \partial_{[i} \Cdef_{|\mu |jk]}$, having used $\D_\mu \gM_{11}  =  D_\mu \gM_{11} - \epsilon^{ijk} \partial_i \Aa_{\mu jk} \gM_{12}$.
We can then identify \eqref{K11} as the Poisson equation for $\varphi \equiv \tfrac{1}{6} \epsilon^{ijk} \Cdef_{ijk}$, as it has the form $\tfrac{1}{\sqrt{g}} D_\mu ( \sqrt{g} D^\mu \varphi ) + \dots =0$.
It is conjugate to the variation $\delta \gM^{11}$. For the non-Riemannian parametrisation, $\gM^{11} = 0$, so allowing this variation corresponds to allowing variations that do not respect the parametrisation.
In terms of the expansion of $\gM^{\alpha \beta}$ in powers of $1/c$, this variation is subleading in origin.
Finally, one can precisely check that this equation \eqref{K11} is indeed exactly the Poisson equation \eqref{CovariantPoisson}, which we found at subleading order in the expansion of the relativistic theory, and here is rewritten in terms of ExFT variables after making the dimensional decomposition of all the fields. (It can be easily checked that the gauge field terms match, using \eqref{32F7F4} to relate the seven-form components appearing here to those of $\widetilde{F}_4$, and a patient calculation shows that inserting the dimensional decomposition of the eleven-dimensional fields matches perfectly.)

\paragraph{Structure of generalised Ricci tensor}

Geometrically, $\mathcal{R}_{\alpha\beta}$ should be thought of as (the $\mathrm{SL}(2)$ part of) a generalised Ricci tensor. 
It is a symmetric generalised tensor of weight 0 and obeys $\gM^{\alpha \beta} \mathcal{R}_{\alpha \beta} = 0$.
When we take the relativistic parametrisation \eqref{32_factorised} of the generalised metric, it can therefore be parametrised as 
\be
\mathcal{R}_{\alpha \beta} = \tfrac{1}{2} \begin{pmatrix} 1 & \hat\Cdef \\ 0 & 1 \end{pmatrix}
\begin{pmatrix}
|\phi|^{1/2} \mathcal{R}_\phi & \mathcal{R}_C \\
\mathcal{R}_C & |\phi|^{-1/2} \mathcal{R}_\phi 
\end{pmatrix}
\begin{pmatrix} 1 & 0\\ \hat\Cdef  & 1 \end{pmatrix}
\ee
with $\mathcal{R}_\phi$ and $\mathcal{R}_C$ tensors of three-dimensional weight 0, such that the variation of the action leads to 
\be
\delta S \supset - \int \dd^8 x\,\dd^6 y \,\sqrt{g} \left( \frac{\delta |\phi|^{1/2}}{|\phi|^{1/2}} \mathcal{R}_\phi + |\phi|^{-1/2} \delta \hat\Cdef \mathcal{R}_C \right)
\label{deltaS}
\ee
Let's examine what happens to the components of $\mathcal{R}_{\alpha \beta}$ in the non-relativistic limit.
We have $|\phi|^{1/2} = \Omega c^3$, $\hat\Cdef = - c^3 \Omega + C + c^{-3} \widetilde C$.
This leads to the expression
\be
\mathcal{R}_{\alpha \beta} = \tfrac{1}{2} \begin{pmatrix} 1 & \Cdef \\ 0 & 1 \end{pmatrix}
\begin{pmatrix}
c^3 \Omega(\mathcal{R}_\phi-\mathcal{R}_C) & \mathcal{R}_C-\mathcal{R}_\phi \\
\mathcal{R}_C-\mathcal{R}_\phi  & c^{-3} \Omega^{-1} \mathcal{R}_\phi 
\end{pmatrix}
\begin{pmatrix} 1 & 0\\ \Cdef  & 1 \end{pmatrix}
\ee
So in principle the independent equations are still $\mathcal{R}_C$ and $\mathcal{R}_\phi$.
However, we already know that this generalised Ricci tensor has no leading parts in $c$ when we take the limit (because none of the ExFT fields contain singular terms).
If we expand
\be
\mathcal{R}_\phi = c^3 \mathcal{R}_\phi^{(3)} + c^0 \mathcal{R}_\phi^{(0)} + c^{-3} \mathcal{R}_\phi^{(-3)} 
\,,\quad
\mathcal{R}_C = c^3 \mathcal{R}_C^{(3)} + c^0 \mathcal{R}_C^{(0)} + c^{-3} \mathcal{R}_C^{(-3)} \,,
\ee
it must be that we have $\mathcal{R}_\phi^{(3)} =\mathcal{R}_C^{(3)}$, $\mathcal{R}_\phi^{(0)}= \mathcal{R}_C^{(0)}$,
viewed as off-shell identities, and the independent equations of motion, i.e. those appearing as the actual finite entries of $\mathcal{R}_{\alpha \beta}$, are actually
\be
\mathcal{R}_\phi^{(3)} = 0 \,,\quad \mathcal{R}_\phi^{(-3)} - \mathcal{R}_C^{(-3)} = 0\,.
\ee
The former is conjugate to $\delta \gM^{22}$ and the latter to the $\delta \gM^{11}$ that is forbidden if we insist on keeping a non-Riemannian parametrisation.
We can go back to the variation \eqref{deltaS} and expand that:
\be
\delta S = - \int  \dd^8 x\,\dd^6 y \,\sqrt{g}  \left( \delta \ln \Omega (\mathcal{R}_\phi - \mathcal{R}_C ) + \Omega^{-1} c^{-3} \delta C \mathcal{R}_C \right)\,,
\ee
hence the first non-zero variations are
\be
\delta S = - \int  \dd^8 x\,\dd^6 y \,\sqrt{g} \left( c^{-3}  \delta \ln \Omega (\mathcal{R}_\phi^{(-3)} - \mathcal{R}_C^{(-3)} ) + \Omega^{-1}\delta C \mathcal{R}_C^{(3)}\right) \,.
\ee
We see again that we get the longitudinal equation of motion for the three-form at finite order, and the extra Poisson equation of motion comes from a subleading variation associated to the variation of the volume factor $\Omega$, which otherwise has no dynamics associated to it in this formulation.

\subsection{Generating non-relativistic generalised metrics via U-duality}
\label{genmetU}

Non-trivial U-duality transformations act as $\mathrm{SL}(2)$ transformations on the generalised metric $\gM_{\alpha \beta}$, via $\gM\rightarrow\gM^\prime= U^T\gM U$ with $\det U = 1$. 
Parametrising $U = \begin{pmatrix} a & b \\ c & d \end{pmatrix}$ the transformation of the non-relativistic parametrisation \eqref{32_nonrelM} gives
\be
\gM_{\alpha \beta}^\prime = \begin{pmatrix} 
 2 a ( a \varphi + c) & 2 a b \varphi + ad + bc \\
 2 a b \varphi + ad +bc & 2 b( b \varphi + d ) 
\end{pmatrix} \,,
\ee
and this remains in the non-relativistic form only if $b=0$, or else if $\varphi$ is constant and $d= - b \varphi$.
In the former case, the effect of the transformation is $\varphi \rightarrow a ( a\varphi +c)$ and so amounts to a scaling and shift of the three-form.

The genuine non-geometric U-dualities correspond to the $\mathrm{SL}(2)$ inversion symmetry with $a=d=0$, $bc=-1$. 
If $\varphi <0$, this takes us from the non-relativistic parametrisation to a relativistic one with
\be
\phi_{ij} = ( - \tfrac{1}{2\varphi} )^{2/3} (\det \tau)^{-1/3} \tau_{ij} \,,\quad
\Cdef_{ijk} = - \tfrac{1}{2\varphi} \epsilon_{ijk} \,.
\ee
These obey $|\det \phi| = \Cdef^2$ which corresponds to a `critical' three-form.

We can apply this to a real supergravity background along the lines of \cite{Lee:2013hma, Berman:2019izh}, namely the M2 brane solution in the form
\be
\dd s^2 = f^{-2/3} \eta_{ij} dy^i dy^j + f^{1/3} \delta_{\mu\nu} dx^\mu dx^\nu \,,\quad 
C_{ijk} = ( f^{-1} + \gamma ) \epsilon_{ijk}\,,\quad 
\ee 
where the harmonic function $f$ obeys $\partial_\mu \partial^\mu f = 0$ and $\gamma$ is a constant.
This has constant exceptional field theory 8-dimensional metric, $g_{\mu\nu} = \delta_{\mu\nu}$, while
\be
\gM_{ij} = \eta_{ij} \,,\quad
\gM_{\alpha \beta} = \begin{pmatrix}
 - \gamma ( f + 2\gamma) & - (1+\gamma f) \\
 -(1+\gamma f) & - f 
\end{pmatrix} \,.
\ee
Sending $f \rightarrow 0$ corresponds exactly to the original limit \eqref{MNCexp_split}.
Alternatively, we can formally U-dualise along the $y^i$ directions (including time) to obtain a solution with
\be
\gM_{\alpha \beta} 
= \begin{pmatrix} 
- f & 1 + \gamma f \\
1+\gamma f & - \gamma ( f+ 2\gamma) 
\end{pmatrix}\,.
\label{transformedM2}
\ee
The standard M2 solution has $\gamma=-1$ and $f = 1 + \tfrac{q}{r^6}$, with $r^2 \equiv \delta_{\mu\nu} x^\mu x^\nu$. 
In this case, the generalised metric \eqref{transformedM2} corresponds to the \emph{negative M2} solution \cite{Dijkgraaf:2016lym}:
\be
\dd s^2 = \widetilde f^{-2/3} \eta_{ij} dy^i dy^j +\widetilde f^{1/3} \delta_{\mu\nu} dx^\mu dx^\nu \,,\quad 
C_{ijk} = ( \widetilde f^{-1} -1 ) \epsilon_{ijk}\,,\quad
\widetilde f = 1 - \tfrac{q}{r^6} \,. 
\ee 
This solution has a naked singularity at $\widetilde f = 0$ $\Leftrightarrow f-2=0$. 
Evidently the generalised metric \eqref{transformedM2} is non-singular everywhere and at $\widetilde{f} = 0$ becomes non-relativistic.
This suggests \cite{Blair:2019qwi} interpreting such backgrounds as containing a singular locus at which the geometry degenerates to a non-relativistic one.

If we alternatively take $\gamma=0$ then the generalised metric \eqref{transformedM2} has the non-relativistic form everywhere, with $\varphi \equiv - \tfrac{1}{2} f$. 
If we now reconsider the equation of motion \eqref{K11} which can only be found by varying the generalised metric before inserting the parametrisation, then this is exactly the equation $\nabla^2 f = 0$ obeyed by the harmonic function.
Finally, we can reconstruct the full 11-dimensional MNC geometry:
\be
\hat \tau_{\hmu}{}^A = ( 0 , \delta_i{}^A ) \,,\quad \hat H^{\hmu \hnu} = \begin{pmatrix} \delta^{\mu\nu} & 0 \\ 0 & 0 \end{pmatrix}\,,\quad
 C_{012} = - \tfrac{1}{2} f \,.
\ee

\subsection{Gauge fields and self-duality in $\Gthree$ ExFT}

Now let's look at what happens in the gauge field sector of the $\Gthree$ ExFT.
Let's repeat the parametrisations \eqref{defFH32} and \eqref{defJ32} now for the field strength components of the non-relativistic theory:
\be
\mathcal{F}_{\mu\nu}{}^{i1} \equiv F_{\mu\nu}{}^i\,,\quad
\mathcal{F}_{\mu\nu}{}^{i2} \equiv \frac{1}{2}\epsilon^{ijk} ( \Fdef_{\mu\nu jk} - \Cdef_{jkl} \Fdef_{\mu\nu}{}^l ) 
\,,\quad
\Fb_{\mu\nu\rho i} \equiv - \Fdef_{\mu\nu\rho i} \,,
\ee
\be
\mathcal{J}_{\mu\nu\rho\sigma}{}^1 \equiv \Fdef_{\mu\nu\rho\sigma}
\,,\quad
\mathcal{J}_{\mu\nu\rho\sigma}{}^2 \equiv \frac{1}{6} \epsilon^{ijk} (  \Fdef_{\mu\nu\rho\sigma ijk} - \Cdef_{ijk} \Fdef_{\mu\nu\rho\sigma} )\,.
\ee
Then the kinetic terms \eqref{Lkin32} in the $\Gthree$ ExFT pseudo-action \eqref{Lexft} are
\be
-  \tfrac{1}{4} \gM_{ij} \gM_{\alpha \beta} \Fa_{\mu\nu}{}^{i\alpha} \Fa^{\mu\nu j \beta} 
- \tfrac{1}{12} \gM^{ij} \Fb_{\mu\nu \rho i} \Fb^{\mu\nu\rho}{}_j 
=
- \tfrac{1}{4} \Omega^{-2/3} \tau_{ij} F^{\mu\nu i} \epsilon^{jkl}  \Fdef_{\mu\nu kl} 
 - \frac{1}{12} \Omega^{2/3} \tau^{ij} \Fdef_{\mu\nu\rho i} \Fdef^{\mu\nu\rho}{}_{j}\,,
\ee
which matches the corresponding terms in the decomposition \eqref{11kin} of the non-relativistic action.

To discuss the three-form gauge field, consider the $\Gthree$ ExFT equation of motion obtained from the pseudo-action by varying $\Ac_{\mu\nu\rho}{}^{\alpha}$:
\be
\begin{split} 
& \D_\sigma ( \sqrt{|g|} \gM_{\alpha \beta} \Fc^{\mu\nu\rho\sigma \beta} )
- 2 \partial_{i\alpha} ( \sqrt{|g|} \gM^{ij} \Fb^{\mu\nu\rho}{}_j )
\\ &- 48 \kappa \epsilon_{\alpha \beta}\epsilon^{\mu\nu\rho \sigma_1\dots\sigma_5} 
\left(
\D_{\sigma_1} \Fc_{\sigma_2 \dots \sigma_5}{}^\beta + 4 \Fa_{\sigma_1\sigma_2}{}^{i\beta} \Fb_{\sigma_3\sigma_4\sigma_5 i}
\right)
=0\,.
\end{split}
\label{Ceom}
\ee
After varying, we must also impose the constraint \eqref{constraintJ}.
This constraint involves the generalised metric, and so it is sensitive to whether we are describing the relativistic or non-relativistic theory.
However, in either case, using the constraint in the equation of motion of $\Ac_{\mu\nu\rho}{}^2$ in fact produces the Bianchi identity \eqref{32BianchiC} for $\Fc_{\mu\nu\rho\sigma}{}^1 = \Fdef_{\mu\nu\rho\sigma}$.
In the relativistic case, with the Riemannian parametrisation \eqref{32_factorised} of the generalised metric (or its Euclidean version), we could go on to use the constraint to eliminate $\Fc_{\mu\nu\rho\sigma}{}^2$ from the equation of motion of $\Ac_{\mu\nu\rho}{}^2$.
The result would be the equation of motion of the three-form $\Cdef_{\mu\nu\rho}$ following from the decomposition of 11-dimensional SUGRA.

Now let's consider the situation where the generalised metric admits the non-relativistic parametrisation \eqref{32_nonrelM}.
In this case, choosing the minus sign for $\kappa$, the constraint \eqref{constraintJ} implies that
\be
\begin{split} 
\sqrt{g}\, \Fdef^{\mu\nu \rho\sigma}  = - \tfrac{1}{ 4!}\epsilon^{\mu\nu\rho\sigma \lambda_1 \dots \lambda_4} \Fdef_{\lambda_1 \dots \lambda_4} \,,\quad
\sqrt{g} \,\Fdef^{\mu\nu \rho\sigma}{}_{ijk} = +\tfrac{1}{ 4!}\epsilon^{\mu\nu\rho\sigma \lambda_1 \dots \lambda_4} \Fdef_{\lambda_1 \dots \lambda_4 ijk} \,.
\end{split}
\ee
So we can no longer eliminate $\Fdef_{\mu\nu\rho\sigma ijk}$ in favour of $\Fdef_{\mu\nu\rho\sigma}$.
This is clearly as expected for the MNC theory for which the former indeed appears explicitly in the action and equations of motion (note it is related to $\widetilde{\Fdef}_{\mu\nu\rho\sigma}$ via \eqref{32F7F4}).
We therefore see that the ExFT constraint gives not only the expected constraint \eqref{magic} that the original four-form field strength becomes self-dual, but also the duality condition with opposite sign which is obeyed by the dual seven-form \eqref{F7constraint}.
Thus the $\Gthree$ ExFT contains the expected degrees of freedom of the non-relativistic theory, and efficiently rearranges them into self-dual and anti-self-dual parts automatically on the non-Riemannian parametrisation.

\section{Conclusions and discussion}
\label{Conc}

\paragraph{Comparison with the Gomis-Ooguri or SNC string} 

The behaviour we found in eleven-dimensional supergravity can be seen to be extremely similar to that which happens on the worldsheet for the Gomis-Ooguri or SNC string \cite{Gomis:2000bd,Andringa:2012uz,Bergshoeff:2019pij}.
To see this, let's revisit the derivation of this worldsheet action by considering the SNC expansion 
\be
\hat g_{\mu\nu}  = c^2 \tau_{\mu\nu} + H_{\mu\nu} \,,\quad
\hat B_{\mu\nu} = - c^2 \epsilon_{\uA \uB} \tau_\mu{}^{\uA} \tau_\nu{}^{\uB} + B_{\mu\nu} \,,
\ee
in the worldsheet Polyakov action
\be
S = - \tfrac{1}{2} \int d^2 \sigma ( \partial_\alpha \hat X^\mu \partial^\alpha \hat X^\nu \hat g_{\mu\nu} - \epsilon^{\alpha \beta} \partial_\alpha \hat X^\mu \partial_\beta \hat X^\nu \hat B_{\mu\nu} )\,.
\ee
We have set the effective tension to one and are in conformal gauge. 
If we also expand the target space coordinates\footnote{This was similarly done in \cite{Harmark:2019upf} however only for the Nambu-Goto action, for which the subleading coordinates do not appear in the finite action as the leading part of the expansion cancels identically.}
\be
\hat X^\mu = X^\mu + c^{-2} \widetilde X^\mu 
\ee
we get
\be
\begin{split}
S = - \tfrac{1}{2} \int d^2 \sigma & \big(
c^2 \tau_\mu{}^{\uA} \partial_\alpha X^\mu ( \eta_{\uA \uB} \tau_\nu{}^{\uB} \partial^\alpha X^\nu + \epsilon^{\alpha \beta} \epsilon_{\uA \uB} \tau_\nu{}^{\uB} \partial_\beta X^\nu ) 
\\ & \quad
+ 2 \tau_\mu{}^{\uA} \partial_\alpha \widetilde X^\mu ( \eta_{\uA \uB} \tau_\nu{}^{\uB} \partial^\alpha X^\nu + \epsilon^{\alpha \beta} \epsilon_{\uA \uB} \tau_\nu{}^{\uB} \partial_\beta X^\nu ) 
\\ & \qquad 
+ \partial_\alpha X^\mu \partial^\alpha X^\nu H_{\mu\nu} - \epsilon^{\alpha \beta} \partial_\alpha X^\mu \partial_\beta X^\nu B_{\mu\nu} 
\big) \,.
\end{split} 
\label{GOpoly}
\ee
This is exactly analogous to the result of the expansion of the 11-dimensional SUGRA action.
Here the Wess-Zumino coupling to the $B$-field plays the role of the Chern-Simons term, and the singular piece can be cancelled by imposing a sort of twisted self-duality constraint on $F_\alpha{}^{\uA} \equiv \tau_\mu{}^{\uA}\partial_\alpha X^\mu$, namely that $\eta_{\uA \uB} F^{\alpha \uB} + \epsilon^{\alpha \beta} \epsilon_{\uA \uB} F_{\beta}{}^{\uB} = 0$.
This then appears as the equation of motion imposed by $\widetilde F_\alpha{}^{\uA} \equiv \tau_\mu{}^{\uA} \partial_\alpha \widetilde X^\mu$, and the latter can be seen to only appear in its anti-self-dual projection.
This corresponds to the familiar rearrangement of the longitudinal directions into chiral and anti-chiral directions (as seen usually in lightcone coordinates). The numbers of degrees of freedom are preserved as the `dual' degrees of freedom in $\widetilde X^\mu$ are similarly constrained.

Normally one derives the finite part of the action \eqref{GOpoly} by rewriting the action in an equivalent form using auxiliary degrees of freedom, such that the limit can be performed without singularities.
After the limit, one finds these auxiliary degrees of freedom correspond to $\widetilde F_\alpha{}^{\uA}$, and impose the chirality/anti-chirality conditions on the longitudinal degrees of freedom.
This is also what happens in the doubled sigma model approach (e.g. \cite{Lee:2013hma, Blair:2020gng}), which starts with coordinates $X$ and duals $\widetilde X$, related by a self-duality constraint involving the generalised metric of double field theory.
Taking the SNC limit in this set-up then leads to the situation as above where the longitudinal $X$ and $\widetilde X$ are no longer related, but separately obey chiral/anti-chirality constraints.
The doubled sigma model action then reproduces the finite terms in \eqref{GOpoly}.
This then is analogous to the exceptional field theory description of the limit of 11-dimensional SUGRA. 

It could be conjectured that the appearance of (self)-duality constraints is a generic feature of non-relativistic limits of theories with topological or Chern-Simons terms, as a requirement for cancelling singular terms arising from the topological term against those arising from the kinetic term.
Schematically given a Lagrangian $\mathcal{L} \sim F \wedge \star F + F \wedge G$ with a non-relativistic expansion leading to a term $c^{n} F \star ( \star F + G)$, then we would take $\star F + G = 0$ as a constraint.
It would be interesting to explore this mechanism in other contexts.

\paragraph{Subleading terms} 

Our derivation of the MNC geometry made use of a field redefinition involving the parameter $c$ which we then sought to send to $\infty$ and interpret as a non-relativistic limit.
This could be extended to a full non-relativistic expansion, including first of all further subleading terms in the metric, with $\hat g_{\mu\nu} = c^2 \tau_{\mu\nu} + c^{-1} H_{\mu\nu} + c^{-4} X_{\mu\nu} + \dots$. It is possible to check that doing so does not affect the expansion of the action up to order $c^0$, and it would be expected on general grounds \cite{Hansen:2019svu} that the first appearance of the first subleading terms simply re-imposes the equations of motion already encountered (as we saw with $\tilde C_3$ and the equations of motion of $C_3$). 
In addition, we could reformulate the expansion by introducing additional one-form gauge fields (as for this case in \cite{Kluson:2019uza}), accompanied by a shift symmetry, such that the three-form $C_{\mu\nu\rho}$ does not transform under boosts.
The resulting more general expansion could then be attacked order-by-order without necessarily sending $c\rightarrow \infty$ or truncating as we did.
Here it would be interesting to compare with the approach of \cite{Hansen:2020pqs}, inputting the eleven-dimensional three-form as matter.
A complicating feature, relative to usual $1/c$ expansions of general relativity leading to Newton-Cartan \cite{VandenBleeken:2017rij,Hansen:2018ofj,Hansen:2019svu} for example, is that the longitudinal vielbein appears in both the metric and three-form and does so at different orders in $c$.

\paragraph{Supersymmetry and non-uniqueness of non-relativistic 11-dimensional supergravity} 

We limited ourselves to an analysis of the bosonic geometry in this paper.
The supersymmetric extension presumably exists and should be constructed.
At the level of supersymmetric double and exceptional field theory, the logic would again be that changing the parametrisation of the generalised vielbein is all that is needed to arrive at the desired theory, and this seems to be possible without obstacles \cite{Blair:2020gng}.

Note that in this paper we started with a non-relativistic expansion tailored to the M2.
There should be a similar expansion based on the M5, in which we have six longitudinal and five transverse directions.
(This should reduce to the dual NSNS six-form expansion discussed in the conclusions of \cite{Bergshoeff:2021bmc}.)
This would then give a \emph{second} non-relativistic version of 11-dimensional supergravity, so although this is the unique maximal supergravity in eleven dimensions, this uniqueness would then no longer hold in the non-relativistic setting. 

\paragraph{Duality web and branes}

An obvious goal for which this paper should be useful is the study of the spacetime actions for the non-relativistic duality web  \cite{Gomis:2000bd} in 11- and 10-dimensions.  
This can proceed both by applying standard dimensional reduction and dualisation to our 11-dimensional action, and by applying similar methods to individual supergravities by taking covariant non-relativistic limits associated to each $p$-brane present in the theory.
Here, we performed a dimensional reduction to type IIA, but we did not discuss the expected T-duality relationship to type IIB, for example.
Similarly, there is presumably a heterotic SNC which could be obtained by reducing non-relativistic M-theory on a longitudinal interval, although it is not immediately obvious what the result of reducing on a transverse interval should be.
Note that the appearance of the original and dual field strength together in the 11-dimensional theory suggests that the appropriate formalism for describing generalisations of Newton-Cartan geometries in type II should be the formalism where the RR $p$-forms are treated `democratically' \cite{Bergshoeff:2001pv}, accompanied by a self-duality constraint. 

Here the double and exceptional field theory formulation may again prove a useful guide. 
Beyond the usual suspects, exceptional field theory also offers a way to handle the vast number of mixed symmetry tensors that appear coupling to exotic branes \cite{Fernandez-Melgarejo:2018yxq, Berman:2018okd}.
It may not be unreasonable to suggest using the $E_{11}$ `master' ExFT recently constructed in \cite{Bossard:2021ebg}, as this presumably provides scope to construct an infinite number of brane scaling limits. Here there is no need to artificially split the coordinates and one can work with 11/10-dimensional quantities throughout, albeit at the obvious price of dealing with a very infinite algebra. 

The ExFT description in this paper demonstrates that the non-relativistic theory is also controlled by the same exceptional Lie algebraic symmetries that appear in the relativistic case.
A distinction can be made between these symmetries as they are used in ExFT and the actual U-duality symmetries present on toroidal reduction. 
As we saw in section \eqref{genmetU}, U-duality transformations can `rotate' between relativistic and non-relativistic theories.
This is also the case for T-duality of non-relativistic strings \cite{Bergshoeff:2018yvt}.
A non-trivial U-duality, corresponding to an $\mathrm{SL}(2)$ inversion transformation in the $\Gthree$ case, acts on three directions in spacetime.
To make a systematic study of U-duality of non-relativistic theories, it would therefore be necessary to consider U-duality transformations acting on 0,1, 2 or 3 longitudinal directions and to check which of these do or do not take you back to a relativistic theory. The $\Gthree$ ExFT description of section \eqref{genmetU} only allowed for U-duality transformations acting on all three longitudinal directions, while the $\Gfour$ ExFT description presented in appendix \ref{AndSL5} would allow for transformations acting on two or three longitudinal directions. A precise group to consider would then be the $\Gsix$ case which can accommodate all possible types of U-dualities acting on the MNC geometry, with some subgroup corresponding to the strict U-duality symmetries of the non-relativistic theory. This analysis is left for future work.

Another interesting question is to understand the consequences of the non-relativistic limit on the brane spectrum of M-theory (and hence also of type IIA, after reducing). 
The `decoupling' of the transverse components of $F_4$ and the longitudinal components $F_7$ presumably means something at the level of the M2 and M5 branes coupling to the three- and six-form: the analysis of \cite{Garcia:2002fa} should be pertinent here.
One could similarly enquire about whether the duality constraint in the type IIA SNC theory can be seen at the level of the string spectrum resulting from the quantisation of the non-relativistic superstring \cite{Kim:2007pc}

Obtaining brane solutions of the non-relativistic theory, whether by directly solving the equations of motion or using U-duality as in section \ref{genmetU}, is also an interesting question. Interestingly, membrane solutions of 11-dimensional SUGRA with transverse self-dual field strength were constructed in \cite{Duff:1997xja} and perhaps can be adapted or used in the non-relativistic setting. 
Even the `flat' spacetime solution may have interesting properties including infinite-dimensional isometries as for the string case \cite{Batlle:2016iel,Bergshoeff:2019pij,Blair:2020gng}.

\section*{Acknowledgements}

CB is supported by the FWO-Vlaanderen through a Senior Postdoctoral Fellowship and also through the project G006119N, and is supported by the Vrije Universiteit Brussel through the Strategic Research Program ``High-Energy Physics''.
DG and NZ are partially supported by the Delta-Institute for Theoretical Physics (D-ITP) funded by the Dutch Ministry of Education, Culture and Science (OCW). DG is also supported in part by CONACyT through the program Fomento, Desarrollo y Vinculacion de Recursos Humanos de Alto Nivel.
We would like to thank Alex Arvanitakis, David Berman, Umut G\"ursoy, Gerben Oling, and Daniel Thompson for helpful discussions.

\appendix
\newcommand{\pmr}{-}


\section{Dimensional decomposition of non-relativistic action for ExFT}
\label{appExDecomp}

\paragraph{Decomposition of $R^{(0)}$}

Consider the part of the scalar curvature $R^{(0)}$ as defined in \eqref{expandRicci} not involving the longitudinal metric, but just the transverse metrics $\hat H_{\hmu\hnu}$ and $\hat H^{\hmu\hnu}$ and the measure factor $\hat\Omega$.
In the dimensional decomposition used in exceptional field theory, the latter two factorise as
\be
\hat H_{\hmu\hnu} = U_{\hmu}{}^{\hrho} U_{\hnu}{}^{\hsigma} \bar H_{\hrho\hsigma} \,,\quad
\hat H^{\hmu\hnu} = (U^{-1})_{\hrho}{}^{\hmu} (U^{-1})_{\hsigma}{}^{\hnu} \bar H^{\hrho\hsigma} 
\,,
\ee
with
\be
U_{\hmu}{}^{\hnu} = \begin{pmatrix} \delta_\mu{}^\nu & A_\mu{}^j \\ 0 & \delta_i{}^j \end{pmatrix} \,,\quad
\bar H_{\hmu\hnu} = \begin{pmatrix} G_{\mu\nu} & 0 \\ 0 & H_{ij} \end{pmatrix} \,,\quad
\bar H^{\hmu\hnu} = \begin{pmatrix} G^{\mu\nu} & 0 \\ 0 & H^{ij} \end{pmatrix} \,.
\ee
Here $G^{\mu\nu}$ is the inverse of $G_{\mu\nu}$, but $H^{ij}$ and $H_{ij}$ are not invertible.
The idea is to completely factor out the matrix $U$ from derivatives of $\hat G$.
Defining
\be
\partial_{\hmu} \hat H_{\hnu \hrho} =  U_{\hmu}{}^{\hsigma} U_{\hnu}{}^{\hlambda} U_{\hrho}{}^{\hat \kappa} \overline{\partial H}_{\hsigma\hlambda\hat\kappa}\,,\quad
\partial_{\hmu} \hat H^{\hnu \hrho} =  U_{\hmu}{}^{\hsigma} (U^{-1})_{\hlambda}{}^{\hnu} (U^{-1})_{\hat \kappa}{}^{\hrho} \overline{\partial h}_{\hsigma}{}^{\hlambda\hat\kappa}
\ee
we have the relatively simple expressions
\be
\overline{\partial H}_{\mu \hnu \hrho} = \begin{pmatrix}
\bar D_\mu G_{\nu \rho} &  H_{kl}\bar D_\mu A_\nu{}^l\\
 H_{jl}\bar D_\mu A_\rho{}^k & \bar D_\mu H_{jk} 
\end{pmatrix} \,,\quad
\overline{\partial H}_{i \hnu \hrho}  = \begin{pmatrix}
\partial_i G_{\nu \rho} &  H_{kl} \partial_i A_\nu{}^l\\
 H_{jl} \partial_i A_\rho{}^k &  \partial_i H_{jk} 
\end{pmatrix} 
\ee
\be
\overline{\partial H}_{\mu}{}^{\hnu \hrho}  = \begin{pmatrix}
\bar D_\mu G^{\nu \rho} &  - G^{\nu\sigma} \bar D_\mu A_\sigma{}^k\\
- G^{\rho\sigma} \bar D_\mu A_\sigma{}^j & \bar D_\mu H^{jk} 
\end{pmatrix} \,,\quad
\overline{\partial H}_{i}{}^{\hnu \hrho} = \begin{pmatrix}
\partial_i G^{\nu \rho} &  - G^{\nu\sigma} \partial_i A_\sigma{}^k\\
- G^{\rho\sigma}\partial_i A_\sigma{}^j & \partial_i H^{jk} 
\end{pmatrix} 
\ee
where $\bar D_\mu \equiv \partial_\mu - A_\mu{}^i \partial_i$.
For instance, consider the following terms in the scalar curvature:
\be
\tfrac{1}{4} \bar H^{\hmu \hnu}\overline{\partial H}_{\hmu \hrho \hsigma} \overline{\partial H}_{\hnu}{}^{\hrho \hsigma} 
- \tfrac{1}{2} \bar H^{\mu\nu} \overline{\partial H}_\mu{}^{\rho \sigma} \overline{\partial H}_{\rho \nu \sigma} \,.
\label{Riccifirstwo}
\ee
A fairly straightforward calculations shows that these equal
\be
\begin{split}
 \tfrac{1}{4} &G^{\mu \nu}    D_\mu G_{\rho\sigma}  D_\nu G^{\rho\sigma} - \tfrac{1}{2} G^{\mu\nu}  D_\mu G^{\rho \sigma}  D_\rho G_{\nu \sigma} 
  -  \tfrac{1}{4} G^{\mu \nu} G^{\rho \sigma} H_{ij}  F_{\mu\rho}{}^i F_{\nu\sigma}{}^j  +  \tfrac{1}{4} G^{\mu \nu}
   D_\mu H_{ij}  D_\nu H^{ij} 
 \\ &\qquad+  \tfrac{1}{4} H^{ij} (  \partial_i G_{\rho\sigma} \partial_j G^{\rho\sigma} 
  + \partial_i H_{kl} \partial_j H^{kl} ) - \tfrac{1}{2} H^{ij} \partial_i H^{kl} \partial_k H_{jl} 
 \\ & \qquad
  - \tfrac{1}{2} ( \delta^i_k+H^{ij}H_{jk})  \bar D_\mu A_\nu{}^k \partial_i G^{\mu\nu}
+ G^{\mu\nu} H^{ij}H_{jk} \partial_l A_\mu{}^k \partial_i A_\nu{}^{l}
\end{split} 
\label{Ricci2}
\ee
where $F_{\mu\nu}{}^i \equiv 2 \bar D_{[\mu} A_{\nu]}{}^i$, $D_\mu = \partial_\mu - L_{A_\mu}$, and acting on $G_{\mu\nu}$ and $G^{\mu\nu}$, we have $D_\mu =\bar D_\mu$. 

Next, consider the part of $R^{(0)}$ that involves $\tau$: 
\be
\tfrac{1}{4} \hat H^{\hmu \hnu}\partial_{\hmu} \hat \tau_{\hrho \hsigma} \partial_{\hnu} \hat \tau^{\hrho \hsigma}
+ \tfrac{1}{4} \hat\tau^{\hmu\hnu} \partial_{\hmu}\tau_{\hrho \hsigma} \partial_{\hnu} \hat H^{\hrho \hsigma} 
- \tfrac{1}{2} \hat\tau^{\hmu \hnu} \partial_{\hnu}  H^{\hrho \hsigma} \partial_{\hrho} \hat\tau_{\hmu \hsigma}
- \tfrac{1}{2} \hat H^{\hmu \hnu} \partial_{\hnu} \hat\tau^{\hrho \hsigma} \partial_{\hrho} \hat\tau_{\hmu \hsigma}
\ee
Similar calculations to above give
\be
\begin{split}
 &
 \tfrac{1}{4} G^{\mu\nu} D_\mu \tau_{ij} D_\nu \tau^{ij} + g^{\mu\nu}  \tau^{ik} \tau_{kj} \partial_i A_\mu{}^l \partial_l A_\nu{}^j 
  - \tfrac{1}{2} \tau^{ik} \tau_{kj} \bar D_\mu A_\nu{}^k \partial_i G^{\mu\nu} 
\\ & \qquad
+\tfrac{1}{4}  H^{i j}\partial_{i}  \tau_{k l} \partial_{j}  \tau^{k l}
+ \tfrac{1}{4} \tau^{ij} \partial_{i}\tau_{k l} \partial_{j}  H^{k l} 
- \tfrac{1}{2} \tau^{i j} \partial_{j}  H^{k l} \partial_{k} \tau_{i l}
- \tfrac{1}{2}  H^{i j} \partial_{j}\tau^{k l} \partial_{k} \tau_{i l}
\end{split}
\label{Riccilong2}
\ee
The terms involving $\tau^{ik} \tau_{kj}$ on the first line here combine with the terms involving $H^{ik} H_{kj}$ in the last line of \eqref{Ricci2} and sum up to give $\delta^i_j = H^{ik} H_{kj} + \tau^{ik} \tau_{kj}$, after which point the rest of the calculation proceeds identically to that normally used in exceptional field theory. 

Finally one has the terms
\be
-  \bar G^{\hmu \hnu} \bar\partial_{\hmu} \ln \hat \Omega \, \bar\partial_{\hnu} \ln \hat \Omega
+ 2 \bar\partial_{\hmu} \ln \hat \Omega \, \overline{\partial G}_{\hnu}{}^{\hmu \hnu}
- \partial_{\hmu} \partial_{\hnu} \hat G^{\hmu \hnu} - \hat G^{\hmu\hnu} \partial_{\hmu} \partial_{\hnu} \ln \hat \Omega
\label{Riccilastfour}
\ee
where $\hat\Omega$ has weight 1, and in the final two terms $\bar \partial_\mu \equiv \bar D_\mu$, $\bar \partial_i \equiv \partial_i$.
Note $D_\mu \ln \hat \Omega = \bar D_\mu \ln \hat \Omega - \partial_i A_\mu{}^i$.
We let $\hat \Omega = \Omega \sqrt{|G|}$, where $\Omega$ has weight 1 under internal diffeomorphisms.
Straightforward manipulations allow one to rewrite \eqref{Riccilastfour} in the decomposition and combine with \eqref{Ricci2} and \eqref{Riccilong2} 
After dropping a total derivative, the final result is:
\be
\begin{split}
R^{(0)}(\hat H, \hat \tau) & =  R_{\text{ext}}(G) + R^{(0)}(H,\tau)  
- \tfrac{1}{4} F_{\mu\nu}{}^{ i} F_{\rho\sigma}{}^{ j} G^{\mu\rho} G^{\nu\sigma} H_{ij}
\\ &\qquad
+ \tfrac{1}{4}G^{\mu\nu} ( D_\mu H_{ij} D_\nu H^{ij} + D_\mu \tau_{ij} D_\nu \tau^{ij} 
+ D_\mu \ln \Omega^2 D_\nu \ln \Omega^2)
\\ &\qquad +\tfrac{1}{4} H^{i j} ( \partial_{i}{ G_{\mu \nu}} \partial_{j}{ G^{\mu \nu} }
+ \partial_{i}{ \ln |G| } \partial_{j}{ \ln |G| })
\end{split} 
\label{Rsplit}
\ee
where
\be
\begin{split}
R_{\text{ext}}(g)  = &
\tfrac{1}{4} G^{\mu \nu} D_\mu G_{\rho \sigma} D_\nu G^{\rho \sigma} 
- \tfrac{1}{2} G^{\mu \nu} D_\mu G^{\rho \sigma} D_\rho G_{\nu \sigma}
- \tfrac{1}{4} G^{\mu \nu} D_\mu \ln |G| D_\nu \ln |G|  \\
&\qquad 
- D_\mu \ln |G| D_\nu G^{\mu \nu} 
- G^{\mu \nu} D_\mu D_\nu \ln |G|
- D_\mu D_\nu G^{\mu \nu} \,,
\end{split}
\label{Ricciext}
\ee
\be
\begin{split}
R^{(0)}(H,\tau) &=
+\tfrac{1}{4}  H^{i j}\partial_{i}  \tau_{k l} \partial_{j}  \tau^{k l}
+ \tfrac{1}{4} \tau^{ij} \partial_{i}\tau_{k l} \partial_{j}  H^{k l} 
- \tfrac{1}{2} \tau^{i j} \partial_{j}  H^{k l} \partial_{k} \tau_{i l}
- \tfrac{1}{2}  H^{i j} \partial_{j}\tau^{k l} \partial_{k} \tau_{i l}
\\ & \qquad +
\tfrac{1}{4} H^{i j} \partial_{i}{ H_{k l}} \partial_{j}{ H^{k l} }
- \tfrac{1}{2} H^{i j} \partial_{j}{ H^{k l}  } \partial_{k}{ H_{i l} }
- \tfrac{1}{4} H^{i j} \partial_{i}{ \ln \Omega^2} \partial_{j}{ \ln \Omega^2 }
\\ & \qquad
- \partial_{i}{ \ln \Omega^2 } \partial_{j}{ H^{i j} }
- \partial_{i}{ \partial_{j}{ H^{i j} } }
- H^{i j} \partial_{i}{ \partial_{j}{\ln \Omega^2} }\,.
\end{split}
\label{Ricciint}
\ee
The measure factor is $\hat \Omega = \Omega \sqrt{|G|}$.
To obtain an Einstein frame action, we let
\be
G_{\mu\nu} = \Omega^{-\tfrac{2}{9-d}} g_{\mu\nu} \,.
\ee

\paragraph{Gauge fields} 

The compact expressions \eqref{Cdecomp_compact} and \eqref{Fdecomp_compact} are equivalent to
\be
\Cdef_{\hmu\hnu\hrho} = (U^{-1})^{\hlambda_1}{}_{\hmu} (U^{-1})^{\hlambda_2}{}_{\hnu} (U^{-1})^{\hlambda_3}{}_{\hrho} C_{\hlambda_1 \dots \hlambda_3}\,,\quad
\Fdef_{\hmu\hnu\hrho\hsigma} = (U^{-1})^{\hlambda_1}{}_{\hmu} (U^{-1})^{\hlambda_2}{}_{\hnu} (U^{-1})^{\hlambda_3}{}_{\hrho} (U^{-1})^{\hlambda_4}{}_{\hsigma}  F_{\hlambda_1 \dots \hlambda_4}\,,
\ee 
giving in components
\be
\begin{split}
\Cdef_{ijk}  & \equiv   C_{ijk} \,, \quad \Cdef_{\mu ij} \equiv  C_{\mu ij} - A_\mu{}^k  C_{ijk} \,,\\
\Cdef_{\mu \nu i } & \equiv   C_{\mu\nu i}- 2 A_{[\mu}{}^j  C_{\nu]ij} + A_\mu{}^j A_\nu{}^k  C_{ijk}\,, \\
\Cdef_{\mu\nu\rho} & \equiv   C_{\mu\nu\rho}- 3A_{[\mu}{}^i  C_{\nu\rho]i} + 3 A_{[\mu}{}^i A_{\nu}{}^j  C_{\rho] ij} - A_\mu{}^i A_\nu{}^j A_\rho{}^k  C_{ijk}
\,,
\label{MNC_splitC_ExFT}
\end{split}
\ee
\be
\begin{split}
\Fdef_{mnpq}  & = 4 \partial_{[m} \Cdef_{npq]}\,, \qquad \Fdef_{\mu mnp}  = D_\mu \Cdef_{mnp} -3 \partial_{[m} \Cdef_{|\mu|np]}\\
\Fdef_{\mu\nu m n}  & = 2 D_{[\mu} \Cdef_{\nu] mn} + \Fdef_{\mu\nu}{}^p \Cdef_{pmn} +2\partial_{[m} \Cdef_{|\mu\nu|n]}\,,\\
\Fdef_{\mu\nu\rho m} & =  3 D_{[\mu} \Cdef_{\nu \rho]m} + 3 \Fdef_{[\mu\nu}{}^n \Cdef_{\rho]mn}- \partial_m \Cdef_{\mu\nu\rho} \,,\\	
\Fdef_{\mu \nu \rho \sigma} & =4 D_{[\mu} \Cdef_{\nu\rho \sigma]} + 6 \Fdef_{[\mu\nu}{}^m \Cdef_{\rho \sigma ] m} \,,
\\
\end{split}
\label{F=dA}
\ee
where $F_{\mu\nu}{}^i$ is as defined in \eqref{kkf}.
The original Bianchi identity $d F_4=0$ becomes a set of equations
\be
\begin{array}{cccl}
&D_\mu \Fdef_{mnpq} & =& 4 \partial_{[m} \Fdef_{npq]}\,, \\
&2 D_{[\mu} \Fdef_{\nu] mnp} & =& - 3 \partial_{[m|} \Fdef_{\mu\nu|np]} - F_{\mu\nu}{}^q \Fdef_{qmnp} \,,\\
& 3 D_{[\mu} \Fdef_{\nu\rho ] m n} & = & 2 \partial_{[m|} \Fdef_{\mu\nu\rho|n]} + 3 F_{[\mu\nu}{}^p \Fdef_{\rho]pmn}\,,\\
&4 D_{[\mu} \Fdef_{\nu\rho\sigma] m} & = & - \partial_m \Fdef_{\mu\nu\rho\sigma} + 6 F_{[\mu\nu}{}^p \Fdef_{\rho\sigma]mp} \,,\\
&5D_{[\mu} \Fdef_{\nu \rho \sigma \lambda]} & =& 10 F_{[\mu\nu}{}^m \Fdef_{\rho\sigma \lambda] m} \,.
\end{array}
\label{FBI_mod}
\ee
The above formulae are applicable to any dimensional reduction.
In particular for the 11-dimensional MNC theory they allow us to easily decompose the terms in the action \eqref{resultaction}.
For example, using the Einstein frame metric to raise indices, the kinetic terms for the field strength are:
\be
\begin{split}
- \tfrac{1}{12} &\hat H^{\hmu_1 \hnu_1} \hat H^{\hmu_2 \hnu_2} \hat H^{\hmu_3 \hnu_3} \hat \tau^{\hmu_4 \hnu_4}  F_{\hmu_1 \hmu_2 \hmu_3 \hmu_4}  F_{\hnu_1 \hnu_2 \hnu_3 \hnu_4}
\\ & = 
- \tfrac{1}{12} \Omega^{{6}/{(9-d)}} \tau^{ij}  \Fdef^{\mu\nu\rho}{}_i \Fdef_{\mu\nu\rho j}
- \tfrac{1}{4} \Omega^{{4}/{(9-d)}}  H^{ij} \tau^{kl}  \Fdef_{\mu \nu ik}  \Fdef^{\mu\nu}{}_{jl}
\\ & \qquad - \tfrac{1}{4} \Omega^{{2}/{(9-d)}} H^{ij} H^{kl} \tau^{pq}  \Fdef_{\mu i k p}  \Fdef^{\nu}{}_{j l q}
 - \tfrac{1}{4}  H^{i j} H^{kl} H^{mn} \tau^{pq}  \Fdef_{ ikmp}  \Fdef_{jlnq	}\,.
\end{split} 
\ee
Similar manipulations apply to the rest of the action.
Let us also indicate how the factorisation applies to an equation of the form
$\partial_{\hsigma} X^{\hmu \hnu\hrho \hsigma} = \Theta^{\hmu \hnu \hrho} $ where $X$ has weight 1, and both $X$ and $\Theta$ admit a factorisation via $U^{-1}$ in terms of quantities $\bar X$ and $\bar \Theta$ independent of bare $A_\mu{}^i$.
This is of course the form of the gauge field equation of motion \eqref{3eomFinite}.
After decomposing, one has the simple expression
\be
D_\sigma \bar X^{\hmu \hnu \hrho \sigma} + \partial_l \bar X^{\hmu \hnu \hrho l}
+ \tfrac{3}{2} F_{\kappa\lambda}{}^l \delta_l^{[\hmu} \bar X^{\hnu \hrho] \kappa \lambda} 
= \bar \Theta^{\hmu\hnu\hrho} \,.
\label{eomuseful}
\ee

\paragraph{Constraint} 
The constraint \eqref{magic} decomposes in terms of the redefined strengths:
\be
\begin{split} 
\sqrt{g}  \Omega^{\tfrac{6}{9-d}}  g^{\mu_1 \nu_1}  \dots  g^{\mu_4\nu_4} \Fdef_{\nu_1 \dots \nu_4} & = - \tfrac{1}{4!} \epsilon^{\mu_1 \dots \mu_4 \hnu_1 \dots \hnu_4 ijk} \tfrac{1}{6} \epsilon_{ABC} \tau_i{}^A \tau_j{}^B \tau_k{}^C    \Fdef_{\hnu_1 \dots \hnu_4}\,,\\
\sqrt{g}\Omega^{\tfrac{4}{9-d}} g^{\mu_1 \nu_1}  \dots  g^{\mu_3\nu_3} H^{ ij}   \Fdef_{\nu_1 \nu_2 \nu_3  j} & = - \tfrac{1}{4!} \epsilon^{\mu_1 \dots \mu_3 i\hnu_1 \dots \hnu_4 pqr} \tfrac{1}{6} \epsilon_{ABC} \tau_p{}^A \tau_q{}^B \tau_r{}^C    \Fdef_{\hnu_1 \dots \hnu_4}\,,\\
\sqrt{g} \Omega^{\tfrac{2}{9-d}} g^{\mu_1\nu_1} g^{\mu_2 \nu_2} H^{ i_1 j_1} H^{ i_2 j_2}  \Fdef_{\nu_1 \nu_2 j_1 j_2 } & = - \tfrac{1}{4!} \epsilon^{ \mu_1 \mu_2 i_1 i_2 \hnu_1 \dots \hnu_4 pqr} \tfrac{1}{6} \epsilon_{ABC} \tau_p	{}^A \tau_q{}^B \tau_r{}^C    \Fdef_{\hnu_1 \dots \hnu_4}\,,\\
\sqrt{g} g^{\mu_1\nu_1}  H^{ i_1 j_1} \dots H^{ i_3 j_3}  \Fdef_{\nu j_1 j_2 j_3} & = - \tfrac{1}{4!} \epsilon^{ \mu i_1 \dots i_3 \hnu_1 \dots \hnu_4 pqr} \tfrac{1}{6} \epsilon_{ABC} \tau_p{}^A \tau_q{}^B \tau_r{}^C    \Fdef_{\hnu_1 \dots \hnu_4}\,,\\
\sqrt{g} \Omega^{-\tfrac{2}{9-d}}   H^{ i_1 j_1} \dots  H^{ i_4j_4}  \Fdef_{j_1j_2j_3j_4} & = - \tfrac{1}{4!} \epsilon^{i_1 \dots i_4 \hnu_1 \dots \hnu_4 pqr} \tfrac{1}{6} \epsilon_{ABC} \tau_p{}^A \tau_q{}^B \tau_r{}^C   \hat \Fdef_{\hnu_1 \dots \hnu_4}\,.
\end{split}
\label{constraints_decomp}
\ee
For instance, when $d=3$ only the first of these is non-zero, giving:
\be
\begin{split}
\sqrt{g} \Omega  g^{\mu_1 \nu_1}  \dots  g^{\mu_4\nu_4} \Fdef_{\nu_1 \dots \nu_4} & 
= - \tfrac{1}{4!} \epsilon^{\mu_1 \dots \mu_4\nu_1 \dots \nu_4 ijk} \tfrac{1}{6} \epsilon_{ABC} \tau_i{}^A \tau_j{}^B \tau_k{}^C    \Fdef_{\nu_1 \dots \nu_4}\,,
\\ & = - \tfrac{1}{4!} \epsilon^{\mu_1 \dots \mu_4\nu_1 \dots \nu_4 } \Omega   \Fdef_{\nu_1 \dots \nu_4}\,.
\end{split}
\ee
When $d=4$ only the first two are non-zero:
\be
\begin{split} 
\sqrt{g}  \Omega^{\tfrac{6}{5}}  g^{\mu_1 \nu_1}  \dots  g^{\mu_4\nu_4} \Fdef_{\nu_1 \dots \nu_4} & = - \tfrac{1}{3!} \epsilon^{\mu_1 \dots \mu_4 \nu_1 \dots \nu_3 l ijk} \tfrac{1}{6} \epsilon_{ABC} \tau_i{}^A \tau_j{}^B \tau_k{}^C    \Fdef_{\nu_1 \nu_2 \nu_3 l}\,,\\
\sqrt{g} \Omega^{\tfrac{4}{5}} g^{\mu_1 \nu_1}  \dots  g^{\mu_3\nu_3} H^{ ij}   \Fdef_{\nu_1 \nu_2 \nu_3  j} & = - \tfrac{1}{4!} \epsilon^{\mu_1 \dots \mu_3 i\nu_1 \dots \nu_4 pqr} \tfrac{1}{6} \epsilon_{ABC} \tau_p{}^A \tau_q{}^B \tau_r{}^C    \Fdef_{\nu_1 \dots \nu_4}\,,\\
\end{split}
\label{Constraints4}
\ee
or if we take $\tfrac{1}{6} \epsilon^{ijkl} \epsilon_{ABC} \tau_i{}^A \tau_j{}^B \tau_k{}^C h_l = \Omega$ these are
\be
\begin{split} 
\sqrt{g}  \Omega^{\tfrac{1}{5}}  g^{\mu_1 \nu_1}  \dots  g^{\mu_4\nu_4} \Fdef_{\nu_1 \dots \nu_4} & = \tfrac{1}{3!} \eta \epsilon^{\mu_1 \dots \mu_4 \nu_1 \dots \nu_3}  h^l \Fdef_{\nu_1 \nu_2 \nu_3 l}\,,\\
\sqrt{g} g^{\mu_1 \nu_1}  \dots  g^{\mu_3\nu_3} H^{ ij}   \Fdef_{\nu_1 \nu_2 \nu_3  j} &  =   \tfrac{1}{4!}  \eta \epsilon^{\mu_1 \dots \mu_3 \nu_1 \dots \nu_4} h^i  \Omega^{\tfrac{1}{5}}\Fdef_{\nu_1 \dots \nu_4}\,.
\end{split}
\label{Constraints4_simpler}
\ee
Here $H^{ij} = h^i h^j$ (as it has rank 1), and so both of these are equivalent.

\paragraph{Result}

Putting everything together, the dimensional decomposition of the finite action $S^{(0)}$ is:
\be
S^{(0)} = \int \dd^{11-d}x \,\dd^{d}y \sqrt{g} ( R_{\text{ext}}(g) + \mathcal{L}_{\text{kin}} + \mathcal{L}_{\text{int}} + \mathcal{L}_{\widetilde F} + \sqrt{g}^{-1} \mathcal{L}_{\text{CS}} ) \,.
\ee
Here, using $g^{\mu\nu}$ to raise $(11-d)$-dimensional indices, we have
\be
\begin{split}
R_{\text{ext}}(g)  = &
\frac{1}{4} g^{\mu \nu} D_\mu g_{\rho \sigma} D_\nu g^{\rho \sigma} 
- \frac{1}{2} g^{\mu \nu} D_\mu g^{\rho \sigma} D_\rho g_{\nu \sigma}
+ \frac{1}{4} g^{\mu \nu} D_\mu \ln g D_\nu \ln g 
+ \frac{1}{2} D_\mu \ln g D_\nu g^{\mu \nu} \,,
\end{split}
\label{11Rext}
\ee
\be
\begin{split} 
\mathcal{L}_{\text{kin}} 
& = 
  \tfrac{1}{4}(  D_\mu H^{ij} D^\mu H_{ij} +  D_\mu \tau^{ij} D^\mu \tau_{ij}  - \tfrac{1}{9-d} D_\mu \ln \Omega^2 D^\mu \ln \Omega^2)
  +   \tfrac{1}{2}  D_\mu \tau_k{}^A \tau_A{}^k  D^\mu \tau_l{}^B \tau_B{}^l
\\ &\qquad + \tfrac{1}{2} H^{ij} \Fdef_{\mu i kl} \epsilon_{ABC} D^\mu \tau_{j}{}^A \tau^{k B} \tau^{l C} 
 - \tfrac{1}{4}  H^{ij} H^{kl} \tau^{pq}  \Fdef_{\mu i k p}  \Fdef^\mu{}_{j l q}
 \\ & \qquad
 +  \tfrac{1}{4} \Omega^{\tfrac{2}{9-d}} \big(- F_{\mu\nu}{}^{ i} F^{\mu\nu}{}^{ j}H_{ij}
+  \Fdef_{\mu\nu kl} F^{\mu\nu m} \epsilon_{ABC}  \tau_{m}^A \tau^{k B} \tau^{l C} 
-   H^{ij} \tau^{kl}  \Fdef_{\mu\nu ik}  \Fdef^{\mu\nu}{}_{jl}\big)
\\ & \qquad 
 - \tfrac{1}{12} \Omega^{\tfrac{4}{9-d}} \tau^{ij} \Fdef_{\mu\nu\rho i}  \Fdef^{\mu\nu\rho}{}_{j} 
\end{split}
\label{11kin}
\ee
and
\be
\begin{split}
\Omega^{\tfrac{2}{9-d}} \mathcal{L}_{\text{int}}
& =  
\tfrac{1}{4} H^{ij} \left(  \partial_i g^{\mu\nu} \partial_j g_{\mu\nu} +  \partial_i \ln g \partial_j \ln g \right)
+ \tfrac{1}{2} \Omega^{\tfrac{2}{9-d}} \partial_i ( H^{ij}\Omega^{-\tfrac{2}{9-d}} ) \partial_j \ln g
\\ &\quad
 +\tfrac{1}{4}  H^{i j}\partial_{i}  \tau_{k l} \partial_{j}  \tau^{k l}
+ \tfrac{1}{4} \tau^{ij} \partial_{i}\tau_{k l} \partial_{j}  H^{k l} 
- \tfrac{1}{2} \tau^{i j} \partial_{j}  H^{k l} \partial_{k} \tau_{i l}
- \tfrac{1}{2}  H^{i j} \partial_{j}\tau^{k l} \partial_{k} \tau_{i l}
\\ & \quad
+\tfrac{1}{4} H^{i j} \partial_i H_{k l} \partial_j H^{k l} 
- \tfrac{1}{2} H^{i j} \partial_j H^{k l} \partial_{k} H_{i l} 
\\ & \quad
 + \tfrac{1}{4} \tfrac{d-7}{(9-d)^2} H^{i j} \partial_i \ln \Omega^2 \, \partial_j \ln \Omega^2
- \tfrac{1}{9-d} \partial_i \ln \Omega^2 \, \partial_j H^{i j} 
\\ & \quad
 - \tfrac{1}{4}  H^{i j} H^{kl} H^{mn} \tau^{pq}  \Fdef_{ ikmp}  \Fdef_{jlnq	}
  +\tfrac{1}{4} H^{im} H^{jn} \Fdef_{ijkl} \epsilon_{ABC} T_{mn}{}^A \tau^{kB}\tau^{lC}
\\ & \quad   + \tfrac{1}{2} H^{ij} T_{ik}{}^A \tau^k{}_A T_{jl}{}^B \tau^l{}_B
\,.
\end{split}
\label{11int}
\ee
The term $\mathcal{L}_{\widetilde F}$ consists of a sum of contractions of $\widetilde \Fdef_{\mu\nu\rho\sigma}$, $\widetilde \Fdef_{\mu\nu\rho i}$, etc. (following analogous redefinition of the components) with the constraints as decomposed in \eqref{constraints_decomp}.
For instance, when $d=3$,
\be
\mathcal{L}_{\widetilde F}
 = - \tfrac{1}{4!} \widetilde \Fdef_{\mu_1 \dots \mu_4} (\sqrt{g} \Omega  g^{\mu_1 \nu_1}  \dots  g^{\mu_4\nu_4} \Fdef_{\nu_1 \dots \nu_4} 
 +\tfrac{1}{4!} \epsilon^{\mu_1 \dots \mu_4\nu_1 \dots \nu_4 } \Omega   \Fdef_{\nu_1 \dots \nu_4} ) \,,
 \label{constraint3d}
\ee
In this case the relationship between the dual seven-form field strength and $\widetilde \Fdef_{\mu\nu\rho\sigma}$ gives
\be
\tfrac{1}{6}\epsilon^{ijk} \Fdef_{\mu_1 \dots \mu_4 ijk} = \Omega ( \widetilde \Fdef_{\mu_1 \dots \mu_4} +\tfrac{1}{4!} \sqrt{g} \epsilon_{\mu_1 \dots \mu_4 \nu_1 \dots \nu_4} \widetilde \Fdef^{\nu_1 \dots \nu_4} ) \,.
\label{32F7F4}
\ee
When $d=4$,
\be
\begin{split}
\mathcal{L}_{\widetilde F}
& =
 -\tfrac{1}{3!} \left(\widetilde \Fdef_{\mu_1\mu_2\mu_3 i} h^i  -  \Omega^{1/5} \epsilon^{\lambda_1\dots\lambda_4\sigma_1\dots\sigma_3} \tfrac{1}{4!} \tfrac{1}{\sqrt{g}} g_{\sigma_1\mu_1} \dots g_{\sigma_3\mu_3} \widetilde \Fdef_{\lambda_1 \dots \lambda_4} \right)
 \times
 \\ & \qquad\quad\times
 \left( \sqrt{g} \Omega^{\tfrac{4}{5}} g^{\mu_1 \nu_1}  \dots  g^{\mu_3\nu_3}   h^j  \Fdef_{\nu_1 \nu_2 \nu_3  j} -    \Omega \tfrac{1}{4!}  \epsilon^{\mu_1 \dots \mu_3 \nu_1 \dots \nu_4}  \Fdef_{\nu_1 \dots \nu_4}\right) \,,
\end{split}
\label{constraint4d}
\ee
Using \eqref{defF7} we can rewrite \eqref{constraint4d} in terms of the dual seven-form field strength directly as
\be
\mathcal{L}_{\widetilde F} = + \tfrac{1}{3! 4!} \Fdef_{\mu_1 \dots \mu_3 ijkl}   \epsilon^{ijkl} 
 \left( \sqrt{g}  \Omega^{-\tfrac{1}{5}} g^{\mu_1 \nu_1}  \dots  g^{\mu_3\nu_3}   h^j  \Fdef_{\nu_1 \nu_2 \nu_3  j} -   \tfrac{1}{4!}  \epsilon^{\mu_1 \dots \mu_3 \nu_1 \dots \nu_4}  \Fdef_{\nu_1 \dots \nu_4}\right) \,.
\label{constraint4dbetter}
\ee
Finally, the Chern-Simons term can be worked out by taking wedge products of \eqref{Fdecomp_compact} and \eqref{Cdecomp_compact}, we do not display this explicitly.


\section{The $\Gfour$ ExFT and its non-relativistic parametrisation}
\label{AndSL5}

In the $d=4$ case, more of the degenerate Newton-Cartan structure is preserved.

\paragraph{Elements of $\Gfour$ ExFT} 

For $d=4$, generalised vectors $V^M = (V^i, V_{ij})$ transform in the $\mathbf{10}$ of $\Gfour$, with $i,j,\dots$ now four-dimensional.
This representation is the antisymmetric representation, and we can see this more clearly as follows.
Let $\fM, \fN, \dots$ denote fundamental five-dimensional indices of $\Gfour$.
Then we can equivalently write a generalised vector as carrying an antisymmetric pair of such indices, $V^M \equiv V^{\fM \fN}=-V^{\fN \fM}$, and on writing $\fM=(i,5)$ we can identify $V^{i5} \equiv V^{i}$, and $V^{ij} \equiv \tfrac{1}{2} \epsilon^{ijkl} V_{kl}$.
The generalised Lie derivative acting on vectors of weight $\lambda_V$ is explicitly
\be
\mathcal{L}_\Lambda V^{\fM \fN} = \tfrac{1}{2} \Lambda^{\fP \fQ} \partial_{\fP\fQ} V^{\fM\fN} + 2 \partial_{\fP\fQ} \Lambda^{\fP[\fM} V^{\fN]\fQ} 
+ \tfrac{1}{2} ( 1 + \lambda_V + \omega ) \partial_{\fP\fQ} \Lambda^{\fP\fQ} V^{\fM\fN}\,.
\ee
The section condition is $\epsilon^{\fM\fN\fP\fQ\fK} \partial_{\fM \fN} \partial_{\fP\fQ}=0$, and below we work with the M-theory solution, where splitting $\fM = (i,5)$ the derivatives $\partial_{ij}$ are viewed as identically zero, and the derivatives $\partial_{i5}$ are identified with the $4$-dimensional partial derivatives.

In this case, the generalised metric admits a factorisation
\be
\gM_{\fM \fN, \fP \fQ} = - ( m_{\fM \fP} m_{\fQ \fN} - m_{\fM \fQ} m_{\fP \fN} ) 
\ee
where the `little metric' $m_{\fM\fN}$ is symmetric and has unit 
determinant. The overall sign in this expression needed for the ExFT action to reproduce SUGRA correctly when we include timelike signatures in the generalised metric, according to the conventions of \cite{Berman:2019izh}.

The gauge fields, $\Aa_\mu{}^M$, $\Ab_{\mu\nu \fM}$, $\Ac_{\mu\nu\rho}{}^{\fM}$ and $\Ad_{\mu\nu\rho\sigma}{}_{M}$ have weights $1/5$, $2/5$, $3/5$ and $4/5$ respectively, with field strengths denoted $\Fa_{\mu\nu}{}^M$,  $\Fb_{\mu\nu\rho \fM}$, $\Fc_{\mu\nu\rho\sigma}{}^{\fN}$ and $\Fd_{\mu\nu\rho\sigma\lambda}{}_{M}$.
Under generalised diffeomorphisms, $\Fa^{M}$ transforms as a generalised vector of weight $1/5$, while $\Fb$ and $\Fc$ transform via the generalised Lie derivative acting as
\be
\begin{split} 
\mathcal{L}_\Lambda \Fb_{\fM}  = \tfrac{1}{2} \Lambda^{\fP\fQ} \partial_{\fP\fQ} \Fb_{\fM} + \Fb_{\fP} \partial_{\fM\fQ} \Lambda^{\fP\fQ}\,,\quad
\mathcal{L}_\Lambda \Fc^{\fM}  = \partial_{\fP\fQ}( \tfrac{1}{2} \Lambda^{\fP\fQ}  \Fc^{\fM} ) - \partial_{\fP\fQ} \Lambda^{\fP\fM} \Fc^{\fQ} 
\,.
\end{split}
\label{5_gldF}
\ee
They obey Bianchi identities:
\be
3 \D_{[\mu} \Fa_{\nu\rho]}{}^{\fM \fN} = \tfrac{1}{2} \epsilon^{\fM \fN \fP \fQ \fK} \partial_{\fP \fQ} \Fb_{\mu\nu\rho \fK}\,,
\label{5BianchiA}
\ee
\be
4 \D_{[\mu} \Fb_{\nu\rho\sigma] \fM} + \tfrac{3}{4}  \epsilon_{\fM \fN \fP \fK \fL} \Fa_{[\mu\nu}{}^{\fN \fP} \Fa_{\rho\sigma ]}{}^{\fK \fL} = \partial_{\fN \fM} \Fc_{\mu\nu\rho\sigma}{}^{\fN}\,,
\label{5BianchiB}
\ee
\be
5 \D_{[\mu} \Fc_{\nu\rho\sigma\lambda]}{}^{\fM} + 10 \Fa_{[\mu\nu}{}^{\fM \fN} \cH_{\rho\sigma \lambda ]\fN}= \tfrac{1}{2}\epsilon^{\fM \fN \fP \fQ \fK}  \partial_{\fN \fP} \Fd_{\mu\nu\rho\sigma\lambda}{}_{\fQ \fK}\,.
\label{5BianchiC}
\ee
The dynamics follow from the variation of an action $S = \int \dd^7x\, \dd^{10}y \mathcal{L}_{\text{ExFT}}$ where $\mathcal{L}_{\text{ExFT}}$ has the same form as \eqref{Lexft},
with $R_{\text{ext}}$ again as defined in \eqref{Rext32}, and \cite{Musaev:2015ces}
\be
\mathcal{L}_{\text{kin}} = + \tfrac{1}{12} \D_\mu \gM_{MN}  \D^\mu \gM^{MN} 
- \tfrac{1}{4} \gM_{MN} \Fa_{\mu\nu}{}^{M} \Fa^{\mu\nu N} 
-  \tfrac{1}{12} m^{\fM \fN} \Fb_{\mu\nu\rho \fM} \Fb^{\mu\nu\rho}{}_{\fN}
\label{Lkin5}
\ee
\be
\begin{split}
\mathcal{L}_{\text{int}}(m,g) &=
\tfrac{1}{12} \gM^{MN} \partial_M \gM^{KL} \partial_N \gM_{KL} - \tfrac{1}{2} \gM^{MN} \partial_M \gM^{KL} \partial_K \gM_{LN} 
+ \tfrac{1}{2} \partial_M \gM^{MN} \partial_N \ln |g|
\\ & \qquad+ \tfrac{1}{4} \gM^{MN} \left( \partial_M g_{\mu\nu} \partial_N g^{\mu\nu} + \partial_M \ln |g| \partial_N \ln |g| \right)  \,.
\end{split}
\label{Lint5}
\ee
The topological term can be defined via its variation (again up to a choice of sign equivalent to changing the sign of $\hat C_3$ in eleven-dimensional SUGRA):
\be
\begin{split}
\delta \mathcal{L}_{\text{top}} = - \tfrac{1}{6\cdot 4!}
\epsilon^{\mu_1\dots\mu_7}
\Big(&
2 \delta \Aa_{\mu_1}{}^{\fM \fN} \Fb_{\mu_2\mu_3\mu_4 \fM}  \Fb_{\mu_5\mu_6\mu_7 \fN}
+ 6 \Fa_{\mu_1 \mu_2}{}^{\fM \fN}  \Delta \Ab_{\mu_3\mu_4 \fM} \Fb_{\mu_5\mu_6\mu_7 \fN}
\\ & \qquad 
  \partial_{\fN \fM} \Delta \Ac_{\mu_1 \mu_2 \mu_3}{}^{\fN}  \Fc_{\mu_4 \dots \mu_7}{}^{\fM}
\Big) \,.
\end{split}
\label{deltastopfive}
\ee
We refer to the original paper \cite{Musaev:2015ces} or the review \cite{Berman:2020tqn} for explicit details.

\paragraph{Review of 11-dimensional SUGRA embedding}

We start with the little metric, $m_{\fM \fN}$.
The parametrisation reproducing \eqref{genmet_usual} is
\be
m_{\fM \fN} = 
|\phi|^{1/10} 
\begin{pmatrix}
|\phi|^{-1/2} \phi_{ij} & - |\phi|^{-1/2} \phi_{ik} \hat \Cdef^k \\
- |\phi|^{-1/2} \phi_{jk} \hat \Cdef^k & |\phi|^{1/2}(-1)^t + |\phi|^{-1/2} \phi_{kl} \hat \Cdef^k\hat \Cdef^l
\end{pmatrix} \,, \quad \hat \Cdef^i \equiv \tfrac{1}{3!} \epsilon^{ijkl} \hat \Cdef_{jkl}\,.
\label{littlemC}
\ee
For the gauge fields, we can again identify $\Aa_\mu^M = (A_\mu{}^i, \hat\Cdef_{\mu ij})$.
However, we already require dualisations when treating the two-forms.
We get four $7$-dimensional two-forms, $\hat \Cdef_{\mu\nu i}$ and a single three-form $\hat \Cdef_{\mu\nu\rho}$. The latter can be dualised into an extra two-form, $\widetilde C_{\mu\nu}$ (identifiable with the components $\hat\Cdef_{\mu\nu ijkl}$ of the six-form in eleven-dimensions) such that $\Ab_{\mu\nu \fM} \sim (\hat\Cdef_{\mu\nu i}, \widetilde C_{\mu\nu})$ gives a five-dimensional representation of $\Gfour$. 
Meanwhile, we can view $\hat \Cdef_{\mu\nu\rho}$ together with the four four-forms $\hat \Cdef_{\mu\nu\rho ijk}$ as comprising the conjugate five-dimensional representation.
The equations of motion of the $\Gfour$ ExFT then imply that the field strengths of these two- and three-forms are related by duality.
This involves the seven-dimensional Hodge star acting on the seven-dimensional indices and the generalised metric acting on the $\Gfour$ indices:
\be
\sqrt{|g|} m^{\fM \fP} \Fb^{\mu\nu\rho}{}_{\fP}   =  -\tfrac{1}{4!} \epsilon^{\mu\nu\rho \sigma_1\dots \sigma_4} \Fc_{\sigma_1\dots\sigma_4}{}^{\fM} 
\ee
Again, the field strengths are all tensors under generalised diffeomorphisms, we may make the (usual) identifications consistent with the Bianchi identities \cite{Berman:2020tqn}
\be
\begin{split} 
\mathcal{F}_{\mu\nu}{}^{i5}& = F_{\mu\nu}{}^i\,,\quad\quad\,\,
\Fa_{\mu\nu}{}^{ij} = \tfrac{1}{2} \epsilon^{ijkl} (\hat\Fdef_{\mu\nu kl} - \hat \Cdef_{klm} \hat\Fdef_{\mu\nu}{}^{m})\,,\\
\Fb_{\mu\nu\rho i} & = -\hat \Fdef_{\mu\nu\rho i} \,,\quad
\Fb_{\mu\nu\rho 5}  = - \tfrac{1}{4!} \epsilon^{ijkl} ( \hat \Fdef_{\mu\nu \rho ijkl} - 4 \hat \Fdef_{\mu\nu\rho i} \hat \Cdef_{jkl})\,,\quad
\\
\Fc_{\mu\nu\rho\sigma}{}^5 & = - \hat\Fdef_{\mu\nu\rho \sigma}\,,\quad\,\,
\Fc_{\mu\nu\rho\sigma}{}^i = + \tfrac{1}{3!} \epsilon^{ijkl}( \hat\Fdef_{\mu\nu\rho\sigma jkl} -\hat \Cdef_{jkl} \hat\Fdef_{\mu\nu\rho\sigma} )
\label{deftildeF}\,.
\end{split} 
\ee

\paragraph{Generalised metric}
The distinction between Riemannian and non-Riemannian parametrisations can be seen at the level of the unit-determinant five-by-five little generalised metric.
For an M-theory parametrisation, this can be written as:
\be
m_{\fM \fN} = \begin{pmatrix} 
m_{ij} & m_{i5} \\
m_{j5} & m_{55}  
\end{pmatrix} \,,\quad
m_{55} \det (m_{ij}) - \tfrac{1}{6} m_{i5} m_{j5} \epsilon^{iklm} \epsilon^{jpqr} m_{kp} m_{lq} m_{mr} = 1 \,.
\label{detConstraint}
\ee
If $\det (m_{ij}) \neq 0$ this leads to the Riemannian parametrisation  \eqref{littlemC} encoding a four-dimensional metric, $g_{ij}$, and a three-form, $\hat \Cdef_{ijk}$.
However, we can also have $\det (m_{ij}) =0$ with $m_{ij}$ of rank 3 and this leads to a non-Riemannian parametrisation which was worked out in \cite{Berman:2019izh}. 
We can rediscover this parametrisation by taking the non-relativistic limit of \eqref{littlemC} using \eqref{simplerlimit}. 
The resulting expression for $m_{\fM \fN}$ is
\be
m_{\fM \fN} 
= \Omega^{-4/5} \begin{pmatrix}
 \tau_{ij} & \tfrac{1}{6} H_{ik} \epsilon^{klmn} \epsilon_{ABC} \tau_l{}^A \tau_{m}{}^B \tau_n{}^C - \tau_{ik} \Cdef^k \\
 \tfrac{1}{6} H_{jk} \epsilon^{klmn} \epsilon_{ABC} \tau_l{}^A \tau_{m}{}^B \tau_n{}^C - \tau_{jk} \Cdef^k  & \tau_{ij} \Cdef^i \Cdef^j - \tfrac{1}{3} \epsilon^{jklm} \epsilon_{ABC} H_{ij} \tau_{k}{}^A \tau_l{}^B \tau_m{}^C \Cdef^i 
\end{pmatrix} \,,
\label{littlemC_limit} 
\ee
in terms of four-dimensional Newton-Cartan variables and $\Cdef^i \equiv \tfrac{1}{3!} \epsilon^{ijkl} \Cdef_{jkl}$.
The unit determinant constraint implies that
\be
- \tfrac{1}{3!} \epsilon^{i_1 \dots i_4} \epsilon^{j_1 \dots j_4} \tau_{i_1 j_1} \tau_{i_2 j_2} \tau_{i_3 j_3} H_{i_4 j_4} = \Omega^2 \,,
\ee
which is the definition of $\Omega^2$ in this case.
As $H_{ij}$ has rank 1, we can introduce a projective vielbein $h_i$ such that $H_{ij} = h_i h_j$ and we take
\be
\tfrac{1}{6} \epsilon^{ijkl} \epsilon_{ABC} \tau_i{}^A \tau_j{}^B \tau_k{}^C h_l =  \Omega\,,
\label{epsOmega}
\ee
choosing to fix an arbitrary sign (by sending $\tau_i{}^A \rightarrow -\tau_i{}^A$ if necessary) which could appear here ($\Omega$ is assumed positive).
Then \eqref{littlemC_limit} can be written as
\be
m_{\fM \fN} 
= \Omega^{-4/5} \begin{pmatrix}
 \tau_{ij} & -  \Omega h_i - \tau_{ik} \Cdef^k \\
-  \Omega h_j - \tau_{jk} \Cdef^k  & \tau_{ij} \Cdef^i \Cdef^j + 2  \Omega h_i \Cdef^i
\end{pmatrix} \,,
\label{littlemC_limit2} 
\ee
which in this form can be checked to correspond to the parametrisation written down in \cite{Berman:2019izh} from first principles. 
Note that the boost invariance, acting as 
\be
\delta h_i = h^j \Lambda_j{}^A \tau_{i A} \,,\quad
\delta \Cdef^i = -  \Omega \Lambda_j{}^A h^j\tau^i{}_A\,,\quad \tau^{i}{}_A \Lambda_i{}^B = 0 \,,
\ee
corresponds to a shift symmetry of the parametrisation \eqref{littlemC_limit2} pointed out in \cite{Berman:2019izh}. This generalises the Milne shift redundancy of the DFT non-Riemannian parametrisation \cite{Morand:2017fnv}.
Here we introduced the inverse vielbeins $h^i$ and $\tau^i{}_A$ obeying the obvious relations
\be
h_i h^i = 1 \,,\quad
\tau^{i}{}_A  \tau_{j}{}^A + h^i h_j = \delta^i_j \,,\quad
\tau^{i}{}_A h_i = 0 \,\quad \tau_{i}{}^A h^i = 0 \,, \quad \tau^i{}_A \tau_i{}^B =\delta_A^B \,.
\ee
The generalised metric in the $10 \times 10$ representation followng from the little metric \eqref{littlemC_limit} can be seen to take the form \eqref{genmetnonrie}, 
after rewriting in the basis where generalised indices run over vector and two-form indices, and using the identities
\be
\begin{split} 
\epsilon^{i_1 \dots i_3 k} \epsilon^{j_1 \dots j_3 l} \tau_{kl} & = - 3! \Omega^2 ( 
\tau^{j_1 [i_1} \tau^{i_2|j_2|} H^{i_3] j_3} + \tau^{j_2 [i_1} \tau^{i_2|j_3|} H^{i_3] j_1}  + \tau^{j_3 [i_1} \tau^{i_2|j_1|} H^{i_3] j_2}
)\,,\\
\epsilon^{i_1 \dots i_3 k} \epsilon^{j_1 \dots j_3 l} H_{kl} & = - 3! \Omega^2 
\tau^{i_1 [j_1|} \tau^{i_2| j_2|} \tau^{i_3| j_3]} \,.
\end{split}
\ee
It is useful to record the explicit expression for the inverse little metric:
\be
m^{\fM\fN}   =
\Omega^{4/5}\begin{pmatrix}
\tau^{ij}  - 2  \Omega^{-1} h^{(i} \Cdef^{j)} &  -  \Omega^{-1} h^i \\
- \Omega^{-1} h^j & 0 
\end{pmatrix} \,.
\ee
Clearly, variations $\delta m^{\fM \fN}$ with $\delta m^{55} \neq 0$ do not preserve this parametrisation.
This means that if we look at the equations of motion $\mathcal{R}_{\fM \fN} = 0$ of the generalised metric, we expect that $\mathcal{R}_{55} = 0$ provides an additional equation of motion that we would not find by varying the action evaluated on the non-relativistic parametrisation.

\paragraph{Field strengths and self-duality in $\Gfour$ ExFT}

Our field strengths \eqref{deftildeF} are now 
\be
\begin{split} 
\mathcal{F}_{\mu\nu}{}^{i5}& = F_{\mu\nu}{}^i\,,\quad\quad\,\,
\Fa_{\mu\nu}{}^{ij} = \tfrac{1}{2} \epsilon^{ijkl} (\Fdef_{\mu\nu kl} -  \Cdef_{klm} \Fdef_{\mu\nu}{}^{m})\,,\\
\Fb_{\mu\nu\rho i} & = - \Fdef_{\mu\nu\rho i} \,,\quad
\Fb_{\mu\nu\rho 5}  = - \tfrac{1}{4!} \epsilon^{ijkl} (  \Fdef_{\mu\nu \rho ijkl} - 4  \Fdef_{\mu\nu\rho i}  \Cdef_{jkl})\,,\quad
\\
\Fc_{\mu\nu\rho\sigma}{}^5 & = - \Fdef_{\mu\nu\rho \sigma}\,,\quad\,\,
\Fc_{\mu\nu\rho\sigma}{}^i = + \tfrac{1}{3!} \epsilon^{ijkl}( \Fdef_{\mu\nu\rho\sigma jkl} - \Cdef_{jkl} \Fdef_{\mu\nu\rho\sigma} )
\label{deftildeFhere}\,.
\end{split} 
\ee
The kinetic terms \eqref{Lkin5} in the $\Gfour$ ExFT action are:
\be
\begin{split}
- \tfrac{1}{4} & \gM_{MN} \Fa^{\mu\nu M} \Fa_{\mu\nu}{}^N  - \tfrac{1}{12} m^{\fM \fN} \Fb^{\mu\nu\rho}{}_{\fM} \Fb_{\mu\nu\rho \fN} 
 \\& = - \tfrac{1}{4} \Omega^{2/5}\left( H_{ij} F^{\mu\nu i} F_{\mu\nu}{}^j 
  -\epsilon^{ABC}\tau_{iA} \tau_B{}^j \tau_C{}^k F^{\mu\nu i} \Fdef_{\mu \nu jk} 
  + \tau^i_C \tau^j{}^C H^{kl} \Fdef^{\mu\nu}{}_{ik} \Fdef_{\mu\nu jl} 
  \right)
 \\ & \quad
   - \tfrac{1}{12} \Omega^{4/5} \tau^{ij}  \Fdef^{\mu\nu\rho}{}_{i} \Fdef_{\mu\nu\rho j} 
  + \tfrac{1}{6}  \Omega^{-1/5}  h^i \Fdef^{\mu\nu\rho}{}_{i}\tfrac{1}{4!}\epsilon^{jklm} \Fdef_{\mu\nu\rho jklm } 
\end{split} 
\ee
which match exactly the corresponding terms in \eqref{11kin} and \eqref{constraint4d}, including the appearance of components of the dual seven-form field strength.

We see again that the ExFT description automatically contains the correct dual fields to reproduce the non-relativistic action immediately.
It's worthwhile to go into some detail about the appearance of dual fields in the relativistic case.
As mentioned above, the decomposition of the 11-dimensional three-form in the $(7+4)$-dimensional split produces four two-forms, $\hat\Cdef_{\mu\nu i}$ and a single three-form, $\hat\Cdef_{\mu\nu\rho}$.
We exchange the latter for an additional two-form, $\hat\Cdef_{\mu\nu}$, in order to obtain the five-dimensional $\Gfour$ multiplet $\Ab_{\mu\nu \fM} = ( \hat\Cdef_{\mu\nu i} ,  \hat\Cdef_{\mu\nu})$.
This is normally done by introducing the two-form into the action as a Lagrange multiplier enforcing the Bianchi identity for $\hat\Fdef_{\mu\nu\rho\sigma}$.
When this is done, the terms involving $\hat\Fdef_4$ in the action are schematically $\hat\Fdef_4 \wedge \star_7 \hat\Fdef_4 - \hat\Cdef_2 \wedge (d \hat\Fdef_4 + \dots) + \hat\Fdef_4 \wedge X_3$, where $X_3$ denotes whatever appears alongside $\hat\Fdef_4$ in the decomposition of the Chern-Simons term. Integrating by parts one defines a field strength $H_3 \sim d\hat\Cdef_2 + X_3$ and treating $\hat\Fdef_4$ then as an independent field, one can integrate that out of the action to produce a kinetic term for $H_3$. The latter is then the $\fM=5$ component of the ExFT field strength $\Fb_{\mu\nu\rho \fM}$, and in this way the ExFT action matches the partially dualised SUGRA action.

In the non-relativistic theory, there is already no kinetic term for $\Fdef_4$ in the decomposed action, as seen from \eqref{11kin}.
It only appears (linearly) in the constraint term \eqref{constraint4d}, schematically in the form $\Fdef_4 \wedge ( \star_7 \widetilde \Fdef_4 + \widetilde \Fdef_{3 i} h^i)$. 
So instead if we carry out the same procedure, we find that $\Fdef_4$ equation of motion sets $H_3 = \star_7 \widetilde\Fdef_4 + \widetilde \Fdef_{3 i} h^i$, which in this case exactly corresponds to the relationship between the dual seven-form and $\widetilde F_4$ as expressed by \eqref{defF7}. 
Hence now it is this $H_3$ that we identify with $\Fb_{\mu\nu\rho ijkl}$ via the above arguments.
All this exactly mirrors what happened for the $\Gthree$ case.

We finish with a brief look at the equations of motion.
The field strength $\Fc_{\mu\nu\rho\sigma}$ of the gauge field $\Ac_{\mu\nu\rho}$ only appears in the topological term.
This gauge field also appears in the field strength $\Fb_{\mu\nu\rho}$.
Its equation of motion has the form $\partial_{\fM \fN} \theta^{\mu\nu\rho\fN}=0$ where
\be
\theta^{\mu\nu\rho \fM}  \equiv \sqrt{g} m^{\fM \fP} \Fb^{\mu\nu\rho}{}_{\fP}  + \tfrac{1}{4!} \epsilon^{\mu\nu\rho \sigma_1\dots \sigma_4} \Fc_{\sigma_1\dots\sigma_4}{}^{\fM} \,.
\label{5duality}
\ee
Meanwhile the equation of motion of $\Ab_{\mu\nu \fM}$ is
\be
\D_\rho ( \sqrt{g} m^{\fM \fN} \Fb^{\mu\nu\rho}{}_{\fN} ) 
+ \tfrac{1}{8} \epsilon^{\fM \fP \fQ \fK \fL} \partial_{\fP \fQ} ( \sqrt{g} \gM_{\fK \fL, \fK^\prime \fL^\prime} \Fa^{\mu\nu \fK^\prime \fL^\prime} )
 - \tfrac{2}{4!}\epsilon^{\mu\nu \lambda_1 \dots \lambda_5} \Fa_{\lambda_1 \lambda_2}{}^{\fM\fN} \Fb_{\lambda_3 \dots \lambda_5 \fN} 
 =0\,.
 \label{Beom5}
\ee
The $\fM=5$ component combines with the $\fM=5$ component of the Bianchi identity \eqref{5BianchiC} to give $\D_\rho \theta^{\mu\nu \rho 5} = 0$.
Hence we integrate and set $\theta^{\mu\nu\rho \fM} = 0$.
Let's examine the content of this constraint. 
Firstly, the $\theta^{\mu\nu\rho 5}$ component implies
\be
  \Omega^{-1/5} \sqrt{g} h^j \Fdef^{\mu\nu\rho}{}_{j} - \tfrac{1}{4!} \epsilon^{\mu\nu\rho \sigma_1\dots \sigma_4} \Fdef_{\sigma_1\dots\sigma_4} = 0
\label{5duality_5}
\ee
This is the 11-dimensional self-duality constraint \eqref{magic} on the transverse part of the four-form field strength, here decomposed as in \eqref{Constraints4}. 
Secondly, setting $\theta^{\mu\nu\rho i}-C^i \theta^{\mu\nu\rho 5}=0$ and projecting gives {
\be
\begin{split}
\sqrt{g}  \Omega^{-1/5}  \Fdef^{\mu\nu\rho}{}_{ijkl} 
+ \tfrac{1}{4!} \epsilon^{\mu\nu\rho\sigma_1 \dots \sigma_4} 4 h_{[i|} \Fdef_{\sigma_1 \dots \sigma_4 |jkl]} =0\,,\\
 \sqrt{g} \Omega^{4/5} \tau^{i A} \Fdef^{\mu\nu\rho}{}_i
-   \tfrac{1}{4!} \epsilon^{\mu\nu\rho\sigma_1 \dots \sigma_4} \tau_i^A \tfrac{1}{3!} \epsilon^{ijkl}  \Fdef_{\sigma_1 \dots \sigma_4 jkl}=0\,.
\end{split} 
\label{5duality_i}
\ee
The first of these is part of the self-duality condition \eqref{F7constraint} obeyed by the totally longitudinal part of the dual-seven form.
The second is part of the duality between the partly longitudinal four-form and the rest of the seven-form.
We see again that the ExFT rearrangement of degrees of freedom exactly captures the novel features of the eleven-dimensional non-relativistic limit.

\bibliography{CurrentBib}

\end{document}